


\font\mayusc=cmcsc10 


      \font \ninebf                 = cmbx9
      \font \ninei                  = cmmi9
      \font \nineit                 = cmti9
      \font \ninerm                 = cmr9
      \font \ninesans               = cmss10 at 9pt
      \font \ninesl                 = cmsl9
      \font \ninesy                 = cmsy9
      \font \ninett                 = cmtt9
      \font \fivesans               = cmss10 at 5pt
						\font \sevensans              = cmss10 at 7pt
      \font \sixbf                  = cmbx6
      \font \sixi                   = cmmi6
      \font \sixrm                  = cmr6
						\font \sixsans                = cmss10 at 6pt
      \font \sixsy                  = cmsy6
      \font \tams                   = cmmib10
      \font \tamss                  = cmmib10 scaled 700
						\font \tensans                = cmss10
    
      \skewchar\ninei='177 \skewchar\sixi='177
      \skewchar\ninesy='60 \skewchar\sixsy='60
      \hyphenchar\ninett=-1
      \def\newline{\hfil\break}%
      \catcode`@=11
      \def\folio{\ifnum\pageno<\z@
      \uppercase\expandafter{\romannumeral-\pageno}%
      \else\number\pageno \fi}
      \catcode`@=12 

      \newfam\sansfam
      \textfont\sansfam=\tensans\scriptfont\sansfam=\sevensans
      \scriptscriptfont\sansfam=\fivesans
      \def\sans{\fam\sansfam\tensans}


      \def\petit{\def\rm{\fam0\ninerm}%
      \textfont0=\ninerm \scriptfont0=
\sixrm \scriptscriptfont0=\fiverm
       \textfont1=\ninei \scriptfont1=
\sixi \scriptscriptfont1=\fivei
       \textfont2=\ninesy \scriptfont2=
\sixsy \scriptscriptfont2=\fivesy
       \def\it{\fam\itfam\nineit}%
       \textfont\itfam=\nineit
       \def\sl{\fam\slfam\ninesl}%
       \textfont\slfam=\ninesl
       \def\bf{\fam\bffam\ninebf}%
       \textfont\bffam=\ninebf \scriptfont\bffam=\sixbf
       \scriptscriptfont\bffam=\fivebf
       \def\sans{\fam\sansfam\ninesans}%
       \textfont\sansfam=\ninesans \scriptfont\sansfam=\sixsans
       \scriptscriptfont\sansfam=\fivesans
       \def\tt{\fam\ttfam\ninett}%
       \textfont\ttfam=\ninett
       \normalbaselineskip=11pt
       \setbox\strutbox=\hbox{\vrule height7pt depth2pt width0pt}%
       \normalbaselines\rm


      \def\bvec##1{{\textfont1=\tbms\scriptfont1=\tbmss
      \textfont0=\ninebf\scriptfont0=\sixbf
      \mathchoice{\hbox{$\displaystyle##1$}}{\hbox{$\textstyle##1$}}
      {\hbox{$\scriptstyle##1$}}{\hbox{$\scriptscriptstyle##1$}}}}}


\def\imag{\mathop{\rm Im}}

.

					\mathchardef\gammav="0100
     \mathchardef\deltav="0101
     \mathchardef\thetav="0102
     \mathchardef\lambdav="0103
     \mathchardef\xiv="0104
     \mathchardef\piv="0105
     \mathchardef\sigmav="0106
     \mathchardef\upsilonv="0107
     \mathchardef\phiv="0108
     \mathchardef\psiv="0109
     \mathchardef\omegav="010A


					\mathchardef\Gammav="0100
     \mathchardef\Deltav="0101
     \mathchardef\Thetav="0102
     \mathchardef\Lambdav="0103
     \mathchardef\Xiv="0104
     \mathchardef\Piv="0105
     \mathchardef\Sigmav="0106
     \mathchardef\Upsilonv="0107
     \mathchardef\Phiv="0108
     \mathchardef\Psiv="0109
     \mathchardef\Omegav="010A



\font\grbfivefm=cmbx5
\font\grbsevenfm=cmbx7
\font\grbtenfm=cmbx10 
\newfam\grbfam
\textfont\grbfam=\grbtenfm
\scriptfont\grbfam=\grbsevenfm
\scriptscriptfont\grbfam=\grbfivefm

\font\calbfivefm=cmbsy10 at 5pt
\font\calbsevenfm=cmbsy10 at 7pt
\font\calbtenfm=cmbsy10 
\newfam\calbfam
\textfont\calbfam=\calbtenfm
\scriptfont\calbfam=\calbsevenfm
\scriptscriptfont\calbfam=\calbfivefm



      \def\bvec#1{{\textfont1=\tams\scriptfont1=\tamss
      \textfont0=\tenbf\scriptfont0=\sevenbf
      \mathchoice{\hbox{$\displaystyle#1$}}{\hbox{$\textstyle#1$}}
      {\hbox{$\scriptstyle#1$}}{\hbox{$\scriptscriptstyle#1$}}}}



\def\pmbf#1{\leavevmode\setbox0=\hbox{#1}%
\kern-.02em\copy0\kern-\wd0
\kern.04em\copy0\kern-\wd0
\kern-.02em\copy0\kern-\wd0
\kern-.03em\copy0\kern-\wd0
\kern.06em\box0 }



						\def\monthname{%
   			\ifcase\month
      \or Jan\or Feb\or Mar\or Apr\or May\or Jun%
      \or Jul\or Aug\or Sep\or Oct\or Nov\or Dec%
   			\fi
							}%
					\def\timestring{\begingroup
   		\count0 = \time
   		\divide\count0 by 60
   		\count2 = \count0   
   		\count4 = \time
   		\multiply\count0 by 60
   		\advance\count4 by -\count0   
   		\ifnum\count4<10
     \toks1 = {0}%
   		\else
     \toks1 = {}%
   		\fi
   		\ifnum\count2<12
      \toks0 = {a.m.}%
   		\else
      \toks0 = {p.m.}%
      \advance\count2 by -12
   		\fi
   		\ifnum\count2=0
      \count2 = 12
   		\fi
   		\number\count2:\the\toks1 \number\count4 \thinspace \the\toks0
					\endgroup}%

				\def\timestamp{\number\day\space
\monthname\space\number\year\quad\timestring}%
				\newskip\abovelistskip      \abovelistskip = .5\baselineskip 
				\newskip\interitemskip      \interitemskip = 0pt
				\newskip\belowlistskip      \belowlistskip = .5\baselineskip
				\newdimen\listleftindent    \listleftindent = 0pt
				\newdimen\listrightindent   \listrightindent = 0pt

				%


\def\petit{\def\rm{\fam0\ninerm}%
\textfont0=\ninerm \scriptfont0=\sixrm \scriptscriptfont0=\fiverm
\textfont1=\ninei \scriptfont1=\sixi \scriptscriptfont1=\fivei
\textfont2=\ninesy \scriptfont2=\sixsy \scriptscriptfont2=\fivesy
       \def\it{\fam\itfam\nineit}%
       \textfont\itfam=\nineit
       \def\sl{\fam\slfam\ninesl}%
       \textfont\slfam=\ninesl
       \def\bf{\fam\bffam\ninebf}%
       \textfont\bffam=\ninebf \scriptfont\bffam=\sixbf
       \scriptscriptfont\bffam=\fivebf
       \def\sans{\fam\sansfam\ninesans}%
       \textfont\sansfam=\ninesans \scriptfont\sansfam=\sixsans
       \scriptscriptfont\sansfam=\fivesans
       \def\tt{\fam\ttfam\ninett}%
       \textfont\ttfam=\ninett
       \normalbaselineskip=11pt
       \setbox\strutbox=\hbox{\vrule height7pt depth2pt width0pt}%
       \normalbaselines\rm
      \def\vec##1{{\textfont1=\tbms\scriptfont1=\tbmss
      \textfont0=\ninebf\scriptfont0=\sixbf
      \mathchoice{\hbox{$\displaystyle##1$}}{\hbox{$\textstyle##1$}}
      {\hbox{$\scriptstyle##1$}}{\hbox{$\scriptscriptstyle##1$}}}}}

      \def\footnoterule{\kern-3pt\hrule width 2cm\kern2.6pt}
      \newdimen\oldparindent\oldparindent=1.5em
      \parindent=1.5em
 
\newcount\footcount \footcount=0
\def\advftncnt{\advance\footcount by1\global\footcount=\footcount}
      \def\fnote#1{\advftncnt$^{\the\footcount}$\begingroup\petit
      \parfillskip=0pt plus 1fil
      \def\textindent##1{\hangindent0.5\oldparindent\noindent\hbox
      to0.5\oldparindent{##1\hss}\ignorespaces}%
 \vfootnote{$^{\the\footcount}$}
{#1\nullbox{0mm}{2mm}{0mm}\vskip-9.69pt}\endgroup}


      \def\item#1{\par\noindent
      \hangindent6.5 mm\hangafter=0
      \llap{#1\enspace}\ignorespaces}
      
      \def\leaderfill{\kern0.5em\leaders
\hbox to 0.5em{\hss.\hss}\hfill\kern
      0.5em}
						\def\hb{\hfill\break}
						
						\def\tdots{$\,\ldots\,$}
    \def\centerrule#1{\centerline{\kern#1\hrulefill\kern#1}}


      \def\boxit#1{\vbox{\hrule\hbox{\vrule\kern3pt
						\vbox{\kern3pt#1\kern3pt}\kern3pt\vrule}\hrule}}

      \def\tightboxit#1{\vbox{\hrule\hbox{\vrule
						\vbox{#1}\vrule}\hrule}}

      \def\looseboxit#1{\vbox{\hrule\hbox{\vrule\kern5pt
						\vbox{\kern5pt#1\kern5pt}\kern5pt\vrule}\hrule}}

      \def\youboxit#1#2{\vbox{\hrule\hbox{\vrule\kern#2
						\vbox{\kern#2#1\kern#2}\kern#2\vrule}\hrule}}




			\def\whitetile#1#2#3{\setbox0=\null
			\ht0=#1 \dp0=#2\wd0=#3 \setbox1=
\hbox{\tightboxit{\box0}}\lower#2\box1}

			\def\nullbox#1#2#3{\setbox0=\null
			\ht0=#1 \dp0=#2\wd0=#3\box0}




\def\fig{\leavevmode Fig.}

\def\equ{\leavevmode Eq.}
\def\eqs{\leavevmode Eqs.}

\def\sect{\leavevmode Sect.}
\def\subsect{\leavevmode Subsect.}
\def\subsects{\leavevmode Subsects.}
\def\sects{\leavevmode Sects.}

\def\equn#1{\ifmmode \eqno{\rm #1}\else \equ~#1\fi}



\def\tev{\ifmmode \mathop{\rm TeV}\nolimits\else {\rm TeV}\fi}
\def\gev{\ifmmode \mathop{\rm GeV}\nolimits\else {\rm GeV}\fi}
\def\mev{\ifmmode \mathop{\rm MeV}\nolimits\else {\rm MeV}\fi}
\def\kev{\ifmmode \mathop{\rm keV}\nolimits\else {\rm keV}\fi}
\def\ev{\ifmmode \mathop{\rm eV}\nolimits\else {\rm eV}\fi}

\def\chidof{\ifmmode
\mathop\chi^2/{\rm d.o.f.}\else $\chi^2/{\rm d.o.f.}\null$\fi}

\def\msbar{\ifmmode
\mathop{\overline{\rm MS}}\else$\overline{\rm MS}$\null\fi}


\def\physmatex{P\kern-.14em\lower.5ex\hbox{\sevenrm H}ys
\kern -.35em \raise .6ex \hbox{{\sevenrm M}a}\kern -.15em
 T\kern-.1667em\lower.5ex\hbox{E}\kern-.125emX\null}%

\def\ref#1{$^{[#1]}$\relax}

\def\prajnyp#1#2#3#4#5{
\frenchspacing{\mayusc #1}, {\sl#2}, {\bf #3}, {#5} {(#4)}}





\def\simeqsub{\mathop{\simeq}\limits}








\def\ddal{\mathop{\vrule height 7pt depth0.2pt
\hbox{\vrule height 0.5pt depth0.2pt width 6.2pt}
\vrule height 7pt depth0.2pt width0.8pt
\kern-7.4pt\hbox{\vrule height 7pt depth-6.7pt width 7.pt}}}
\def\sdal{\mathop{\kern0.1pt\vrule height 4.9pt depth0.15pt
\hbox{\vrule height 0.3pt depth0.15pt width 4.6pt}
\vrule height 4.9pt depth0.15pt width0.7pt
\kern-5.7pt\hbox{\vrule height 4.9pt depth-4.7pt width 5.3pt}}}
\def\ssdal{\mathop{\kern0.1pt\vrule height 3.8pt depth0.1pt width0.2pt
\hbox{\vrule height 0.3pt depth0.1pt width 3.6pt}
\vrule height 3.8pt depth0.1pt width0.5pt
\kern-4.4pt\hbox{\vrule height 4pt depth-3.9pt width 4.2pt}}}




\mathchardef\lap='0001


\def\lsim{\mathop{\setbox0=\hbox{$\displaystyle 
\raise2.2pt\hbox{$\;<$}\kern-7.7pt\lower2.6pt\hbox{$\sim$}\;$}
\box0}}
\def\gsim{\mathop{\setbox0=\hbox{$\displaystyle 
\raise2.2pt\hbox{$\;>$}\kern-7.7pt\lower2.6pt\hbox{$\sim$}\;$}
\box0}}

\def\gsimsub#1{\mathord{\vtop to0pt{\ialign{##\crcr
$\hfil{{\mathop{\setbox0=\hbox{$\displaystyle 
\raise2.2pt\hbox{$\;>$}\kern-7.7pt\lower2.6pt\hbox{$\sim$}\;$}
\box0}}}\hfil$\crcr\noalign{\kern1.5pt\nointerlineskip}
$\hfil\scriptstyle{#1}{}\kern1.5pt\hfil$\crcr}\vss}}}

\def\lsimsub#1{\mathord{\vtop to0pt{\ialign{##\crcr
$\hfil\displaystyle{\mathop{\setbox0=\hbox{$\displaystyle 
\raise2.2pt\hbox{$\;<$}\kern-7.7pt\lower2.6pt\hbox{$\sim$}\;$}
\box0}}
\def\gsim{\mathop{\setbox0=\hbox{$\displaystyle 
\raise2.2pt\hbox{$\;>$}\kern-7.7pt\lower2.6pt\hbox{$\sim$}\;$}
\box0}}\hfil$\crcr\noalign{\kern1.5pt\nointerlineskip}
$\hfil\scriptstyle{#1}{}\kern1.5pt\hfil$\crcr}\vss}}}

\def\ii{{\rm i}}
\def\dd{{\rm d}}

\def\ee{{\rm e}}





\def\frac#1#2{{#1\over#2}}
\def\dfrac#1#2{{\displaystyle{#1\over#2}}}
\def\tfrac#1#2{{\textstyle{#1\over#2}}}
\def\ffrac#1#2{\leavevmode
   \kern.1em \raise .5ex \hbox{\the\scriptfont0 #1}%
   \kern-.1em $/$%
   \kern-.15em \lower .25ex \hbox{\the\scriptfont0 #2}%
}%



\def\brochureb#1#2#3{\pageno#3
\headline={\ifodd\pageno{\rheadline}
\else\lheadline\fi}
\def\rheadline{\hfil -{#2}-\hfil}
\def\lheadline{\hfil-{#1}-\hfil}
\footline={\hss -- \number\pageno\ --\hss}
\voffset=2\baselineskip}

\def\nada{\phantom{M}\kern-1em}
\def\brochureendcover#1{\vfill\eject\pageno=1{\nada#1}\vfill\eject}





\def\chapterb#1#2#3{\pageno#3
\headline={\ifodd\pageno{\ifnum\pageno=#3\hfil\else\rheadline\fi}
\else\lheadline\fi}
\def\rheadline{\hfil -{#2}-\hfil}
\def\lheadline{\hfil-{#1}-\hfil}
\footline={\hss -- \number\pageno\ --\hss}
\voffset=2\baselineskip}


\def\bookendchapter{\ifodd\pageno
 \vfill\eject\footline={\hfill}\headline={\hfill}\null \vfill\eject
 \else\vfill\eject \fi}

\def\obookendchapter{\ifodd\pageno\vfill\eject
 \else\vfill\eject\null \vfill\eject\fi}


\def\booksection#1{
\setbox0=\vbox{\hsize=0.85\hsize\tolerance=500\raggedright\hfuzz=6mm
\noindent{\medfib #1}\medskip}\goodbreak\vskip0.6cm\box0
\nobreak
\noindent}
\def\booksubsection#1{
\setbox0=\vbox{\hsize=0.85\hsize\tolerance=400\raggedright\hfuzz=4mm
\noindent{\fib #1}\smallskip}\goodbreak\vskip0.45cm\box0
\nobreak
\noindent}




\def\figurasc#1#2{\petit{\noindent\sc#1}\ #2}

\def\captiontype{\tolerance=800\hfuzz=1mm\raggedright\noindent}



\def\abstracttype#1{
\hsize0.7\hsize\tolerance=800\hfuzz=0.5mm \noindent{\fib #1}\par
\medskip\petit}


\def\hb{\hfill\break}


\font\twelverm=cmr12 
\font\smallsc=cmcsc10 at 9pt 
\font\fib=cmfib8
\font\medfib=cmfib8 at 9pt


\font\sc=cmcsc10 

\font\addressfont=cmbxti10 at 9pt


\catcode`@=11 

\newdimen\pagewidth \newdimen\pageheight \newdimen\ruleht
 \maxdepth=2.2pt  \parindent=3pc
\pagewidth=\hsize \pageheight=\vsize \ruleht=.4pt
\abovedisplayskip=6pt plus 3pt minus 1pt
\belowdisplayskip=6pt plus 3pt minus 1pt
\abovedisplayshortskip=0pt plus 3pt
\belowdisplayshortskip=4pt plus 3pt

\newinsert\margin
\dimen\margin=\maxdimen




\newdimen\paperheight \paperheight = \vsize
\def\topmargin{\afterassignment\@finishtopmargin \dimen0}%
\def\@finishtopmargin{%
  \dimen2 = \voffset		
  \voffset = \dimen0 \advance\voffset by -1in
  \advance\dimen2 by -\voffset	
  \advance\vsize by \dimen2	
}%
\def\advancetopmargin{%
  \dimen0 = 0pt \afterassignment\@finishadvancetopmargin \advance\dimen0
}%
\def\@finishadvancetopmargin{%
  \advance\voffset by \dimen0
  \advance\vsize by -\dimen0
}%
\def\bottommargin{\afterassignment\@finishbottommargin \dimen0}%
\def\@finishbottommargin{%
  \@computebottommargin		
  \advance\dimen2 by -\dimen0	
  \advance\vsize by \dimen2	
}%
\def\advancebottommargin{%
  \dimen0 = 0pt\afterassignment\@finishadvancebottommargin \advance\dimen0
}%
\def\@finishadvancebottommargin{%
  \advance\vsize by -\dimen0
}%
\def\@computebottommargin{%
  \dimen2 = \paperheight	
  \advance\dimen2 by -\vsize	
  \advance\dimen2 by -\voffset	
  \advance\dimen2 by -1in	
}%
\newdimen\paperwidth \paperwidth = \hsize
\def\leftmargin{\afterassignment\@finishleftmargin \dimen0}%
\def\@finishleftmargin{%
  \dimen2 = \hoffset		
  \hoffset = \dimen0 \advance\hoffset by -1in
  \advance\dimen2 by -\hoffset	
  \advance\hsize by \dimen2	
}%
\def\advanceleftmargin{%
  \dimen0 = 0pt \afterassignment\@finishadvanceleftmargin \advance\dimen0
}%
\def\@finishadvanceleftmargin{%
  \advance\hoffset by \dimen0
  \advance\hsize by -\dimen0
}%
\def\rightmargin{\afterassignment\@finishrightmargin \dimen0}%
\def\@finishrightmargin{%
  \@computerightmargin		
  \advance\dimen2 by -\dimen0	
  \advance\hsize by \dimen2	
}%
\def\advancerightmargin{%
  \dimen0 = 0pt \afterassignment\@finishadvancerightmargin \advance\dimen0
}%
\def\@finishadvancerightmargin{%
  \advance\hsize by -\dimen0
}%
\def\@computerightmargin{%
  \dimen2 = \paperwidth		
  \advance\dimen2 by -\hsize	
  \advance\dimen2 by -\hoffset	
  \advance\dimen2 by -1in	
}%

\def\onepageout#1{\shipout\vbox{ 
    \offinterlineskip 
    \vbox to 3pc{ 
      \iftitle 
        \global\titlefalse 
        \setcornerrules 
      \else\ifodd\pageno \rightheadline\else\leftheadline\fi\fi
      \vfill} 
    \vbox to \pageheight{
      \ifvoid\margin\else 
        \rlap{\kern31pc\vbox to\z@{\kern4pt\box\margin \vss}}\fi
      #1 
      \ifvoid\footins\else 
        \vskip\skip\footins \kern-3pt
        \hrule height\ruleht width\pagewidth \kern-\ruleht \kern3pt
        \unvbox\footins\fi
      \boxmaxdepth=\maxdepth
      } 
    }
  \advancepageno}

\def\setcornerrules{\hbox to \pagewidth{\vrule width 1pc height\ruleht
    \hfil \vrule width 1pc}
  \hbox to \pagewidth{\llap{\sevenrm(page \folio)\kern1pc}%
    \vrule height1pc width\ruleht depth\z@
    \hfil \vrule width\ruleht depth\z@}}
\newbox\partialpage




\hyphenation{ha-ya-ka-wa acha-sov}


\input epsf.sty
\raggedbottom
\footline={\hfill}
\rightline{Revised: \timestamp}
\smallskip
\rightline{FTUAM 01-08 }
\rightline{May,  2001}
\rightline{(hep-ph/0106025)}
\bigskip
\hrule height .3mm
\vskip.6cm
\centerline{{\twelverm Precision
 Determination of  the Pion Form Factor  and Calculation of 
the Muon $g-2$}}
\medskip
\centerrule{.7cm}
\vskip1cm

\setbox9=\vbox{\hsize65mm {\noindent\fib 
J. F. de Troc\'oniz and F. J. 
Yndur\'ain} 
\vskip .1cm
\noindent{\addressfont Departamento de F\'{\i}sica Te\'orica, C-XI,\hb
 Universidad Aut\'onoma de Madrid,\hb
 Canto Blanco,\hb
E-28049, Madrid, Spain.}\hb}
\smallskip
\centerline{\box9}
\bigskip
\setbox0=\vbox{\abstracttype{Abstract} 
We perform a new calculation of the hadronic contributions, $a({\rm Hadronic})$
 to the anomalous magnetic moment of the muon, $a_\mu$. 
For the low energy contributions of order $\alpha^2$ we carry over an analysis of
 the pion form factor $F_\pi(t)$ using recent data both on $e^+e^-\to\pi^+\pi^-$ and 
$\tau^+\to \bar{\nu}_\tau \pi^+\pi^0$. In this analysis we take into account that the phase of 
the form factor is equal to that of $\pi\pi$ scattering. 
This allows us to profit fully from analyticity properties 
so we can  use also experimental information on $F_\pi(t)$ at spacelike $t$. 
At higher energy we use QCD to supplement experimental 
data, including the recent measurements of $e^+e^-\to\;{\rm hadrons}$ 
both around 1 GeV and near the $\bar{c}c$ threshold. 
This yields a precise determination of the $O(\alpha^2)$   and  $O(\alpha^2)+O(\alpha^3)$ 
hadronic part of the photon vacuum polarization pieces,
$$10^{11}\times a^{(2)}({\rm h.v.p.})=6\,909\pm64;\quad
10^{11}\times  a^{(2+3)}({\rm h.v.p.})=7\,002\pm66$$
As  byproducts we also get the masses and widths of the $\rho^0,\,\rho^+$, and 
very accurate values for the charge radius and second coefficient of the pion.

Adding the remaining order $\alpha^3$ hadronic contributions we find
$$10^{11}\times a^{\rm theory}(\hbox{Hadronic})=
6\,993\pm69\quad(e^+e^-\,+\,\tau\,+\,{\rm spacel.})$$ 
The error above includes
 statistical, systematic and estimated  theoretical errors. 
The figures given are obtained including $\tau$ decay data; 
if we restrict ourselves to $e^+e^-$ data, 
slightly lower values and somewhat higher errors are found. 

This is to be compared with the figure obtained by subtracting pure electroweak 
contributions from the recent experimental value,
obtained from measurements of the muon gyromagnetic ratio ($g-2$), which reads
$$10^{11}\times a^{\rm exp.}(\hbox{Hadronic})=7\,174\pm150.$$ 
}
\centerline{\box0}
\brochureendcover{Typeset with \physmatex}
\pageno=1
\brochureb{\smallsc j. f. de troc\'oniz and f. j.  yndur\'ain}{\smallsc precision
 determination of  the pion form factor  and calculation of 
the muon $g-2$}{1}

\booksection{1. Introduction}
The appearance of a new, very precise measurement of the muon magnetic
 moment\ref{1} has triggered the interest in {\sl theoretical} calculations 
of this quantity. 
Particularly, because the experimental figure (we give the result 
for the anomaly, averaged with older determinations\ref{2})
$$10^{11}\times a_\mu(\hbox{exp.})=116\,592\,030\pm150
\equn{(1.1)}$$
lies slightly above theoretical evaluations based on the standard 
model, as much as $2.6\sigma$ in some cases.

It should be noted that all modern\fnote{By modern we here mean, somewhat arbitrarily,  
those obtained  since 1985. 
A more complete list of references, including earlier work, may be found in ref.~7.} 
theoretical determinations\ref{3-7} 
are compatible among themselves within errors (of order $100\times10^{-11}$) 
and that, with few exceptions, they are 
also compatible with the experimental 
result, (1.1), at the  level of $1.5\sigma$ or less. Because of this, it is our feeling 
that a new, {\sl complete} evaluation would be welcome since, 
in fact, there exists as yet no calculation  
that takes fully into account all theoretical constraints and
 all the new experimental data. These experimental 
data allow an improved evaluation of the low energy hadronic contributions 
to $a_\mu$, both directly from $e^+e^-$
 annihilations (in the $\rho$ region\ref{8} and around the $\phi$ 
resonance\ref{9})  
and, indirectly, from $\tau$ decays\ref{10} and, also indirectly, 
from measurements of the pion form factor in the spacelike region.\ref{11} 
Moreover,  the BES\ref{12} data, covering 
$e^+e^-$ annihilations in the vicinity of $\bar{c}c$ threshold,  
permit a reliable evaluation of the corresponding hadronic pieces. 
In fact, the main improvements of the present paper are the calculation of
 the two pion contribution to the hadronic part of $a_\mu$,  
using all available experimental information and fulfilling 
compatibility with all our theoretical knowledge, 
and the pinning down of the multipion, $KK$ and $\bar{c}c$ contributions. 
This we do in \sects~3,~4 (in \sect~2 we formulate the problem). 
In \sect~5 we discuss other hadronic corrections, 
including one that, as far as we know, has been hitherto neglected, 
and which, though small ($\sim46\times10^{-11}$) 
is relevant at the level of accuracy for which we are striving.
 Finally, in \sect~6 we discuss our results and compare them with  experiment.

The main outcome of our analysis is an accurate and reliable determination of the  
hadronic contributions to $a_\mu$ at order $\alpha^2$. 
In fact, in all regions where there are difficulties 
we perform at least two evaluations, and take into account  
their consistency (or lack thereof). 
Furthermore, we discuss in some detail 
(including ambiguities) the $O(\alpha^3)$ hadronic contributions.

As a byproduct of the low energy calculations we 
can also give precise values for the $\rho^0$, $\rho^+$ 
masses and widths,
$$\eqalign{
m_{\rho^0}=772.6\pm0.5\;\mev,\quad \gammav_{\rho^0}=147.4\pm0.8\,\mev;\cr
m_{\rho^+}=773.8\pm0.6\;\mev,\quad \gammav_{\rho^+}=147.3\pm0.9\,\mev;\cr
}\equn{(1.2)}$$
for the P-wave $\pi\pi$ scattering 
length,
$$a_1^1=(41\pm2)\times10^{-3}\,m_{\pi}^{-3},\equn{(1.3)}$$
and for the pion mean squared charge radius and second coefficient:
$$\eqalign{\langle r^2_\pi\rangle=0.435\pm0.002\;{\rm fm}^2,\;c_\pi=3.60\pm0.03\;\gev^{-4}&
\quad (e^+e^-\,+\,\tau\,+\,{\rm spacelike}).\cr
\langle r^2_\pi\rangle=0.433\pm0.002\;{\rm fm}^2,\;c_\pi=3.58\pm0.04\;\gev^{-4}&
\quad (e^+e^-\,+\,{\rm spacelike}).\cr
}\equn{(1.4)}$$
We give results both using only direct results on $F_\pi$, from 
$e^+e^-$ annihilations, or involving also the  
decay $\tau^+\to\bar{\nu}_\tau\pi^+\pi^0$, which last we consider to be our best estimates.
 
So we write,
$$10^{11}\times a(\hbox{Hadronic})=
\cases{
6\,993\pm69\quad(e^+e^-\,+\,\tau\,+\,{\rm spacel.})\cr
6\,973\pm99\quad(e^+e^-\,+\,{\rm spacel.})\cr
}
\equn{(1.5)}$$  

Note that 
in $a(\hbox{Hadronic})$ we include {\sl all} 
hadronic contributions, $O(\alpha^3)$ as well as $O(\alpha^2)$.
 The errors include  the statistical errors, as well as the  
  estimated systematic and theoretical ones. 
This is
to be compared with the value deduced from (1.1) and electroweak corrections 
$$10^{11}\times a^{\rm exp.}(\hbox{Hadronic})=7\,174\pm150,$$
from which (1.5) differs by  $1.1\sigma$.

\booksection{2. Contributions to $a_\mu$}
We divide the various contributions to $a_\mu$ as follows:
$$a_\mu=a(\hbox{QED})+a(\hbox{Weak})+a(\hbox{Hadronic}).
\equn{(2.1)}$$
Here $a(\hbox{QED})$ denote the pure quantum electrodynamics corrections, and 
$a(\hbox{Weak})$ are the ones due to $W,\,Z$ and Higgs exchange. 
The hadronic contributions can, in turn, be split as
$$a(\hbox{Hadronic})=a^{(2)}(\hbox{h.v.p.})+a(\hbox{Other hadronic, $O(\alpha^3)$}).
\equn{(2.2)}$$ 
$a^{(2)}(\hbox{h.v.p.})$ are the corrections due to the hadronic
 photon vacuum polarization  contributions  (\fig~1), 
nominally of  order $\alpha^2$ (see \subsects~3.3 and 5.2  for a qualification of this statement). 
We will discuss in detail the ``Other hadronic,  $O(\alpha^3)$" in \sect~5.
 
\topinsert{
\setbox0=\vbox{\hsize9.truecm{\epsfxsize 7.4truecm\epsfbox{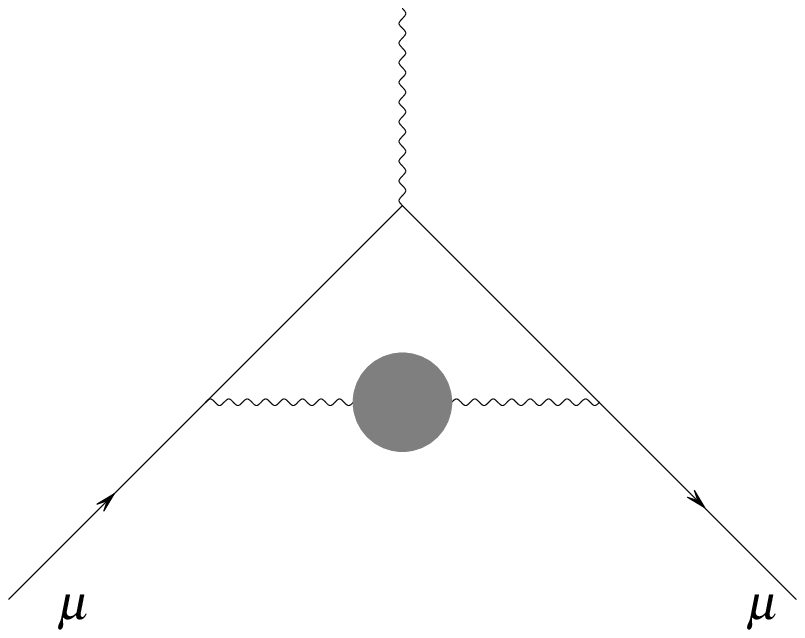}}} 
\setbox6=\vbox{\hsize 5.5truecm\captiontype\figurasc{Figure 1. }{The order $\alpha^2$
hadronic contributions to the muon magnetic moment. 
The blob represents an arbitrary hadronic state.\hb
\phantom{XX}}\hb
\vskip.1cm} 
\medskip
\line{\box0\hfil\box6}
\medskip
}\endinsert

According to the review of Hughes and Kinoshita\ref{13} one has
$$\eqalign{10^{11}\times a(\hbox{QED})=116\,584\,705&\pm1.8\cr
10^{11}\times a(\hbox{Weak})=151&\pm4.\cr
}
\equn{(2.3)}$$

There is no dispute about these numbers. 
If we combine them with (1.1), we can convert this into 
a measurement of the hadronic part of the anomaly:
$$10^{11}\times a^{\rm exp.}(\hbox{Hadronic})=7\,174\pm150.\equn{(2.4)}$$ 
Our task in the present paper is the evaluation of this quantity.

We now say a few words about the  piece $a^{(2)}(\hbox{h.v.p.})$, which is 
the most important component of $a(\hbox{Hadronic})$.
 As  Brodsky and de~Rafael\ref{14} have shown,  
 it can be written as
$$\eqalign{a^{(2)}(\hbox{h.v.p.})=&12\pi\int_{4m^2_\pi}^\infty \dd t\,K(t)\imag\piv(t),\cr
K(t)=&\dfrac{\alpha^2}{3\pi^2t}\hat{K}(t);\quad \hat{K}(t)=
\int_0^1\dd x\,\dfrac{x^2(1-x)}{x^2+(1-x)t/m^2_\mu}.\cr}
\equn{(2.4a)}$$
Here $\piv$ is the hadronic part of the photon 
vacuum polarization function. 
An alternate formula is obtained by expressing $\imag \Piv$ in terms of 
the ratio
$$R(t)=\dfrac{\sigma^{(0)}(e^+e^-\to{\rm hadrons})}{\sigma^{(0)}(e^+e^-\to\mu^+\mu^-)},
\quad \sigma^{(0)}(e^+e^-\to\mu^+\mu^-)\equiv\dfrac{4\pi\alpha^2}{3t}:$$
$$a^{(2)}(\hbox{h.v.p.})=\int_{4m^2_\pi}^\infty \dd t\,K(t) R(t).
\equn{(2.4b)}$$
The superindex (0) here means `lowest order in the electromagnetic 
interactions'.
 
At low energy ($t\leq 0.8\,\gev^2$) we can separate the contribution from three pion states
 and that from two pions. The first will be discussed in  \sect~4. 
The two pion contribution in turn can be expressed in terms of the 
pion form factor, $F_\pi$,
$$\imag \piv_{2\pi}(t)=\dfrac{1}{48\pi}\left(1-\dfrac{4m^2_\pi}{t}\right)^{3/2} 
|F_\pi(t)|^2,\equn{(2.5)}$$
so that,
for the two-pion contribution up to energy squared $t_0$,
$$a_\mu(2\pi; t_0)=\tfrac{1}{4}\int_{4m^2_\pi}^{t_0} \dd t\;
\left(1-\dfrac{4m^2_\pi}{t}\right)^{3/2}K(t)|F_\pi(t)|^2.
\equn{(2.6)}$$

\booksection{3. The pion form  factor}
\vskip-0.7truecm
\booksubsection{3.1. Theory}
The evaluation of the pion form factor is slightly complicated by the phenomenon 
of $\omega-\rho$ interference. This can be solved by considering only the 
isospin $I=1$ component, and adding later the $\omega\to2\pi$ 
and interference 
in the standard Gounaris--Sakurai way. 
This is equivalent to neglecting, in a first approximation, the breaking of isospin 
invariance. We will also neglect for now  electromagnetic corrections. 
In this approximation the properties of $F_\pi(t)$ are the following: 
\item{(i) }{$F_\pi(t)$ is an analytic function of $t$, with a cut 
from $4m^2_\pi$ to infinity.}
\item{(ii) }{On the cut, the phase of $F_\pi(t)$ is, because of unitarity, identical to 
that of the P-wave, $I=1$, $\pi\pi$ scattering, $\delta_1^1(t)$, and 
this equality 
holds until the opening of the inelastic threshold at $t=t_0$
 (Fermi--Watson final state interaction 
theorem).}
\item{(iii) }{For large $t$, $F_\pi(t)\simeq 1/t$. Actually, one knows the coefficient of this 
behaviour, but we will not need it here.}
\item{(iv) }{$F(0)=1$.}
 
The inelastic threshold occurs, rigorously speaking, at $t=16m^2_\pi$. 
However, it is an experimental fact that inelasticity is negligible 
until the quasi-two~body channels $\omega\pi,\,a_1\pi\,\dots$ are open. 
In practice we will take
$$t_0\simeq 1\;\gev^2,$$
and fix the best value for $t_0$  empirically. 
It will be $t_0=1.1\,\gev^2$, and we will see that, if we keep close to 
 this value, the dependence on  $t_0$ is very slight. 

The properties (i-iv) can be taken into account with the well-known 
Omn\`es-Muskhelishvili method. We construct a 
function $J(t)$ with the proper phase by defining 
$$J(t)=\exp\left\{\dfrac{t}{\pi}\int_{4m^2_\pi}^{t_0} \dd s\;
\dfrac{\delta_1^1(s)}{s(s-t)}+
\dfrac{t}{\pi}\int^{\infty}_{t_0} \dd s\;\dfrac{\bar{\delta}_1^1(s)}{s(s-t)}\right\}.
\equn{(3.1a)}$$
We have written the dispersion relation with one subtraction to ensure 
that $J(0)=1$. 
The singular integrals are  understood to be calculated replacing $t\to t+\ii\epsilon$, 
$\epsilon>0$, and letting then $\epsilon\to0$. 
In particular, we have       
$$|J(t)|=\exp\left\{\dfrac{t}{\pi}{\rm P.P.}\int_{4m^2_\pi}^{t_0} \dd s\;
\dfrac{\delta_1^1(s)}{s(s-t)}+
\dfrac{t}{\pi}\int^{\infty}_{t_0} \dd s\;\dfrac{\bar{\delta}_1^1(s)}{s(s-t)}\right\}, 
\quad 4m^2_\pi\leq t\leq t_0. 
\equn{(3.1b)}$$
Defining then the function $G$ by
$$F_\pi(t)=G(t)J(t),\eqno{(3.2)}$$ 
it follows from properties i-ii that $G(t)$ is analytic with
 only the exception of a cut from $t_0$ to infinity, 
as we have already extracted the correct phase 
below $t=t_0$. 

We can, in \equn~(3.1a), take any value we like for the phase $\bar{\delta}_1^1(t)$, as a change of
 it only results in a redefinition of $G$; but it is convenient to 
choose  $\bar{\delta}_1^1(t)$ so that it joins smoothly $\delta_1^1(t)$ at $t=t_0$ 
to avoid spurious singularities that would deteriorate the convergence, and 
so that  $J$ has the correct behaviour at infinity. 
Both properties are ensured if we take, simply,
$$\bar{\delta}_1^1(t)=\pi+\left[\delta_1^1(t_0)-\pi\right]\dfrac{t_0}{t}$$
so that $\bar{\delta}_1^1(t_0)=\delta_1^1(t_0)$ and,
 for large $t$, $\bar{\delta}_1^1(t)\to \pi$ and we recover the 
behaviour $1/t$ of the form factor. 
Then we can rewrite more explicitly (3.1) by integrating the piece with $\bar{\delta}_1^1$:
$$J(t)=\ee^{1-\delta_1^1(t_0)/\pi}
\left(1-\dfrac{t}{t_0}\right)^{[1-\delta_1^1(t_0)/\pi]t_0/t}
\left(1-\dfrac{t}{t_0}\right)^{-1}
\exp\left\{\dfrac{t}{\pi}\int_{4m^2_\pi}^{t_0} \dd s\;
\dfrac{\delta_1^1(s)}{s(s-t)}\right\}.
\equn{(3.3)}$$
\medskip
\noindent{3.1.1. \sl The phase $\delta^1_1$}
\medskip 
We can apply  the effective range theory to the phase  $\delta^1_1$. According to 
this, the function
$$\psi(t)\equiv \dfrac{2k^3}{t^{1/2}}\cot \delta_1^1(t),\quad k=\dfrac{\sqrt{t-4m^2_\pi}}{2}
\equn{(3.4a)}$$
is analytic in the variable $t$ except for two cuts: a cut from $-\infty$ to $0$, and 
a cut from $t=t_0$ to $+\infty$.
To profit from the analyticity properties of $\psi$ we will 
make a conformal transformation.\fnote{The 
method of conformal transformations is rigorous, simpler and produces better results than 
that employed in ref.~4.} We define 
$$w=
\dfrac{\sqrt{t}-\sqrt{t_0-t}}{\sqrt{t}+\sqrt{t_0-t}}.
\equn{(3.4b)}$$
When $t$ runs the cuts, $w$ goes around the unit circle.
 We may therefore expand $\psi$ in a power series 
 convergent inside
the unit disc. However, the existence of the $\rho$ resonance implies that 
we must have   $\cot \delta_1^1(m^2_\rho)=0$. It is therefore convenient to
 incorporate this piece of knowledge and expand not $\psi$ itself but the ratio   
$\psi(t)/(m^2_\rho-t)\equiv\hat{\psi}(t)$: so we write,
$$\psi(t)=(m^2_\rho-t)\hat{\psi}(t)=(m^2_\rho-t)\left\{b_0+b_1w+
b_2w^2+\cdots\right\}.
\equn{(3.4c)}$$

The P-wave, $I=1$ $\pi\pi$ scattering length,\fnote{For details on $\pi\pi$
 scattering, including analyticity 
properties and the Fermi--Watson theorem, see e.g. ref.~15. 
More details on the solution of the Omn\`es--Muskhelishvili 
equation can be found in N.~I.~Muskhelishvili, {\sl Singular
 Integral Equations}, Nordhoof, 1958.} $a_1^1$, is related to $\psi$ by
$$a_1^1=\dfrac{1}{m_\pi\psi(4m^2_\pi)}.
\equn{(3.5a)}$$

Likewise, from the relation
$$\dfrac{1}{\cot \delta_1^1(t)-\ii}\;\simeqsub_{t\simeq m^2_\rho}\;
\dfrac{\hbox {const.}}{m^2_\rho-t-2k^3\ii/t^{1/2}\hat{\psi}(t)}$$
we find the expression for the rho width:
$$\gammav_\rho=\dfrac{2k^3_\rho}{m^2_\rho\hat{\psi}(m^2_\rho)},
\quad
k_\rho=\tfrac{1}{2}\sqrt{m^2_\rho-4m^2_\pi}.
\equn{(3.5b)}$$
Experimentally,\ref{15}  
 $a_1^1\simeq (0.038\pm0.003) m^{-3}_\pi$, 
and, according to the Particle Data Group Tables,\ref{16}     
$m_\rho=769.3\pm0.8;\;\gammav_\rho=150.2\pm0.8\,\mev$. Note, however, that we do {\sl not} 
assume the values of $m_\rho,\,\gammav_\rho$. 
We only require that $\psi$ has a zero, and will let the fits 
fix its location and residue. 

It turns out that, to reproduce the width and scattering length, and to fit 
the pion form factor as well (see below), only two terms in the 
expansion are needed, so we approximate
$$\delta_1^1(t)={\rm Arc\; cot}\left\{\dfrac{t^{1/2}}{2k^3}
(m^2_\rho-t)\left[b_0+b_1\dfrac{\sqrt{t}-\sqrt{t_0-t}}{\sqrt{t}+\sqrt{t_0-t}}\right]\right\};
\equn{(3.6)}$$
$m_\rho,\,b_0,\,b_1$ are free parameters in our fits.
\medskip
\noindent 3.1.2. {\sl The function $G(t)$}
\medskip 
Because we have already extracted the correct phase up to $t=t_0$, 
it follows that the function 
$G(t)$ is analytic except for a cut from $t=t_0$ to $+\infty$. 
The conformal transformation 
$$z=\dfrac{\tfrac{1}{2}\sqrt{t_0}-\sqrt{t_0-t}}{\tfrac{1}{2}\sqrt{t_0}+\sqrt{t_0-t}}
\eqno{(3.7a)}$$
maps this cut plane into the unit circle. 
So we may write the 
expansion,
$$G(t)=1+A_0+c_1z+c_2 z^2+c_3 z^3+\cdots
\equn{(3.7b)}$$
that will be convergent for all $t$ inside the cut plane. 
We can implement the condition $G(0)=1$, necessary to ensure 
$F_\pi(0)=1$ 
to each order, by writing
$$A_0=-\left[c_1z_0+c_2 z_0^2+c_3 z_0^3+\cdots\right],\quad
z_0\equiv z|_{t=0}=-1/3.$$

The expansion then reads,

$$G(t)=1+c_1(z+1/3)+c_2 (z^2-1/9)+c_3 (z^3+1/27)+\cdots.
\eqno{(3.8)}$$
We will need two-three terms in the expansion, so we will 
approximate
$$G(t)=1+
c_1\left[\dfrac{\tfrac{1}{2}\sqrt{t_0}-\sqrt{t_0-t}}{\tfrac{1}{2}\sqrt{t_0}+\sqrt{t_0-t}}
+\tfrac{1}{3}\right]+
c_2\Bigg[\left(\dfrac{\tfrac{1}{2}\sqrt{t_0}-\sqrt{t_0-t}}{\tfrac{1}{2}\sqrt{t_0}+\sqrt{t_0-t}}\right)^2
-\tfrac{1}{9}\Bigg],$$
$c_1,\,c_2$ free parameters.

An interesting feature of our method is that, even if we only kept {\sl one} term in each of 
the expansions (3.6,~8), that is to say, if we took $b_1=c_1=c_2=0$, 
we could reproduce the experimental data with only a 15\% error; so 
we expect (and this is the case) fast convergence of the series. 
It is important  also that our expression for $F_\pi(t)$ is valid in the spacelike as well as in the  
timelike region, provided only $t<t_0$. 
What is more,  (3.6,~8) represent the more general expressions compatible with 
analyticity,  the Fermi--Watson theorem 
and the effective range theory, which follow 
only from the requirements of unitarity and causality. 
Therefore, by employing our expansions,
 we do not introduce uncontrolled biases in the analysis, and hence we minimize
the model dependent errors.\fnote{The remaining approximations are 
 neglect of the inelasticity between $16m^2_\pi$ 
and $t_0$, experimentally known to be at the $10^{-3}$ 
level or below, and  we have the errors due to the truncation 
of the expansions;  we will also check that they are small. 
By contrast, other functional forms used in in the literature 
 are either incompatible with the phase of $F_\pi$, or with its 
analyticity properties (or both), which  
causes  biases in the fits. \hb
 The errors due to breaking of isospin and electromagnetic corrections will be 
discussed below.}

\booksubsection{3.2. Fits}
In order to fit $F_\pi$, and hence get the $2\pi$ low energy ($4m^2_\pi\leq t\leq0.8\;\gev^2$) 
contribution to $a^{(2)}(\hbox{h.v.p.})$, we have 
available three sets of data: 
\item{$\bullet$}{$e^+e^-\to\pi^+\pi^-$, $t$ timelike (Novosibirsk, ref.~8).}
\item{$\bullet$}{$F_\pi(t)$, $t$ spacelike (NA7, ref.~11).}
\item{$\bullet$}{In addition, one can use data from the decay
 $\tau^+\to\bar{\nu}_\tau \pi^+\pi^0$ 
(Aleph and Opal, ref.~10).}
 
For this last
 we have to assume isospin invariance, 
{\sl and} neglect the 
isospin $I=2$ component of $\pi^+\pi^0$, to write the form factor $v_1$ 
for $\tau$ decay 
 in terms of $F_\pi$:
$$v_1=\tfrac{1}{12}\left(1-\dfrac{4m^2_\pi}{t}\right)^{3/2}|F_\pi(t)|^2,
\equn{(3.9a)}$$
where, in terms of the weak vector current $V_\mu=\bar{u}\gamma_\mu d$, 
and in the exact isospin approximation,
$$\Piv^V_{\mu\nu}=\left(-p^2g_{\mu\nu}+p_\mu p_\nu\right)\Piv^V(t)=
\ii\int\dd^4x\,\ee^{\ii p\cdot x}\langle0|{\rm T}V^+_\mu(x)V_\nu(0)|0\rangle;\quad
v_1=2\pi\imag \Piv^V.
\equn{(3.9b)}$$

Before presenting the results of the 
fits a few matters have to be discussed. 
A first point to clarify is that we will {\sl not} include in the fits the 
old data on $F_\pi$ in the 
spacelike or timelike regions, or  
on pion-pion phase shifts\ref{17}. 
We have checked that, if we add the first two sets, the results of the fit 
vary very little (see below); 
but they cause a bias. This is so because there is incompatibility\fnote{At 
the level of $1.5$ to $2\sigma$.} 
between old spacelike and timelike data,
and also with data on $\pi\pi$ 
phase shifts, already noticed in CLY.\ref{4}
Doubtlessly, this is due to the fact that 
most old data for spacelike momentum were extracted from processes
 with one pion off its mass shell 
so that models were necessary to extrapolate to the physical form factor.
In fact, a very important feature of the NA7\ref{11} data
 is that they are obtained from scattering of real pions off electrons, 
hence we do not require models to extract $F_\pi$ from data.
 
The reason for the model dependence of $\pi\pi$ phase shift analyses is that 
 these are extracted from fits to $\pi N\to\pi\pi N$ scattering and thus 
require a model for the pseudoscalar form factor of the nucleon, a model for the  
interactions of the nucleon and the final state pions, 
and 
a model for the dependence of $\pi\pi$ scattering on the mass of 
an external pion. 
Indeed, different methods of extrapolation result in different sets of 
 phase shifts, 
as can be seen in the experimental papers of Hyams et al. and Protopopescu et al., ref~17, 
where five different determinations are given. 
However, we will use the scattering length, $a^1_1$, and 
employ the  $\pi\pi$ phase shifts as a very important {\sl 
a posteriori} test of our results. 

We could consider, besides this 
information, to include as input the values of several quantities that can be 
estimated  
with chiral perturbation theory methods, such as $\langle r^2_\pi\rangle$ and 
$a^1_1$. We do not do so because
 the problem with these calculations is the estimate of their errors, 
a difficult matter; so we have preferred to avoid possible biases and 
instead {\sl obtain} these quantities as 
byproduct of our calculations. 
Then we check that the results we get for all of them are in agreement, within errors, with 
the chiral perturbation theory results; see below. 
With respect to $a^1_1$  we actually constrain it to the region obtained
 from  $\pi\pi$ scattering
 experimental data only; its error is chosen such that it encompasses the various values 
given in the different experimental determinations (ref.~17). 
We take, 
$$a^1_1=(38\pm3)\times 10^{-3}\;m_\pi^{-3}; $$ 
we will see that the value our best fit returns for this quantity is 
satisfactorily close to this, as indeed we get $(41\pm2)\times 10^{-3}\;m_\pi^{-3}$.

Another remark concerns the matter of isospin breaking, due to 
electromagnetic interactions or the  mass difference between $u,\;d$
 quarks, that would spoil the equality (3.9a). 
It is not easy to estimate this. A large part of the breaking, the 
$\omega\to2\pi$ contribution and $\omega-\rho$ mixing,  
are taken into account by hand, but this does not exhaust the effects. 
For example, merely changing the quark masses from $m_{\pi^+}+m_{\pi^-}$ 
to  $m_{\pi^0}+m_{\pi^0}$, in a Breit--Wigner model for the $\rho$,  
shifts $a^{(2)}(\hbox{h.v.p.})$ by $\sim50\times10^{-11}$, so 
a deviation of this order should not be surprising.\fnote{The 
relevance of isospin breaking in this context was pointed out 
by V.~Cirigliano, G.~Ecker and H.~Neufeld, hep-ph/0104267, 
2001.}

As stated above, \eqs~(3.9) were obtained neglecting 
the mass difference $m_u-m_d$ 
and electromagnetic corrections, 
in particular the $\pi^0 - \pi^+$ mass difference. 
We can take the last partially into 
account by distinguishing between the pion masses in the 
phase space factor in (3.9a). 
To do so, write now (3.9b) as
$$\Piv^V_{\mu\nu}=
\ii\int\dd^4x\,\ee^{\ii p\cdot x}\langle0|{\rm T}V^+_\mu(x)V_\nu(0)|0\rangle=
\left(-p^2g_{\mu\nu}+p_\mu p_\nu\right)\Piv^V(t)+p_\mu p_\nu \piv^{S};\quad
v_1\equiv2\pi\imag \Piv^V.
\equn{(3.10a)}$$  
We get 
$$v_1=\tfrac{1}{12}
\left\{\left[1-\dfrac{(m_{\pi^+}-m_{\pi^0})^2}{t}\right]
\left[1-\dfrac{(m_{\pi^+}+m_{\pi^0})^2}{t}\right] \right\}^{3/2}|F_\pi(t)|^2.
\equn{(3.10b)}$$
To compare with the experimentally measured quantity, 
which involves all of $\imag \piv^V_{\mu\nu}$, we have to 
neglect the scalar component $\piv^S$, which is proportional to $(m_d-m_u)^2$, 
and thus very small.

To understand the situation we will proceed by steps. First of all, we start by fitting 
 {\sl separately} $e^+e^-$ and $\tau$ 
data, in the timelike region, using (3.9a) 
(we remark that although in $a(2\pi;\,t\leq0.8\;\gev^2)$ only enter the 
values of $F_\pi(t)$ for $4m^2_\pi$ to $0.8\,\gev^2$, 
we fit the whole range up to $t=t_0=1.1\,\gev^2$). 
Then, we get quite different results:
$$a(2\pi;\,t\leq0.8\;\gev^2)=
\cases{4\,715\pm67\quad (e^+e^-;\;\chidof=106/109=0.96)\cr
4\,814\pm26\quad (\tau;\;\chidof=52/48=1.09).\cr}
\equn{(3.11a)}$$
This takes into account statistical errors only 
for $e^+e^-$, but includes systematic ones 
for $\tau$ decay as these are  incorporated in the 
available data. 

The slight advantage of the first figure in (3.11a)
 in what regards the \chidof\ 
makes one wonder that the difference is really caused by 
isospin breaking 
(in which case the value obtained from $\tau$ decay should be 
rejected) or is due to random fluctuations of the data, as 
well as to the systematics of the experiments. 
The second explanation has in its favour that, if we include the
{\sl spacelike} data into the fit (but still use (3.9a)) 
the discrepancy is softened, and we get compatible results:
$$a(2\pi;\,t\leq0.8\;\gev^2)=
\cases{4\,754\pm55\quad (e^+e^-\,+\;{\rm spacelike};\;\chidof=179/154)\cr
4\,826\pm23\quad (\tau\,+\;{\rm spacelike};\;\chidof=112/93).\cr}
\equn{(3.11b)}$$
This last result allows us to draw the following conclusion: that  
  part of the discrepancy 
between results obtained with $e^+e^-$ and $\tau$ decay is still of statistical origin,
 but also it would seem that part is genuine.

\topinsert{
\setbox0=\vbox{\hsize15.truecm\hfil{\epsfxsize 13.4truecm\epsfbox{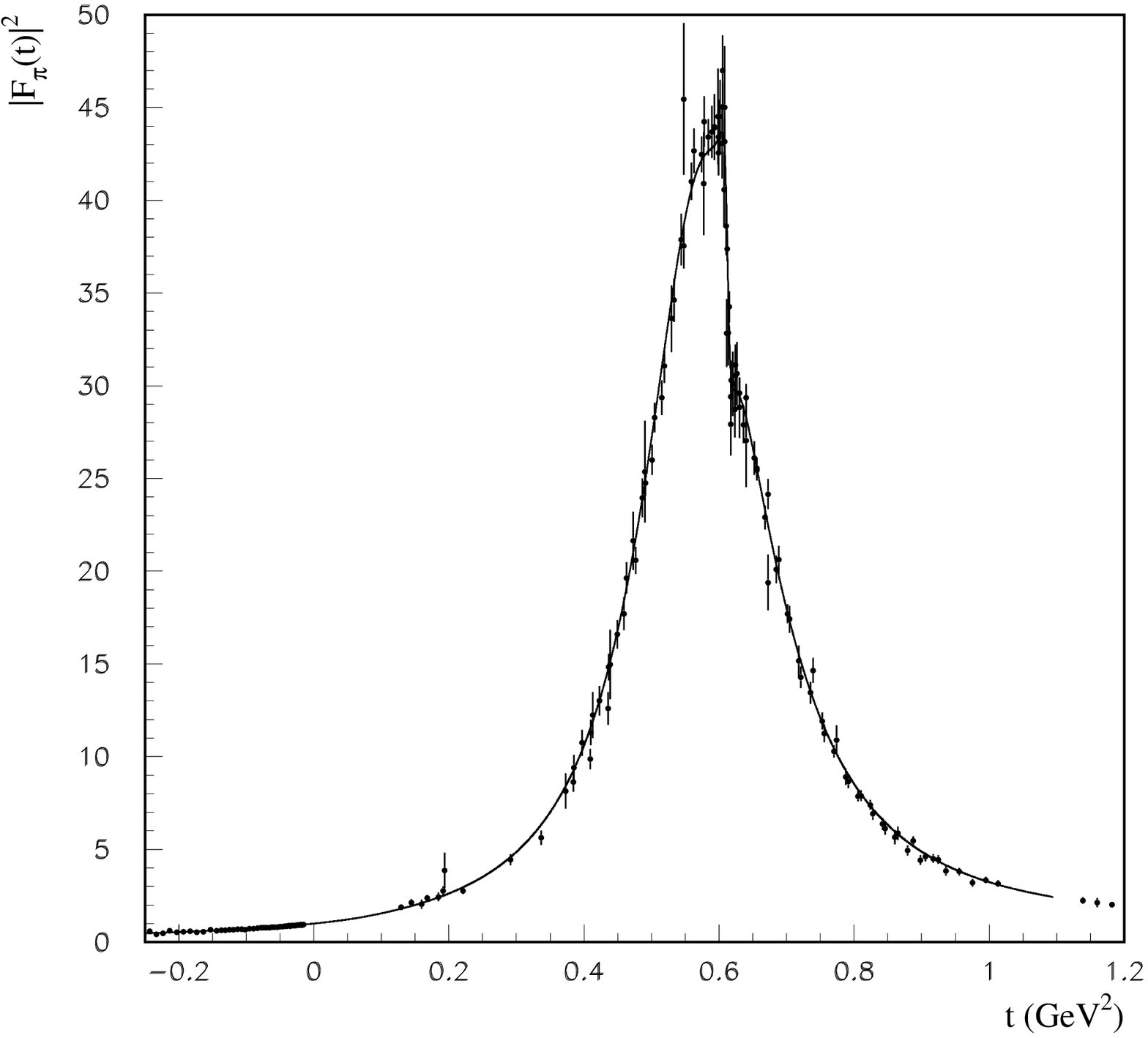}}\hfil} 
\setbox6=\vbox{\hsize 14truecm\captiontype\figurasc{Figure 2. }{Plot of 
the fit to $|F_\pi(t)|^2$, timelike (ref. 8) and spacelike (ref.~11) data. The 
theoretical curve actually drawn is that obtained by fitting also $\tau$ data, but 
the curve obtained fitting only $e^+e^-$
 could not be distinguished from that drawn if we plotted it. 
A blowup of the fit in the spacelike region may be seen in Fig.~4.}\hb
\vskip.1cm} 
\medskip
\centerline{\box0}
\centerline{\box6}
\medskip
}\endinsert

In an attempt to take into account at least some of the  
isospin breaking effects, we have fitted simultaneously
 $e^+e^-$, $\tau$ decay, both including spacelike data, allowing for different 
values of the mass and width of the 
rho (but keeping other parameters, 
in particular $c_1$, $c_2$, common for both  $e^+e^-$ 
and  $\tau$ fits). We, however, still use (3.9a). 
 In this case we find convergence of the results; we have\fnote{When 
evaluating $a(2\pi;\,t\leq0.8\;\gev^2)$ we of course use the 
parameters $m_\rho$, $b_0$, $b_1$ corresponding 
to $\rho^0$; see below.} 

$$a(2\pi;\,t\leq0.8\;\gev^2)=
4\,779\pm30,\quad \chi^2/{\rm d.o.f.}=
248/204;\qquad (e^+e^-\,+\tau\,+\;{\rm spacelike}),\equn{(3.12)}$$
which is compatible (within errors) with both numbers in (3.11b).

\topinsert{
\setbox9=\vbox{
\setbox0=\vbox{\hsize16.4truecm\line{\hfil{\epsfxsize 7.5truecm\epsfbox{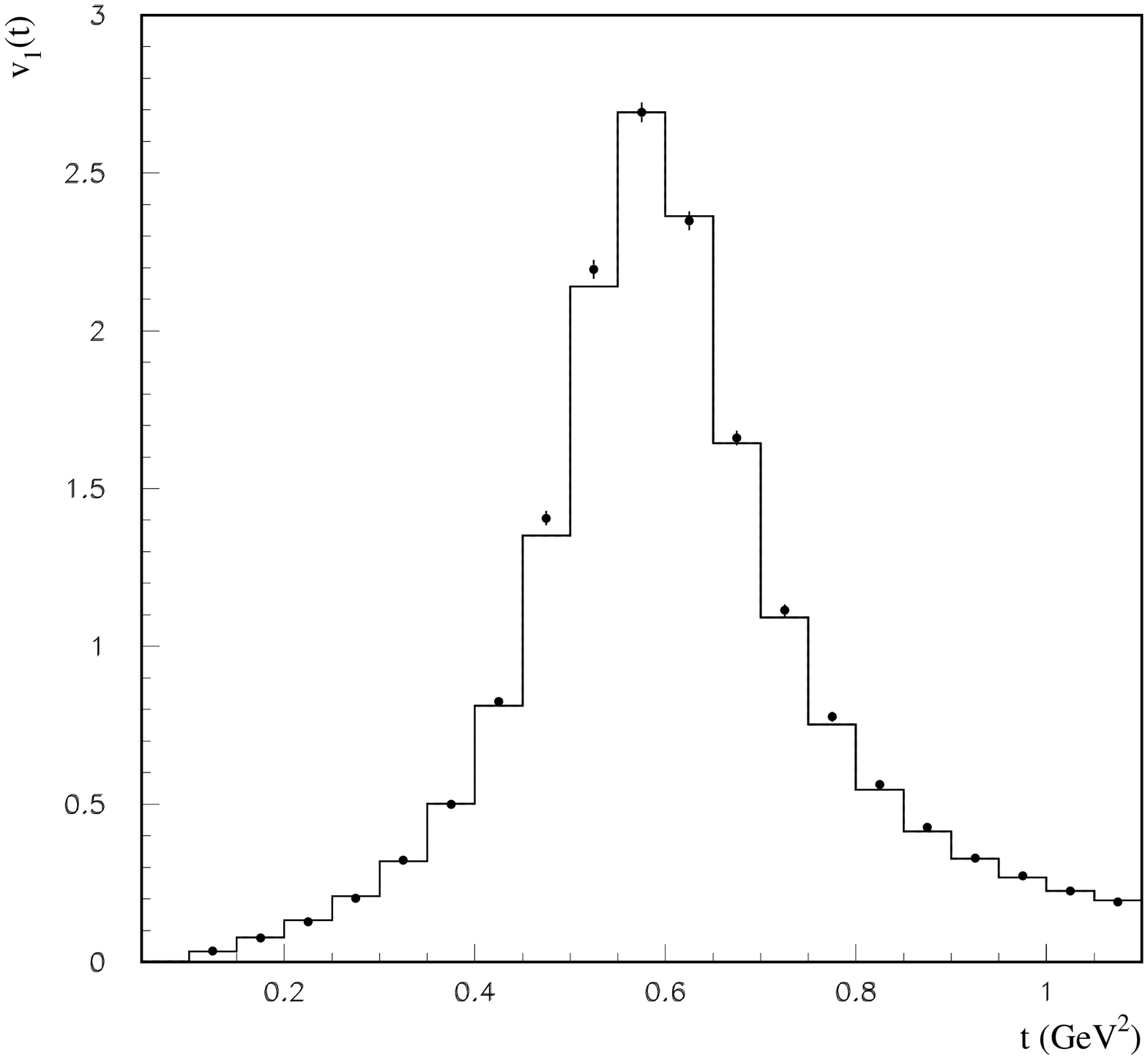}}\hfil
{\epsfxsize 7.5truecm\epsfbox{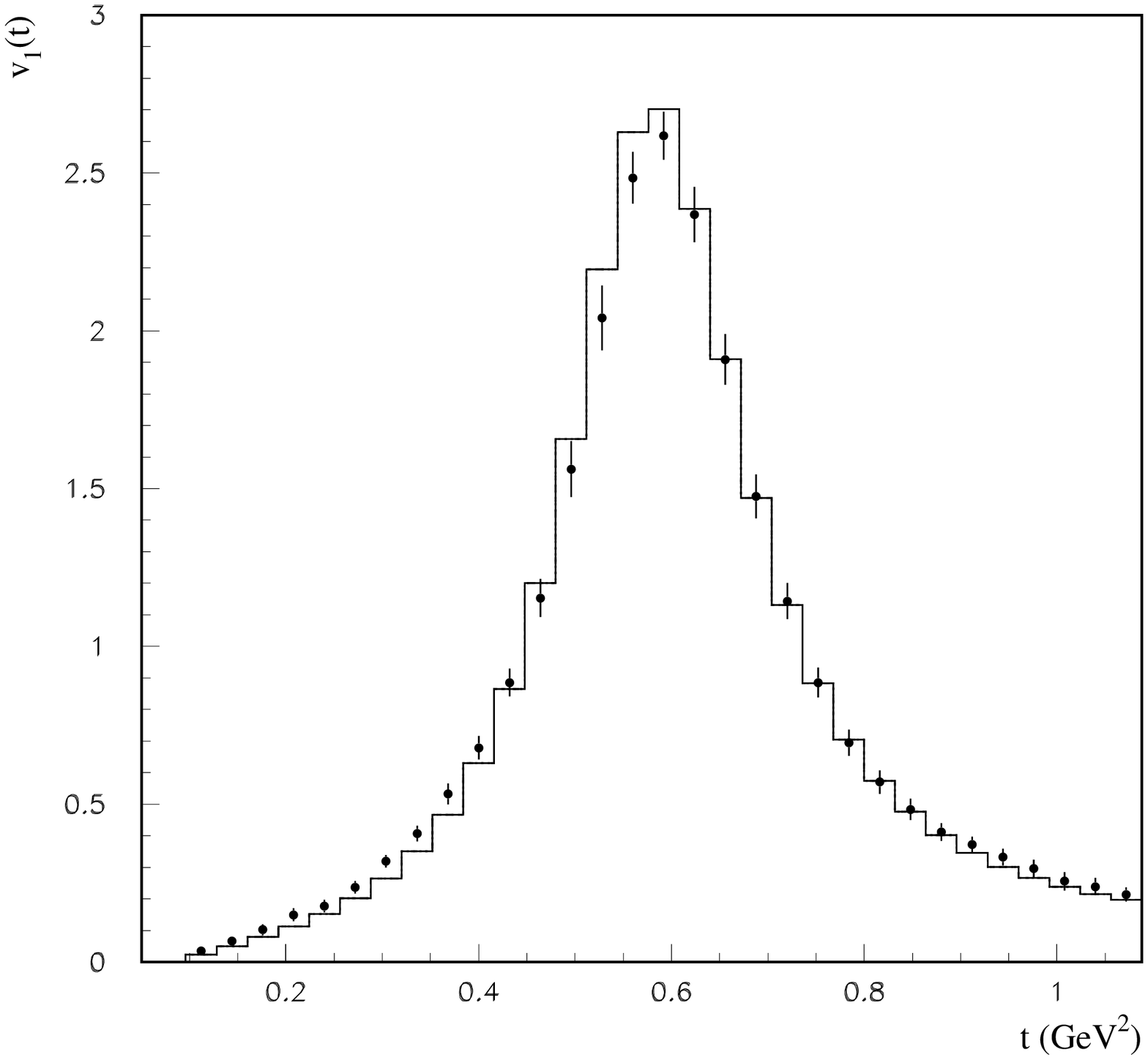}}\hfil}}
\setbox6=\vbox{\hsize 13truecm\captiontype\figurasc{Figure 3. }{Plot 
of the fits to $v_1(t)$ (histograms), and data from $\tau$ decay (black dots).\hb
Left: Aleph data. Right: Opal data. (ref.~10).}
Note that the theoretical values (histograms) are results of the {\sl same} 
calculation, with the same parameters, so the differences between the two fits merely 
reflect the slight variations between the two experimental determinations.} 
\bigskip
\centerline{\box0}
\centerline{\box6}
\bigskip}
\box9
}\endinsert  

It is to be noted that, if we had not allowed for different masses
 and widths for the neutral and charged rho, we would have obtained, in this common fit,
$$10^{11}\times a(2\pi;\,t\leq0.8\;\gev^2)=4\,822\pm30,\qquad \chi^2/{\rm d.o.f.}=264/206; 
\qquad (e^+e^-\,+\,\tau\,+\,{\rm spacelike})$$
i.e., a  larger \chidof\ and a value quite different from that obtained 
with only $e^+e^-$ and spacelike data. 
So it would appear that allowing for different parameters for the neutral and charged rho really 
takes into account a good part of the isospin breaking effects. 

Finally, we take into account the kinematical effects of the $\pi^\pm$, $\pi^0$ 
mass difference repeating the fit using (3.10b) now.\fnote{For consistency 
we should also have taken the expresion
 $k=\tfrac{1}{2}\{[t-(m_{\pi^+}-m_{\pi^0})^2][t-(m_{\pi^+}+m_{\pi^0})^2]\}^{1/2}$, 
altered the threshold to $t=(m_{\pi^+}+m_{\pi^0})^2$ 
for tau decay and allowed for different scattering lengths. 
We have checked that the effect of this on the contribution to $a$ leaves it 
well inside our error bars; 
we will discuss the results one gets in a separate paper. Note that it makes sense 
to still consider the same $c_1$, $c_2$ for $e^+e^-$ and 
tau decay as these parameters are associated 
with $G$ whose imaginary part vanishes below $t=s_0\sim 1\,\gev^2$ 
where the effects of isospin breaking should be negligible.}
The result of the fit with $e^+e^-$ data only is of course unchanged, 
but we reproduce it to facilitate the comparison and for ease of reference. 
We find what we consider our best results:
$$\eqalign{
10^{11}\times a(2\pi;\,t\leq0.8\;\gev^2)=&\;4\,774\pm31,\quad \chi^2/{\rm d.o.f.}=246/204; 
\qquad (e^+e^-\,+\,\tau\,+\,{\rm spacelike}).\cr
10^{11}\times a(2\pi;\,t\leq0.8\;\gev^2)=&\;4\,754\pm55,\quad \chi^2/{\rm d.o.f.}=179/154;
\qquad (e^+e^-\,+\,{\rm spacelike}).\cr
}
\equn{(3.13)}$$
We remark that the results for the evaluation including $\tau$ decays are rather insensitive to 
the use of (3.10b), 
but what change there is, it goes in the right direction:
 the \chidof\ has improved slightly, and the values for the anomaly including the $\tau$ 
have become slightly 
more compatible with the figure obtained using $e^+e^-$ data only. 
This makes us confident that most of the effects due to isospin breaking, 
both from $u,\,d$ mass differences and from electromagnetic effects (about which we will 
say more in 
\subsects~3.3 and 5.2) have already been incorporated in our calculation.
The fit may be seen depicted in \fig~2 for $|F_\pi|^2$, 
with  timelike and 
spacelike data, and in \fig~3 for the quantity $v_1$ in $\tau$ decay.

The $\chidof$ of the fits is slightly above unity; in next subsection
 we will see that including {\sl systematic} 
errors cures the problem. 
For example, just adding the systematic normalization error for the spacelike data\ref{11} 
gives a shift of the central value of  $31\times10^{-11}$ and the 
\chidof\ decreases to $152/153$ for the evaluation with
$e^+e^-$ data only.  The quality of the fit to the spacelike data
 is shown in \fig~4, which is a blowup 
of the corresponding part of \fig~2.

\topinsert{
\setbox0=\vbox{\hsize16.4truecm
\line{\hfil{\epsfxsize 7.4truecm\epsfbox{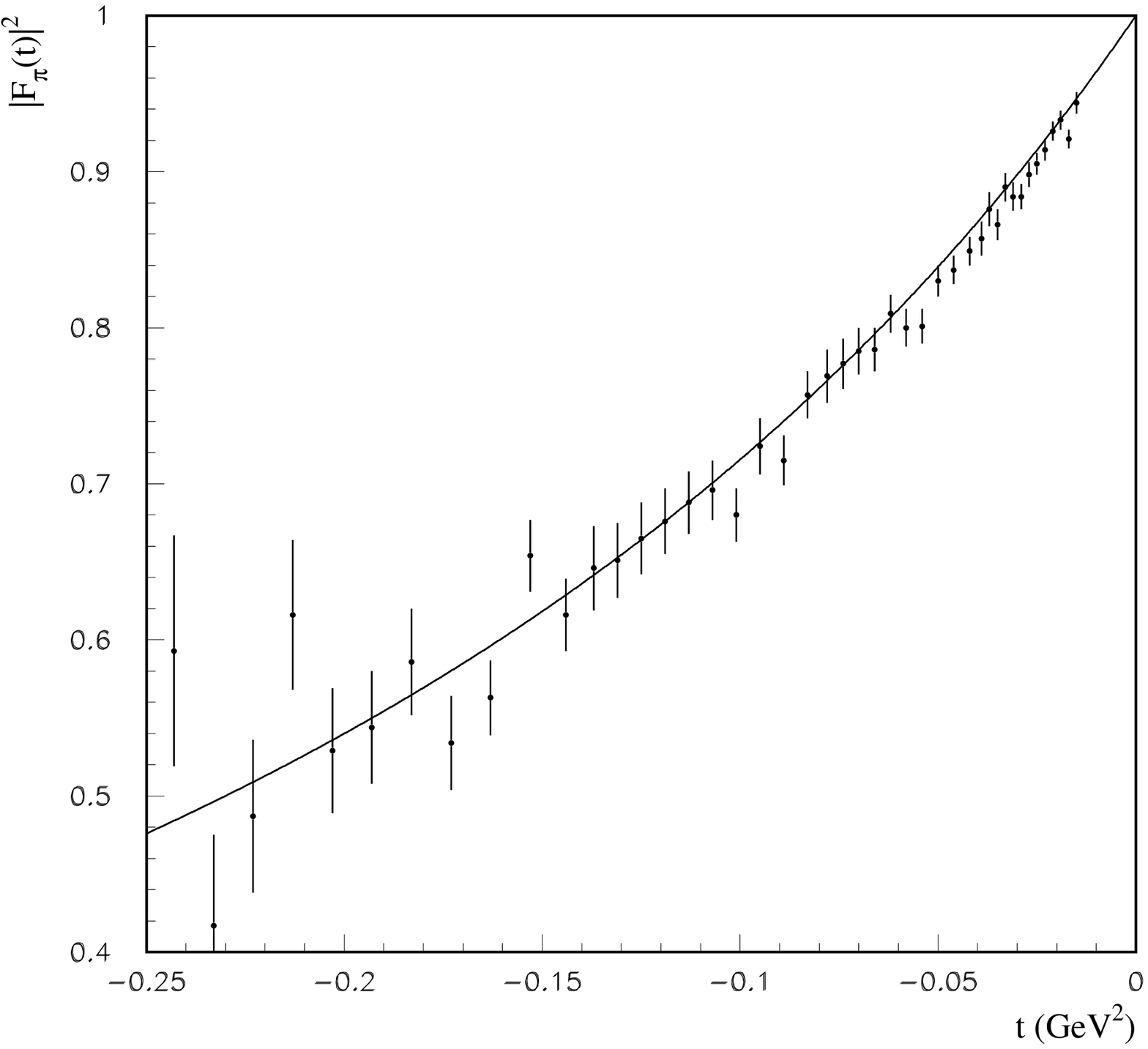}}\hfil
{\epsfxsize 7.4truecm\epsfbox{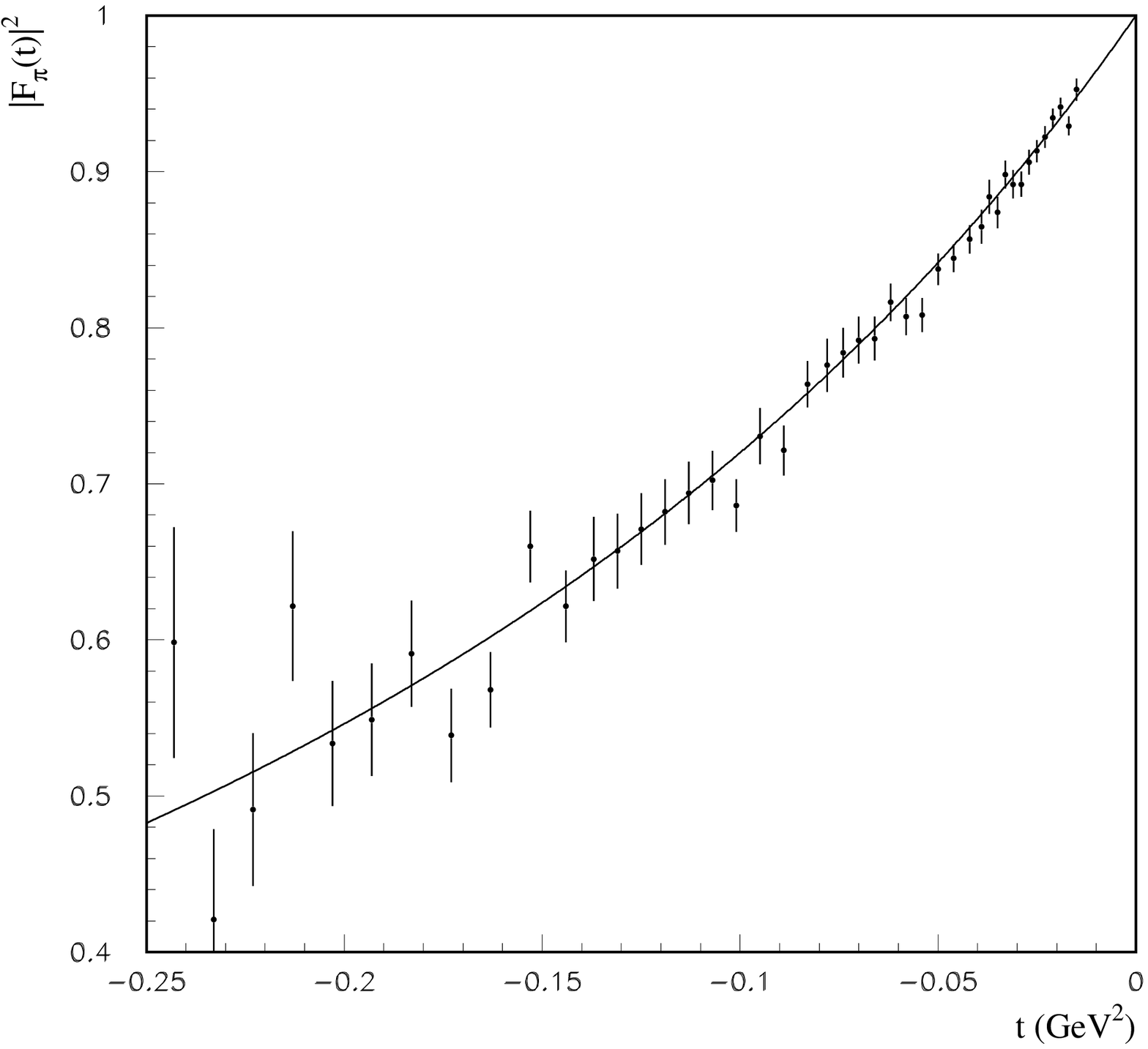}}\hfil}} 
\setbox6=\vbox{\hsize 13truecm\captiontype\figurasc{Figure 4. }{Plot of 
the fit to $|F_\pi(t)|^2$ in the spacelike region. 
With only statistical errors (left) and including 
systematic experimental errors (right).}\hb
\vskip.1cm} 
\medskip
\centerline{\box0}
\centerline{\box6}
\medskip
}\endinsert

The parameters of the fits are also compatible. 
We have,
$$\eqalign{
c_1=0.23\pm0.02,\;c _2=-0.15\pm0.03;\;b_0=1.062\pm0.005,\;b_1=0.25\pm0.04
&\quad (e^+e^-\,+\,\tau\,+\,{\rm spacelike});\cr
c_1=0.19\pm0.04,\;c_2=-0.15\pm0.10;\;b_0=1.070\pm0.006,\;b_1=0.28\pm0.06
&\quad (e^+e^-\,+\,{\rm spacelike}).\cr
}
\equn{(3.14)}$$
In the first line the parameters $c_1$, $c_2$ are common for 
$\rho^0$, $\rho^+$. 
$b_0$ and $b_1$ vary very little; the ones shown 
correspond to the values of $m_{\rho^0}$, 
$\gammav_{\rho^0}$ as 
given below in \equn{(3.15)}.
The values $b_0=\hbox{Constant}$, $b_1=0$ would
 correspond to a perfect Breit--Wigner shape for the $\rho$.
Another fact to be mentioned is that including the corrected phase space factor (3.10b) 
helps a little to make compatible the parameters for both fits; if we had used (3.9a) we 
would have obtained
$$c_1=0.23\pm0.01,\;c _2=-0.16\pm0.03;\;b_0=1.060\pm0.005,\;b_1=0.24\pm0.04
\quad (e^+e^-\,+\,\tau\,+\,{\rm spacelike}).$$ 

An important feature of our fit is that the 
coefficients decrease with increasing order. This, together with the fact that the 
conformal variables $w,\,z$ are of modulus well below unity in the 
regions of interest ($4m^2_\pi\leq t\leq0.8\,\gev^2$ for 
$w$, $-0.25\,\gev^2\leq t\leq0.8\,\gev^2$ for $z$):
$$-0.57\leq w\leq0.24,\quad -0.38\leq z\leq -0.02,$$ 
ensures good convergence of the expansions. 
We have also checked that including extra terms in the 
expansions does not improve the quality of the fits  significantly.

Besides the results for the anomaly we obtain reliable determination of a set of parameters. 
We have those pertaining to the rho,
$$\eqalign{m_{\rho^0}=&\;772.6\pm0.5\;\mev,\quad \gammav_{\rho^0}=147.4\pm0.8\,\mev;\cr
m_{\rho^+}=&\;773.8\pm0.6\;\mev,\quad \gammav_{\rho^+}=147.3\pm0.9\,\mev.\cr
}
\equn{(3.15)}$$
The figures are in reasonable agreement with the Particle Data Group values\fnote{It 
should be noted that the various  determinations
 for $m_\rho$ reported by the PDG\ref{16} actually cluster around {\sl several} 
different values.} given before.
 
The value for the scattering length the fit returns is comfortably close to 
the one obtained from $\pi\pi$ phase shifts; we get
$$a_1^1=(41\pm2)\times 10^{-3}\;m_\pi^{-3}.$$
This  value of  $a^1_1$ is slightly larger, but compatible with recent determinations based on
an analysis of $\pi\pi$ scattering (ACGL)   or chiral perturbation theory 
(CGL, ABT) that give (ref.~18)
$$a^1_1  =(37.9\pm0.5)\times 10^{-3}\;m_\pi^{-3} {\rm (CGL)};\;
a^1_1  =(37\pm2)\times 10^{-3}\;m_\pi^{-3}  {\rm (ACGL)};\;
a^1_1  =(38\pm2)\times 10^{-3}\;m_\pi^{-3}  {\rm (ABT)}.$$

Also from  our fits we obtain the low energy coefficients  of the pion form factor,
$$F^2_\pi(t)\simeqsub_{t\to0}1+\tfrac{1}{6}\langle r^2_\pi\rangle t+c_\pi t^2:$$
$$\eqalign{
\langle r^2_\pi\rangle=0.435\pm0.002\;{\rm fm}^2,\;c_\pi=3.60\pm0.03\;\gev^{-4}&
\quad (e^+e^-\,+\,\tau\,+\,{\rm spacelike});\cr
\langle r^2_\pi\rangle=0.433\pm0.002\;{\rm fm}^2,\;c_\pi=3.58\pm0.04\;\gev^{-4}&
\quad (e^+e^-\,+\,{\rm spacelike}).\cr}
\equn{(3.16)}$$
These figures are also compatible with, but much more precise than, the current 
estimates:\ref{18}
$$\langle r^2_\pi\rangle=0.431\pm0.026\;{\rm fm}^2,\;c_\pi=3.2\pm1.0\;\gev^{-4}.$$

Another remark is that in all these fits we took $t_0=1.1\;\gev^2$. 
The dependence of the results on 
this parameter, $t_0$, is very slight, provided we remain around this value. 
Thus, for example, if we take $t_0=1.2\,\gev^2$ the value of 
$a(2\pi;\,t\leq0.8\;\gev^2)$ 
only increases by $4\times10^{-11}$, and the global $\chi^2$ 
only varies by one unit.

As further checks of the stability and reliability of our results we mention 
the following two. 
First, we could, as discussed above, have imposed the more stringent values for 
 $a^1_1$ as given in ref.~18. Now, if for example we take, in accordance with (ACGL) in 
this reference, the 
value $a^1_1=(37\pm2)\times 10^{-3}\times m_\pi^{-3}$, 
instead of the value $a^1_1=(38\pm3)\times 10^{-3}\times m_\pi^{-3}$
that follows from only {\sl experimental} $\pi\pi$ data, the fit deteriorates. 
The fit returns the value   $a^1_1=(39\pm1)\times 10^{-3}\times m_\pi^{-3}$ 
for the scattering length, 
in (slightly) better agreement with the input; 
but we do not consider this an improvement as the global $\chi^2$ 
 increases by 2 units. 

The corresponding value for the contribution to the anomaly changes very little, from the value 
 $(4774\pm31)\times 10^{-11}$ (Eq.~{(3.13)}) to  $(4768\pm 32)\times 10^{-11}$ now, 
i.e., a shift of only $6\times 10^{-11}$ with a small increase of the error. 
Thus, the results are insensitive to a more stringent input for  $a^1_1$ 
but, because the quality of the fit deteriorates,   
 we still consider the result with the more relaxed input 
 $a^1_1=(38\pm3)\times 10^{-3}\times m_\pi^{-3}$ to be less biased. 
    
Secondly, we have {\sl not} used the experimental phase shifts as input
 (except for the values of the scattering length). 
So, the values that follow from our expression (3.6), with 
 the parameters given in (3.14), constitute really a {\sl prediction} 
 for 
$\delta_1^1(t)$. 
This can be compared with the existing experimental values for this 
quantity,\ref{17} a comparison that may be found in \fig~5. 
The agreement is remarkable. 
The result one would have obtained if {\sl including} 
the phase shifts in the fit may be found after Table~1.

\topinsert{
\setbox0=\vbox{\hsize10.8truecm{\noindent\epsfxsize 9.5truecm\epsfbox{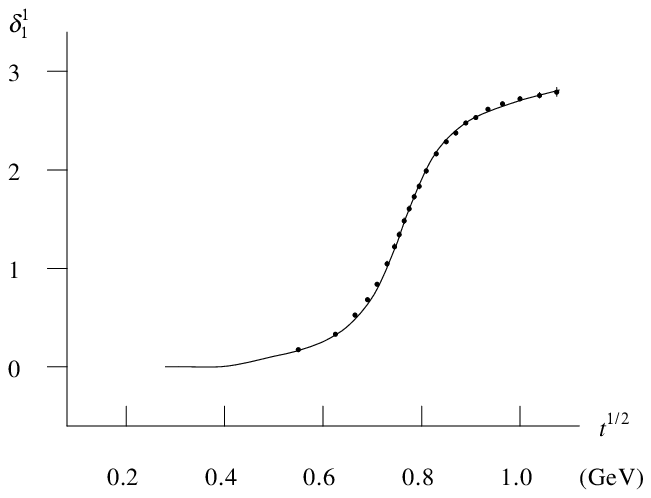}}} 
\setbox6=\vbox{\hsize 5.truecm\captiontype\figurasc{Figure 5. }{Our 
predicted phase $\pi\pi$ shift, $\delta_1^1$ (in radians), compared 
to the experimental values for the same (Solution 1 from 
Protopopescu et al., ref.~17).  
The experimental errors are of the order of the size of the black dots.
\vskip0.6truecm
}} 
\line{{\tightboxit{\box0}}\hfil\box6}
}
\endinsert

Before finishing this section we have to clarify the matter of 
 the $\omega$ and $\omega - \rho$ contribution to $a(2\pi;\,t\leq0.8\;\gev^2)$. 
Our fits to $e^+e^-$ data have actually been made including in the function $F_\pi$ as
given above, \equn{(3.2)}, a coefficient to take into account 
the $\omega\to2\pi$ contribution. To be precise, we have used the expression
$$F^{\rm all}_\pi(t)=F_\pi(t)\times
\dfrac{1+\sigma\dfrac{m^2_\omega}{m^2_\omega-t-\ii m_\omega\gammav_\omega}}{1+\sigma}
\equn{(3.17)}$$
where the notation is obvious. We take from the PDG\ref{16} the values for 
the mass and width of the $\omega$,
$$m_\omega=782.6\pm0.1\;\mev,\quad\gammav_\omega=8.4\pm0.9\;\mev,$$ 
and the fit gives a mixing parameter $\sigma=(16\pm1)\times10^{-4}$.
 
As is known, this Gounnaris--Sakurai\ref{18} parameterization is only 
valid for $t\simeq m_{\omega,\rho}^2$ and, 
in particular, its extrapolation to $t\sim0$ is not 
acceptable. 
This effect is very small, less than one part in a thousand. 
However,  to play it safe, we have also adopted the following alternate procedure: 
we have obtained a first approximation to $F_\pi$ by fitting the experimental data 
{\sl excluding} the region $0.55\;\gev^2\leq t\leq0.65\;\gev^2$. 
Then we have fitted only this region adding there also the $\omega$ piece, as in 
(3.17). The resulting value for $a(2\pi;\,t\leq0.8\;\gev^2)$ varies very little; 
it decreases by something between $2$ and $12\times10^{-11}$, 
depending on the fit. 
We may consider this as part of the theoretical error of our calculation, 
to be discussed in next subsection.

To finish this subsection, we present in Table~1 a comparison
 both with old results that also use 
analyticity properties, and a recent one (which does not).\fnote{A 
different case is the analysis of A.~Pich 
and J.~Portoles,
Phys. Rev.  {\bf D63} 093005,  (2001), which also uses the 
Omn\`es equation method. 
This paper  presents unfortunately a number of weak points.
For example, the authors use an incorrect 
analyticity structure for $\delta^1_1$ (their equations (8) and (A2), 
without left hand cut or inelasticity cut); they also employ
 a mere Breit--Wigner to describe the  phase in the rho region, 
while it is known  that the rho deviates from this 
(e.g., our term $b_1$).
They also forget the cut at high 
energy in their equivalent of our $G$ function. 
In what respects their results,  the sitation is as follows. 
The value Pich and Portol\'es give (in units of $10^{-11}$ and
for the contribution of $2\pi$ 
at energies below $1.2 \gev^2$) is
$5110\pm60 (PP)$, with a \chidof=33.8/21. 
This is substantially higher than all other results:
$5027\pm61\;$ ({ Narison}, $\tau$ and $e^+e^-$ data); 
$5044\pm67\; \hbox{(Our result, only timelike}\; \tau$ data), \chidof=53/48,  
and 
$5004\pm51$ (Our best result, including $e^+e^-$ and spacelike data),
 \chidof=213/204.
In the last three we have taken the piece $0.8\leq t\leq 1.2\;\gev^2$, equal
 to $230\pm5$, from $e^+e^-$ data. 
The result of Davier and H\"ocker (ref.~5), using tau data only, is essentially like
ours.  No doubt the bias introduced by the wrong parametrization of 
the paper of Pich and Portol\'es  is 
responsible for their discrepancy.}
\bigskip
\setbox0=\vbox{\petit
\medskip
\setbox1=\vbox{\offinterlineskip\hrule
\halign{
&\vrule#&\strut\hfil#\hfil&\quad\vrule\quad#&
\strut\quad#\quad&\quad\vrule#&\strut\quad#\cr
 height2mm&\omit&&\omit&&\omit&\cr 
&\hfil$4\,795\pm61$ \hfil&&\hfil N1\hfil&
&\hfil $\tau+e^+e^-$\hfil& \cr
 height1mm&\omit&&\omit&&\omit&\cr
\noalign{\hrule} 
height1mm&\omit&&\omit&&\omit&\cr
&\phantom{\Big|}$4\,730\pm100$&&\hfil N2\hfil&&\hfil$e^+e^-$ \hfil& \cr
\noalign{\hrule}
&\phantom{\Big|}$4\,846\pm50$&&\hfil CLY, AY\ref{4}\hfil&&\hfil$e^+e^-$\hfil& \cr
\noalign{\hrule}
&\phantom{\Big|}$4\,794\pm50$&&\hfil CLY-II\ref{4}\hfil&&\hfil $e^+e^-\,+\,\pi\pi$ ph. shifts
\phantom{l}\hfil& \cr
\noalign{\hrule}
&\phantom{\Big|}$4\,774\pm31$&&\hfil TY1\hfil&&\hfil$\tau+e^+e^-$\hfil&\cr
\noalign{\hrule}
&\phantom{\Big|}$4\,754\pm55$&&\hfil TY2\hfil&&\hfil only $e^+e^-$\hfil&\cr
 height1mm&\omit&&\omit&&\omit&\cr
\noalign{\hrule}}
\vskip.05cm}
\centerline{\box1}
\smallskip
\centerline{\petit Table 1}
\centerrule{6cm}
\medskip
\centerline{\petit Comparison of evaluations of $10^{11}\times a(2\pi;\,t\leq0.8\;\gev^2)$.}
 \centerline{\petit N1, N2 are in ref.~7. TY denotes our result here (statistical errors only 
for the $e^+e^-$ and spacelike data).}
\medskip}
\box0
\medskip

The difference  between the old CLY, AY and the new determinations 
is due to a large extent to the influence of the new Novosibirsk and 
NA7 data which allow us in particular to obtain a robust 
result: the CLY evaluation used only 18 
data points!   
In this respect, we note that, if we had included the $\pi\pi$ phase shifts in the fit (with 
also $\tau$ decay data) we would have obtained $4781\pm29$ for this $2\pi$ contribution,
 i.e., a shift of only 
7 units (as compared with a shift of 48 in the CLY-II evaluation). 
The value of the scattering length 
would be $a^1_1=(43\pm3)\times 10^{-3}m^{-3}_\pi$ now. 
 The corresponding $\chi^2/{\rm d.o.f.}$,
 276/227 with only statistical errors, is also  good.
This	 is an important proof of the stability of our results against introduction of 
extra data.  
(However, as explained above, we prefer the result without fitting phase shifts 
because of the  model-dependence of the last).    

The difference between the 
results of Narison (N),  who does not take into account the Fermi--Watson theorem or 
the spacelike data and TY, who do, is due in good part  to, precisely, the influence of the 
spacelike data which also help reduce the errors.

\booksubsection{3.3. Systematic and theoretical errors for the 
pion form factor contribution}
Errors included in this work are divided into statistical and systematic.

Evaluation of the statistical errors is standard: the fit procedure
(using the program MINUIT) provides the full error (correlation)
matrix at the $\chi^2$ minimum. This matrix is used when calculating the
corresponding integral for $a_{\mu}$, therefore incorporating
automatically all the correlations among the various fit parameters.

In addition, for every energy region, we have considered the errors
that stem from experimental systematics, as well as those originating
from deficiencies of the theoretical analysis.
The experimental systematics covers the errors given by the individual
experiments included in the fits. Also, when conflicting sets
of data exist, the calculation has been repeated, and the given systematic
error bar enlarged to encompass all the possibilities.

In general, errors (considered as uncorrelated) have been added in quadrature.
The exceptions are explicitly discussed along the text.

We next discuss the errors that stem from  experimental systematics,  
as well as those originating from deficiencies of the theoretical analysis 
for the $2\pi$ contribution,  
in the low energy region 
$4m^2_\pi\leq t\leq 0.8\,\gev^2$. 
We start with the  systematic errors of the data.  
They are evaluated by taking them into 
account in a new fit. 
In this way we find, in units of $10^{-11}$, and neglecting the mass differences corrections 
(i.e., using (3.9a) for tau data)
$$\eqalign{\hbox{Exp. Sys.}=&\pm40 \quad (e^+e^-\,+\,\tau)\cr
\hbox{Exp. Sys.}=&\pm66 \quad (e^+e^-).\cr
}$$
To estimate the degree of correlation of the systematic errors pertaining to  
several experiments is a difficult task; we choose to consider the full 
range from $0$ to $1$. 
The error bars given cover all the possibilities.
 The \chidof\ of the fit 
improves when taking these systematic errors into account to
$$\eqalign{
\chi^2/{\rm d.o.f.}=214/204&\quad (e^+e^-\,+\,\tau)\cr
\chi^2/{\rm d.o.f.}=145/154& \quad (e^+e^-).\cr
}$$

The error given for the 
case in which we include the decays of the tau 
would be smaller, and the \chidof\ would improve, if we used the correct kinematical formula for 
phase space, \equn{(3.10b)}; 
we would have obtained
$$\hbox{Exp. Sys.}=\pm30;\quad \chi^2/{\rm d.o.f.}=213/204\quad (e^+e^-\,+\,\tau).$$
In spite of this, we choose to accept the larger error ($\pm40$) as we feel that it includes 
residual effects of isospin breaking and electromagnetic corrections, 
different for the tau and $e^+e^-$ cases, that we will discuss at the end of this subsection.

 We discuss the systematic and theoretical errors in the higher energy regions  
 in next section, but we mention here that 
the systematic error 
($4\times10^{-11}$) for $2\pi$ between $t=0.8$ and $1.2\,\gev^2$ is  added coherently 
to the lower energy $2\pi$ piece. 

In addition to this we have several theoretical sources of error. 
First, that originating in the approximate character of the Gounnaris--Sakurai 
method for including the $\omega$. 
This we estimate as discussed at the end of \subsect~3.2, getting 
on the average $\pm7\times10^{-11}$. 
Then, the dependence of our results on $t_0$ can 
be interpreted as a theoretical uncertainty, 
that we take equal to $4\times10^{-11}$.  
Composing these errors
 quadratically, we can complete (3.13) to
$$a(2\pi;\,t\leq0.8\;\gev^2)=
\cases{
4\,774\pm31\;(\hbox{St.})\pm41\;(\hbox{Sys. +th.})=
4\,774\pm51\quad (e^+e^-\,+\tau\,+\;{\rm spacelike})\cr
4\,754\pm55\;(\hbox{St.})\pm66\;(\hbox{Sys. +th.})=
4\,754\pm86\quad (e^+e^-\,+\;{\rm spacelike}).\cr}
\equn{(3.19)}$$

To finish this subsection we will discuss in some detail some matters concerning  
to radiative corrections and 
isospin breaking in as much as they affect 
the error analysis. 
We will start with the analysis based on $e^+e^-$ data. 
When evaluating the pion form factor we have used formulas, deduced in particular from unitarity 
and analyticity, 
that only hold if we neglect electromagnetic (e.m.) interactions. 
However, experimentalists measure the pion form factor in the real world, 
where the $\pi^+\pi^-$ certainly interact electromagnetically. 
Not only this, but the initial particles ($e^+e^-$) also interact between themselves, 
and with the pions.

These last electromagnetic interactions, however, can be evaluated
 and they are indeed subtracted 
when presenting experimental data on $F_\pi$; the uncertainties this
 process generates are estimated 
and included in the errors provided with the data. 
We will thus only discuss the uncertainties associated to 
e.m. interactions of the $\pi^+\pi^-$ alone. 
These particles may exchange a photon, or radiate a soft photon that is not detected 
(see the corresponding figures in \subsect~5.2). 
We may then define two quantities: $F^{(0)}_\pi$, 
which is the form factor we would have if there were no e.m. interactions; 
and $F^{\rm(real)}_\pi$, which is the measured quantity, even after removal of 
radiative corrections for initial states or mixed ones. 
Actually, $F^{\rm(real)}_\pi$ depends on the experimental setup through 
the cuts applied to ensure that no ({\sl hard}\/) photon is emitted. 

Our formalism, as developed in \subsect~3.1, applies to $F^{(0)}_\pi$, 
but we fit $F^{\rm(real)}_\pi$. 
Therefore, we are introducing an ambiguity
$$F^{\rm(real)}_\pi-F^{(0)}_\pi$$
which is of order $\alpha$. 

The error induced by this ambiguity should be small. 
In fact, what enters into $a_\mu$ is the sum of 
the contribution of $F_\pi^{(\rm real)}$, 
which is what we actually fit, and that of 
the state $\pi^+\pi^-\gamma$, which can be obtained from the process
$$e^+e^-\to(\gamma)\to \pi^+\pi^-\gamma.$$ 
For this reason, we believe that the  error 
due to the mismatch of $F_\pi^{(\rm real)}$ and $F^{(0)}$   
is included in the errors to our fits here;\fnote{In 
particular for the evaluation including $\tau$ decay data, see below} 
the 
estimated error for the $\pi^+\pi^-\gamma$ contribution, 
$9\times10^{-11}$,  will be evaluated in  \subsect~5.2.

Tau decay presents the same difficulties, and we expect a similar uncertainty 
as for $e^+e^-$ collisions. 
But apart from the effect $F^{\rm(real)}_\pi\neq F^{(0)}_\pi$ discussed, 
it poses extra problems when relating it to $F_\pi$. To discuss this, we 
take first $m_u\neq m_d$, $\alpha=0$; then we choose  
 $\alpha\neq0$, but $m_u= m_d$. Higher effects, proportional to $\alpha(m_u- m_d)$, 
$\alpha^2$ and $(m_u- m_d)^2$  
shall be neglected.

For $\alpha=0$, the masses of $\pi^+$ and $\pi^0$ become equal, 
but isospin invariance is still broken. 
This means that, in particular, the quantity $\piv^S$ in (3.10a) is nonzero
 and therefore the experimentally measured $v_1$ does not coincide with that in 
(3.10b). We expect this effect to be small, since it is of second order, $(m_d-m_u)^2$. 
If the scale is  
the QCD parameter $\lambdav$, then this will be of relative size $10^{-4}$;
but other effects need not be so small. 
We have tried to take them into account by allowing for 
different parameters for $\rho^+$, $\rho^0$; this produced a
 substantial shift, of about $40\times10^{-11}$ for $a_\mu$. 

Then we set $m_u=m_d$ and take e.m. interactions to be nonzero. 
Apart from the effects already discussed, this produces
 the mass difference between charged and neutral pions. 
This we took (partially) into account 
when using the modified phase space of (3.10b).  
The ensuing effect for $a_\mu$ turned out to 
be small, $\sim4\times10^{-11}$.

Remnants of $\alpha\neq0$ and $m_u\neq m_d$ will likely still 
affect the determination of $F_\pi$ from $e^+e^-$ and $\tau$ decay data; 
but we feel that accepting the systematic/theoretical error 
of $40\times10^{-11}$ covers the related 
uncertainty.

\booksection{4. Contributions to $a^{(2)}({\rm h.v.p.})$ from
  $t>0.8\,\gev^2$. The full $a^{(2)}({\rm h.v.p.})$}
\vskip-0.7truecm
\booksubsection{4.1. The higher energy contributions, and
 the $3\pi$ contribution}
At higher energies we will get a substantial 
improvement over  determinations based on old data\ref{20} 
because of the existence of 
  very precise measurements from Novosibirsk\ref{9} and Beijing,\ref{12} 
gathered in the last two--three years,  
which will help remove a large part of the existing errors. 
This is particularly true of the region up to $t=3\;\gev^2$ which caused 
an important part of the total errors in pre-1998 calculations of $a^{(2)}({\rm h.v.p.})$. 
We turn to it next.
\bigskip
\noindent4.1.1. {\sl The region up to  $t= 3\;\gev^2$}
\medskip
We consider first the contribution of two, three, four pion, \dots,  and
$KK$ intermediate states for $0.8\leq t\leq1.2\,\gev^2$. 
In what follows n.w.a. will 
mean {\sl narrow width approximation},  r.d.a. {\sl resonance dominance approximation} 
(but not narrow approximation) and 
s.o.i.c. {\sl sum over individual channels}. 
For the n.w.a. we use the standard formula. 
Denoting by $\gammav_{ee}(V)$ to the width into $e^+e^-$ of a vector resonance $V$ with 
mass $M$, its contribution to $a^{(2)}(\hbox{h.v.p.})$ is given in this approximation by
$$a(V)=\dfrac{3\gammav_{ee}(V)\hat{K}(M^2)}{\pi M}.
\equn{(4.1)}$$
The uncertainty on $a(V)$ is calculated by Gaussian error propagation for
the parameters in (4.1). In practice, it is dominated by the
experimental error of the electronic width.

\topinsert
{
\setbox0=\vbox{\hsize10.truecm{\epsfxsize 8.5truecm\epsfbox{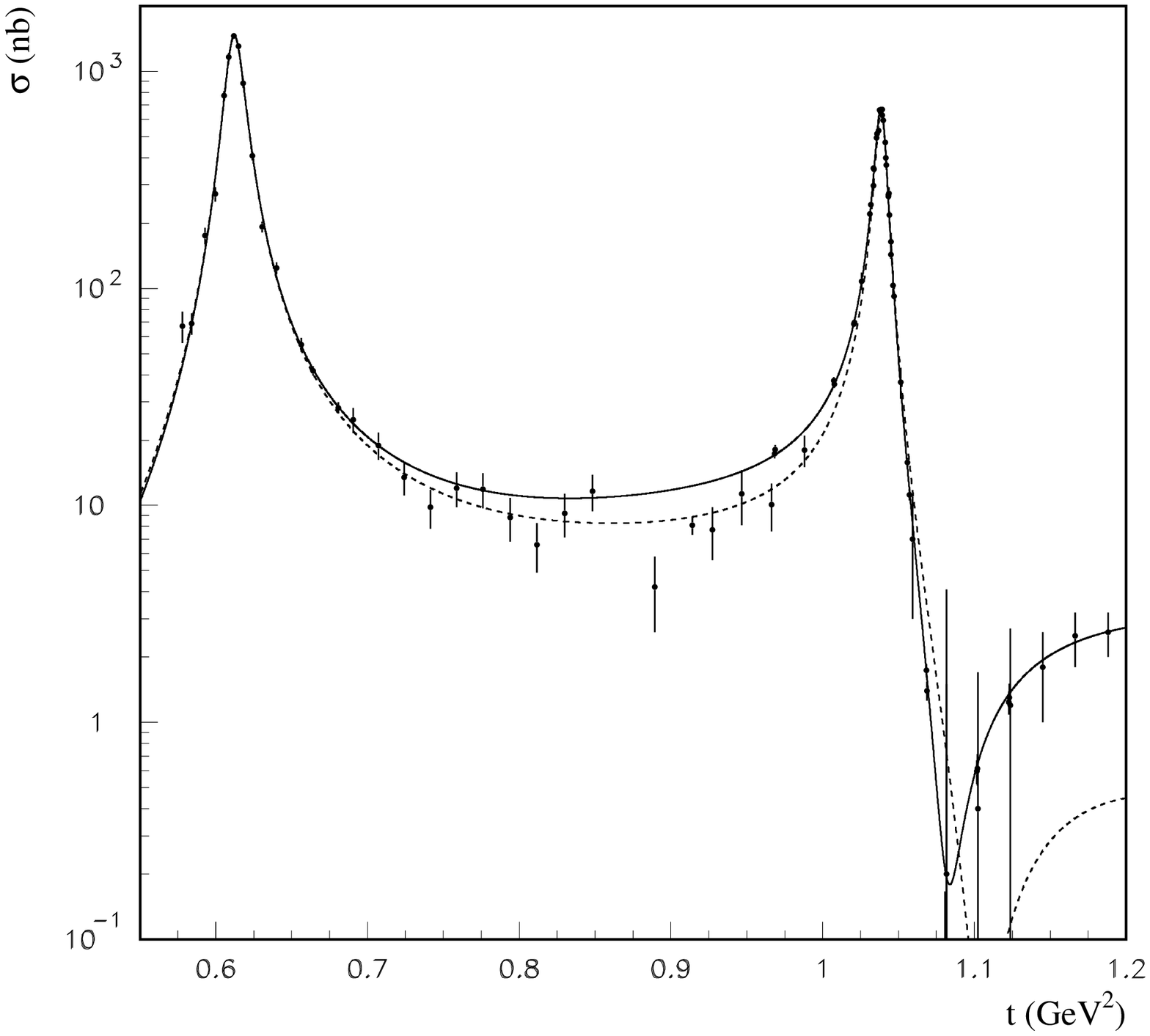}}\hfil} 
\setbox6=\vbox{\hsize 6truecm\captiontype\figurasc{Figure 6. }{Plot of 
the fit to the $e^+e^-\to3\pi$ cross section up to\hb $t=1.2\;\gev^2$, 
with data from ref.~9.
Continuous line: fit to CMD2 and SND data.
Dashed line: fit to CMD2 and ND.}\hb
\vskip.1cm} 
\medskip
\line{\box0\hfil\box6}
}
\endinsert

\medskip
\setbox1\vbox{\hsize 5truecm \noindent $3\pi$ states, $9m^2_\pi\leq t\leq1.2\;\gev^2$\hb
\hrule\medskip}
\box1
  
In the narrow width approximation one  gets  the $\omega$, $\phi$ 
contributions:
$$\eqalign{10^{11}\times a(3\pi;\omega)=&348\pm13\cr
10^{11}\times a(3\pi;\phi)=&62\pm3,\cr}
\equn{(4.2)}$$
but this misses the region between $\omega$ and $\phi$, 
and the interference effect just above the last. 
So we will use experimental data.\ref{9} This gives
$$10^{11}\times a(3\pi;t\leq1.2\;\gev^2)=438\pm4\,(\hbox{Stat.})\pm11\,(\hbox{Sys.}).
\equn{(4.3)}$$

The fit to the $3\pi$ experimental cross section, with data from ref.~9, may be found in \fig~6. 
The upper curve (continuous line in \fig~6) is a fit to the CMD2 and SND data. 
We have used a Breit--Wigner parameterization for the $\omega$ and $\phi$ 
resonances, plus a constant term 
and the exact threshold factor for $3\pi$ states. 
 The \chidof\ is $63/60$; 
we consider this our central result here.  The 
dashed  curve fits instead the data from CMD2 and ND (Dolinsky et al, ref.~20);
 the quality of the fit is poorer 
($\chidof=52/37$). It fits better the region between the $\omega$ and $\phi$, but fails to reproduce 
the data beyond $1.06\,\gev^2$. In fact, we include the second fit 
 to estimate the corresponding systematic uncertainty;
the small difference in terms of the integrals between the two fits, $8\times10^{-11}$, 
 is included into the systematic error.

\bigskip
\setbox1\vbox{\hsize 5.8truecm \noindent $2\pi$ states, $0.8\;\gev^2\leq t\leq1.2\;\gev^2$\hb
\hrule\medskip}
\box1
This $2\pi$ state contribution is
$$10^{11}\times a(2\pi;t\leq1.2\;\gev^2)=230\pm3\pm4.
\equn{(4.4)}$$

\topinsert
{
\setbox0=\vbox{\hsize9.truecm\hfil{\epsfxsize 7.2truecm\epsfbox{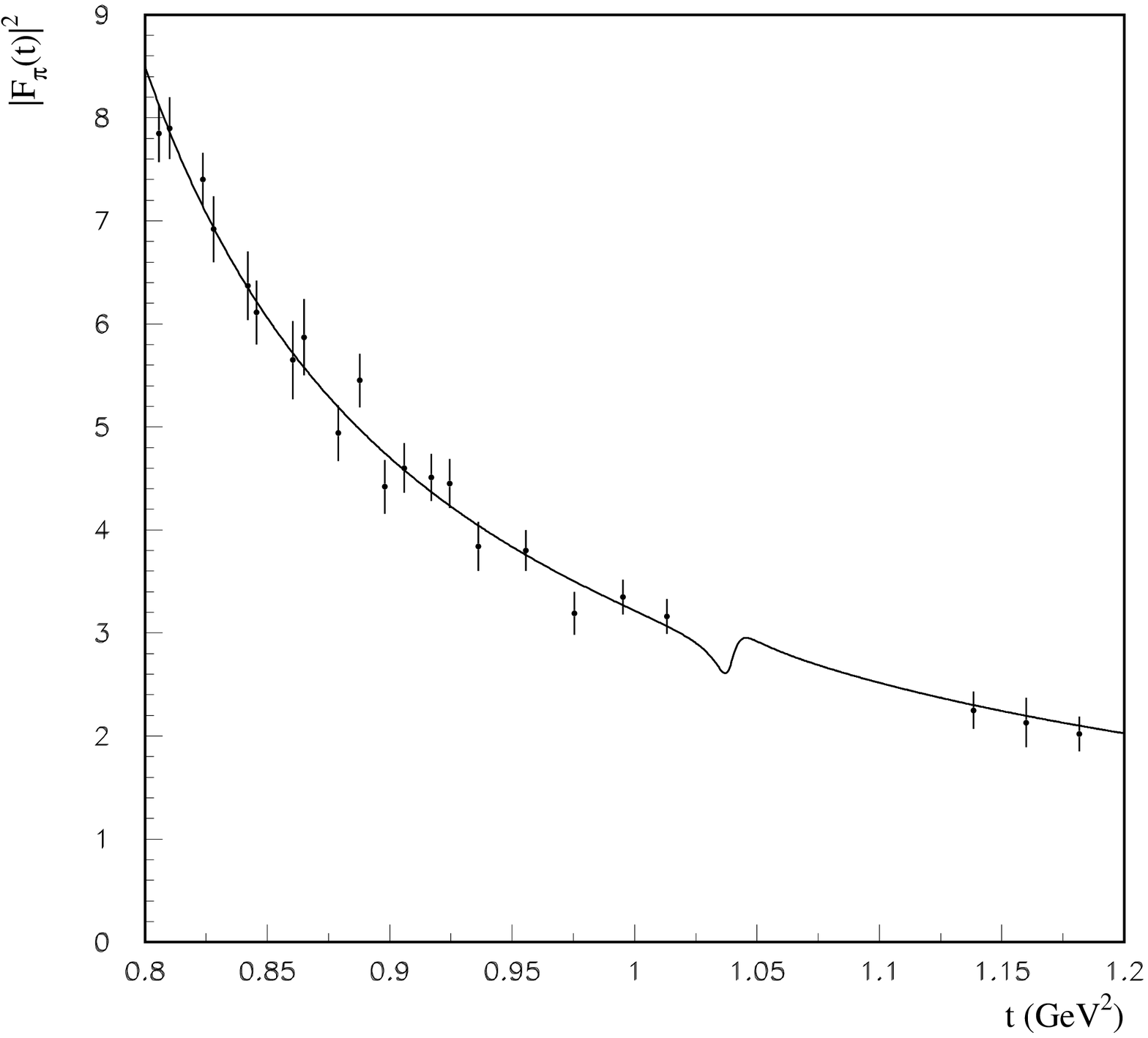}}\hfil} 
\setbox6=\vbox{\hsize 5truecm\captiontype\figurasc{Figure 7. }{Plot of 
the fit to $|F_\pi(t)|^2$ in the region $0.8\leq t\leq1.2\;\gev^2$. }\hb
\vskip.1cm} 
\line{\box0\hfil\box6}
}
\endinsert

The evaluation of the contribution of 
the $2\pi$ state has  greatly improved (with respect to older calculations) 
because of the information from recent Novosibirsk\ref{8} data 
on $e^+e^-\to2\pi$. 
We have fitted the  experimental value of $|F_\pi|^2$  with an expression $1/(bt+a)$, 
$a,\,b$ completely free parameters; 
the result of this  fit may be seen depicted in \fig~7. 
(A similar result is obtained if we extended our earlier calculation 
of $F_\pi(t)$ to $t\sim1.2\,\gev^2$ by setting 
$t_0=1.2$; but we prefer the result based only on experimental data.)  
Of the two errors given for the $2\pi$ contribution the first is statistical 
and the second, systematic, has been added {\sl coherently} 
to the systematic error on the low energy $2\pi$ contribution, as discussed in \subsect~3.3. 

The wiggle in \fig~7 for $t\sim m^2_\phi$ is due to the 
interference of the decay $\phi\to2\pi$. This is similar to 
the $\omega - \rho$ effect, and has been  
treated in a similar manner;  we have incorporated 
it using the formulas and parameters given by Achasov et al.\ref{9} 
The influence of this effect on $a_\mu$ is minute.

\bigskip
\setbox1\vbox{\hsize 6.truecm \noindent $KK$ states, $0.8\;\gev^2\leq t\leq1.2\;\gev^2$\hb
\hrule\medskip}
\box1

An important contribution is that of $KK$ states. In the n.w.a., this is given by 
the $\phi$: 
$$10^{11}\times a(KK; \phi)=332\pm9,\equn{(4.5)}$$
but this is a dangerous procedure here; the vicinity of the $KK$ threshold 
distorts the shape of the resonance. 
We thus have to calculate this $KK$ contribution directly from experiment. 
We have used two fitting procedures. 
In the first, we fit simultaneously the $K^+K^-$ and $K_LK_S$ data of Achasov et al.,\ref{9} 
 with the same 
parameters for the $\phi$. We get, 

$$10^{11}\times a(K^+K^-;\,t\leq 1.2\,\gev^2)=185.5\pm1.5\,(\hbox{Stat.})\pm13\,(\hbox{Sys.})
$$
and
$$10^{11}\times a(K_LK_S;\,t\leq 1.2\,\gev^2)=129.5\pm0.7.$$
The quality of the fit, shown in \fig~8, is good ($\chidof=84/82$).

In the second fitting procedure, we add the    
 $K_LK_S$  data of  Akhmetshin el al.,\ref{9}
 obtaining the result   
$$10^{11}\times a(K_LK_S;\,t\leq 1.2\,\gev^2)=128.4\pm0.5\,(\hbox{Stat.})\pm2.6\,(\hbox{Sys.}).$$
The fit is now less good,
 but the integrals are essentially identical for both fits. 
Adding the $KK$ results together we find
$$10^{11}\times a(KK;\,t\leq 1.2\,\gev^2)=314\pm2\,(\hbox{Stat.})\pm13\,(\hbox{Sys.}).
\equn{(4.6)}$$
The systematic errors have been obtained repeating 
the fits, including now the systematic errors given by the experiments.

We mention in passing that the ratio of contributions of $K^+K^-$ and
 $K_LK_S$, 1.44, 
agrees well with the   ratio\ref{16}
$$\dfrac{\gammav(\phi\to K^+K^-)}{\gammav(\phi\to K_LK_S)}=1.46\pm0.03.$$

\topinsert{
\setbox9=\vbox{
\setbox0=\vbox{\hsize16.4truecm\line{\hfil{\epsfxsize 7.8truecm\epsfbox{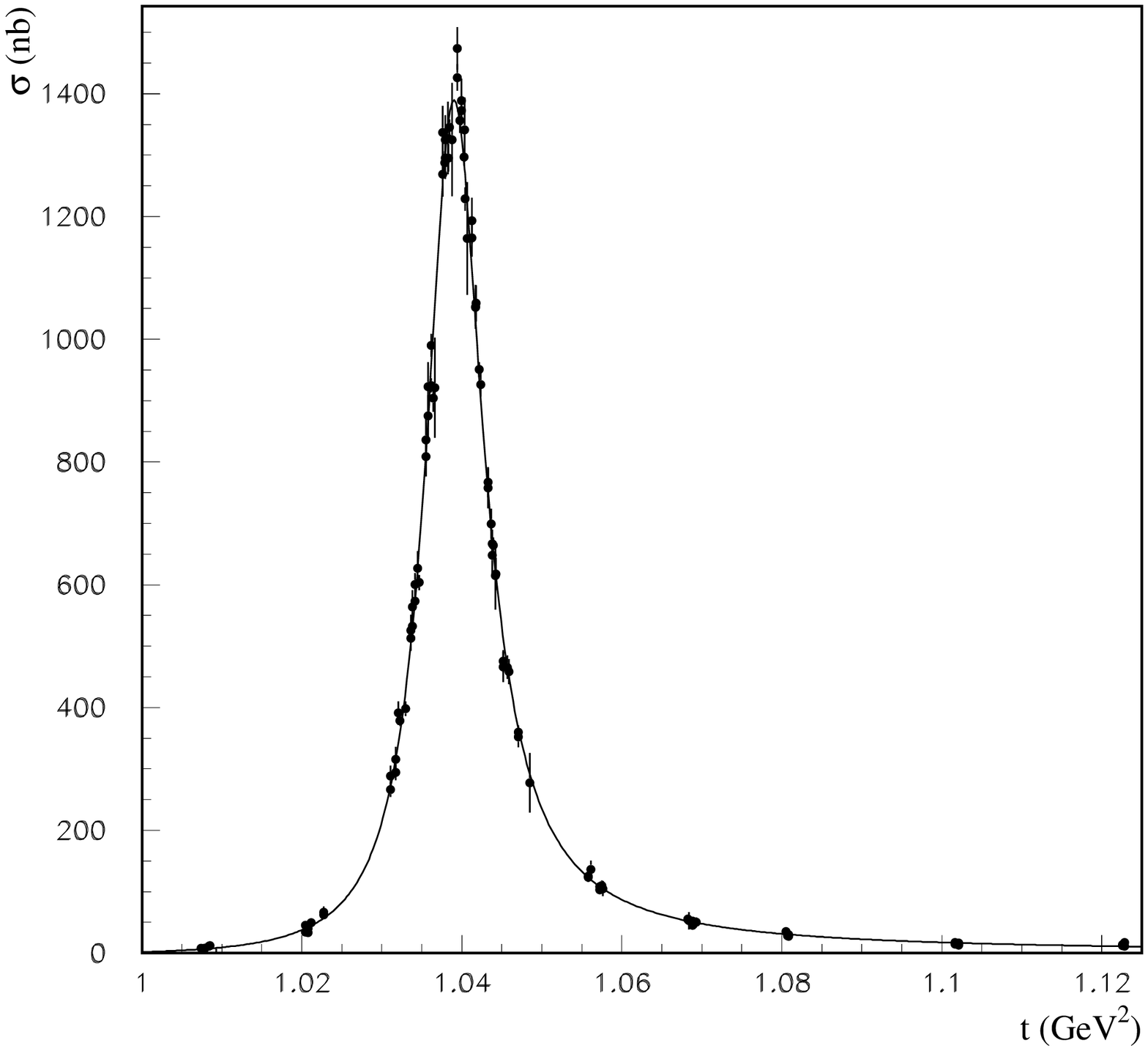}}\hfil
{\epsfxsize 7.8truecm\epsfbox{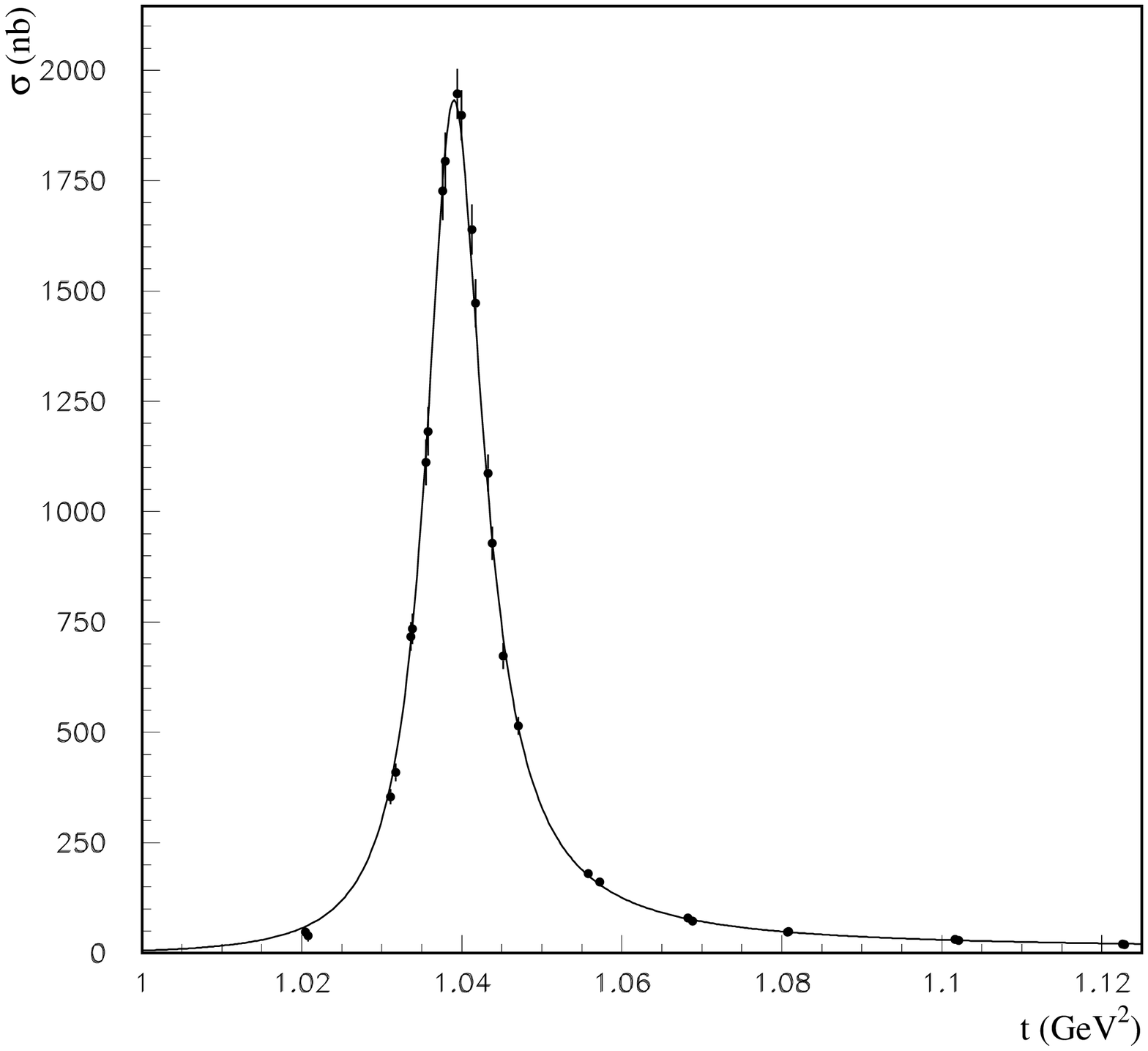}}\hfil}}
\setbox6=\vbox{\hsize 12truecm\captiontype\figurasc{Figure 8. }{Plot 
of the fit to the cross section   $e^+e^-\to K_LK_S$ (left), and to
  $e^+e^-\to K^+K^-$ (right). 
 Data from ref.~9.}} 
\bigskip
\centerline{\box0}
\centerline{\box6}
\bigskip}
\box9
}\endinsert 

\bigskip
\setbox1\vbox{\hsize 9.truecm \noindent Other states: $4\pi,\,5\pi,\,\eta\pi^0\pi^0\cdots$;
 $0.8\leq t\leq1.2\;\gev^2$\hb
\hrule\medskip}
\box1

The four pion contribution, including the quasi-two body state $\omega\pi$, may be obtained from 
recent Novosibirsk data,\ref{9} or from the compilation of 
Dolinsky et al.\ref{20} If we use the last we get
$$10^{11}\times a(4\pi;t\leq1.2\;\gev^2)=25\pm4;$$
if we fit the data of Akhmetshin et al.\ref{9} we find
$$10^{11}\times a(4\pi;t\leq1.2\;\gev^2)=18\pm3.$$
Of the $4\pi$ contribution most  is due to the $\omega\pi^0$ channel; only a small fraction 
($2.4\times10^{-11}$) comes from the $\pi^+\pi^-\pi^+\pi^-$ 
states. 
We take, for this $4\pi$ contribution, the figure
$$10^{11}\times a(4\pi;t\leq1.2\;\gev^2)=20\pm5,\equn{(4.7)}$$
which covers all possibilities.

The five, six, \tdots, pions as well as $\omega\to\eta+2\pi^0$ contributions are very small.\ref{16,20} 
Altogether, they give
$$10^{11}\times a(5\pi,\,6\pi,\,\eta\pi^0\pi^0,\cdots,\;t\leq1.2\;\gev^2)=4\pm2.
\equn{(4.8)}$$

We present the summary of our results 
in the important region $0.8\;\gev^2\leq t\leq 1.2$ plus the $3\pi$ 
contribution below $1.2\;\gev^2$ in the following Table~2:
\bigskip
\setbox0=\vbox{\petit
\medskip
\setbox1=\vbox{\offinterlineskip\hrule
\halign{
&\vrule#&\strut\hfil#\hfil&\quad\vrule\quad#&
\strut\quad#\quad&\quad\vrule#&\strut\quad#\cr
 height2mm&\omit&&\omit&&\omit&\cr 
&\hfil\phantom{l} Channels\hfil&&\hfil\hfil&
&\hfil Comments\hfil& \cr
 height1mm&\omit&&\omit&&\omit&\cr
\noalign{\hrule} 
height1mm&\omit&&\omit&&\omit&\cr
&\phantom{\Big|}$\pi^+\pi^-$&&\hfil$230\pm3\pm4$\hfil&&\hfil$0.8\leq t\leq 1.2\gev^2$ \hfil& \cr
\noalign{\hrule}
&\phantom{\Big|}$3\pi$&&\hfil$438\pm4\pm11$\hfil&&\hfil$9m^2_\pi\leq t\leq 1.2\gev^2$\phantom{l}\hfil& \cr
\noalign{\hrule}
&\phantom{\Big|}$K^+K^-$&&\hfil$186\pm2\pm13$\hfil&&\hfil\hfil& \cr
\noalign{\hrule}
&\phantom{\Big|}$K_LK_S$&&\hfil$128\pm1\pm2$\hfil&&\hfil\hfil&\cr
\noalign{\hrule}
&\phantom{\Big|}$4\pi$&&\hfil$20\pm5$\hfil&&\hfil including $\omega\pi^0$\hfil&\cr
\noalign{\hrule}
&\phantom{\Big|}Multipion, $\eta+2\pi, \cdots$&&\hfil$4\pm2$\hfil&&\hfil\hfil&\cr
\noalign{\hrule}
&\phantom{\Big|}&\omit&\hfil\hfil&\omit&\hfil\hfil&\cr
\noalign{\hrule}
&\phantom{\Big|}{\sc Total}&&\hfil$1\,006\pm19$\hfil&&Syst. error for $2\pi$ 
not included &\cr
 height1mm&\omit&&\omit&&\omit&\cr
\noalign{\hrule}}
\vskip.05cm}
\centerline{\box1}
\smallskip
\centerline{\petit Table 2}
\centerrule{6cm}
\medskip
\centerline{Contribution to $a^{(2)}$ of various channels up
 to $t=1.2\;\gev^2$ ($2\pi$ below $0.8\,\gev^2$ {\sl not} included).}
\medskip}
\box0
\medskip

\setbox1\vbox{\hsize 4.cm \noindent $1.2\;\gev^2\leq t\leq2\;\gev^2$\hb
\hrule\medskip}
\box1

We consider three determinations:
$$\eqalign{
270\pm27&\quad \hbox{(Here)}\cr
278\pm25&\quad \hbox{(s.o.i.c., CLY\ref{4})}\cr
265\pm22&\quad \hbox{(VMD+QCD; AY)}.\cr}\equn{(4.9)}$$

The first is obtained from a numerical integration of the data,\ref{20} 
with a parabolic fit.  
The method referred to as ``VMD+QCD; AY", details of which can be found 
in the AY\ref{4} paper, consists in interpolating between a vector meson dominance 
(VMD) calculation for quasi-two~body processes ($\omega\pi,\;\rho\pi, \dots$),
 plus a Breit--Wigner expression for
 two-body channels ($\pi\pi,\,KK$, \dots) at the lower end,
 and perturbative QCD at the upper end, 
the interpolation being obtained by fitting experimental data (see \fig~9). 
Because we want to present a result as  model-independent 
 as possible, we will take as our preferred figure that obtained here from experimental data: 
$$10^{11}\times a(1.2\;\gev^2\leq t\leq2\,\gev^2)=270\pm27.\equn{(4.10)}$$

\topinsert{
\setbox0=\vbox{\hsize9.truecm{\epsfxsize 7.4truecm\epsfbox{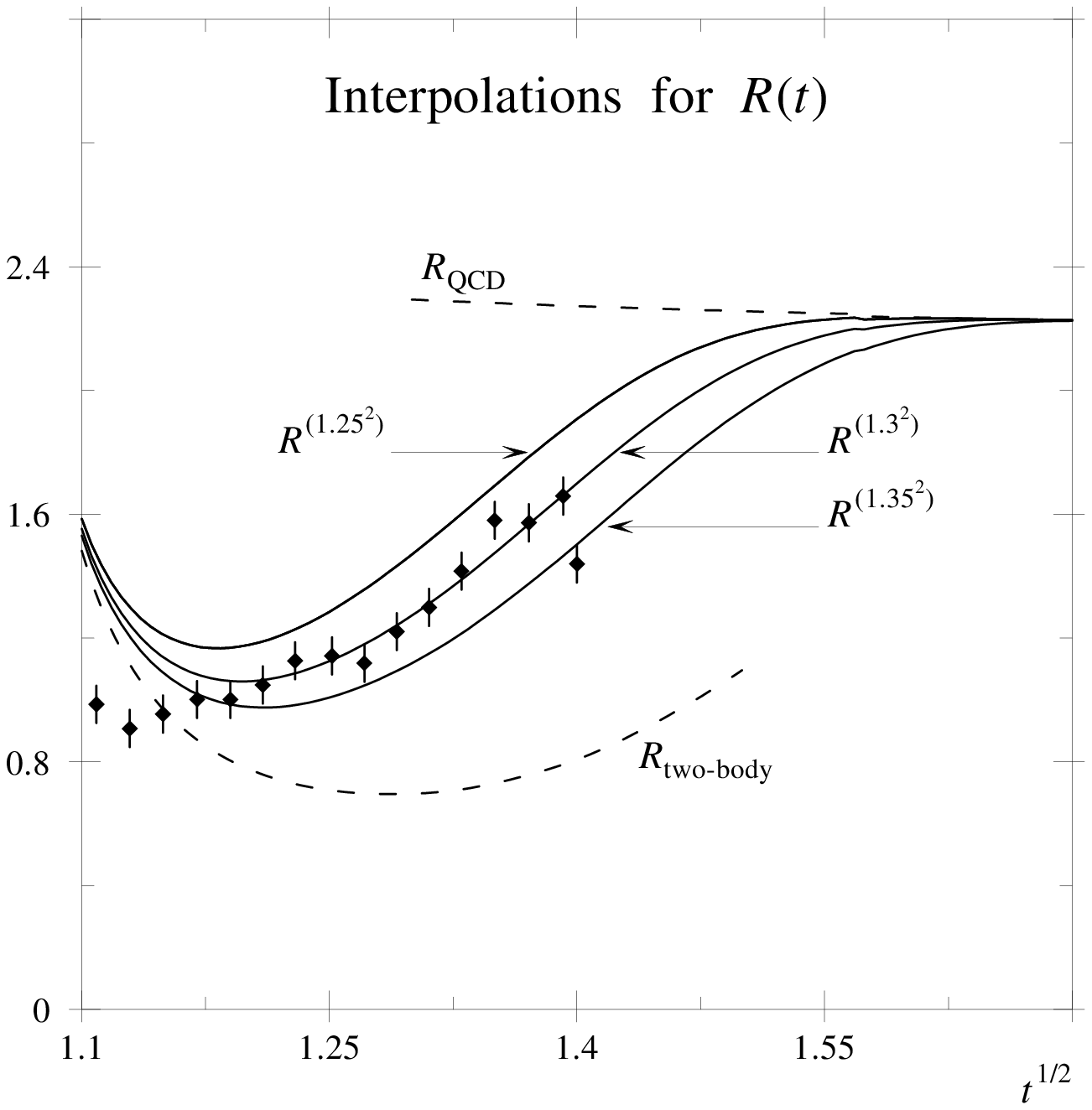}}} 
\setbox6=\vbox{\hsize 6.truecm\captiontype\figurasc{Figure 9. }{Experimental 
data and various interpolations between the VMD calculation, for small $t$, 
and QCD for larger $t$. From AY, ref.~4; data from ref.~20.}\hb
\vskip.1cm} 
\medskip
\line{\box0\hfil\box6}
\medskip
}\endinsert
 
\bigskip

\setbox1\vbox{\hsize 4cm \noindent $2\;\gev^2\leq t\leq3\;\gev^2$\hb
\hrule\medskip}
\box1

$$\eqalign{240&\pm3\,(\lambdav)\pm3\,(\hbox{Cond.) ({\rm QCD})}\cr
250&\pm19\qquad \hbox{(N\ref{7}, r.d.a.; only}\;e^+e^-)\cr
276&\pm36\qquad \hbox{(N\ref{7}, r.d.a.};\;e^+e^-\,+\,\tau)\cr
222&\pm5\;\hbox{(St.)}\pm15\;\hbox{(Sys.)}\quad \hbox{(J, exp. data}).\cr}
\equn{(4.11)}$$
(J) here denotes an evaluation, integrating with the trapezoidal rule, of a compilation 
 of experimental data 
supplied by F.~Jegerlehner.

For the QCD calculations we take the following approximation: for $n_f$ massless quark 
flavours, 
with charges $Q_f$, we write
$$\eqalign{R^{(0)}(t)=&\;3\sum_fQ_f^2\Bigg\{1+\dfrac{\alpha_s}{\pi}+
(1.986-0.115n_f)\left(\dfrac{\alpha_s}{\pi}\right)^2\cr
+&\;\Big[-6.64-1.20n_f-0.005n_f^2-1.240\dfrac{(\sum_f Q_f)^2}{3(\sum_f Q_f^2)}\Big]
\left(\dfrac{\alpha_s}{\pi}\right)^3\Bigg\}.\cr}
$$
To this one adds mass and nonperturbative corrections. 
We take into account the $O(m^{2})$ effect for quarks with 
running mass $\bar{m}_i(t)$, which correct $R^{(0)}$ by the amount\fnote{The 
 corrections of order $m^4$ may be found in 
the paper of Narison,\ref{7} together with references. 
We have checked that the effect of this correction is smaller than 
the errors of the leading terms.} 
$$-3\sum_iQ_i^2\bar{m}^2_i(t)\left\{6+28\dfrac{\alpha_s}{\pi}
+(294.8-12.3n_f)\left(\dfrac{\alpha_s}{\pi}\right)^2\right\}t^{-1}.$$
Finally, for the condensates we add
$$\dfrac{2\pi}{3}\left(1-\tfrac{11}{18}\dfrac{\alpha_s}{\pi}\right)\langle\alpha_s G^2\rangle
\sum_f Q^2_f$$
and
$$24\pi^2\left[1-\tfrac{23}{27}\dfrac{\alpha_s}{\pi}\right]m_i\langle\bar{\psi}_i\psi_i\rangle.$$
We neglect the condensates corresponding to heavy quarks ($c,\,b$) and 
express those for $u,\,d,\,s$ in terms of 
$f^2_\pi m^2_\pi$, $f^2_K m^2_K$ using the well-known PCAC relations.

In the QCD calculation, the error labeled ``Cond." is found by 
inserting the variation obtained  
setting quark and gluon condensates to zero, and that labeled 
$\lambdav$ by varying the QCD parameter. 
For this parameter 
we take the recent determinations\ref{21} 
that correspond to the value
$$\alpha_s(M_Z^2)=0.1172\pm 0.003;$$
to be precise, we have taken (in \mev, and to four loops),
$$\lambdav=373\pm80,\;t\leq m_c^2;\quad \lambdav=283\pm50, \;m_c^2\leq t\leq m_b^2;\quad
\lambdav=199\pm30,\;t\geq m_b^2.$$
For the gluon condensate we take $\langle\alpha_s G^2\rangle=0.07\,\gev^4$.

The four evaluations give comparable results, with the r.d.a. ones  
larger,  and presenting also larger errors.  
As proved by the reliability of QCD calculations of semileptonic $\tau$ decays, 
a similar process in a similar energy range, we think  perturbative QCD can be 
trusted here, 
so we select the corresponding value as our best result. 
We write thus
$$10^{11}\times a(2\;\gev^2\leq t\leq3\,\gev^2)=240\pm6,\equn{(4.12)}$$
where we have added linearly the errors due to $\lambdav$ and the condensates.

As a verification of the reliability of 
the calculation, as well as the improvement it presents when compared with earlier determinations, in the 
rather involved energy range $0.8\leq t\leq 3\,\gev^2$ (including 
here the full $3\pi$ contribution), 
we compare our value here (adding, for the occasion, 
the channels $\omega,\,\phi\to\pi^0\gamma,\,\eta\gamma$, see \subsect~5.2)
 with that obtained by Narison\ref{7} who uses
resonance dominance and s.o.i.c., and to the old CLY\ref{4} evaluation, with  s.o.i.c. 
and QCD:
$$\eqalign{
10^{11}\times a(0.8\;\gev^2\leq t\leq3\,\gev^2 +\, \omega\to3\pi)=&1\,559\pm34\;\hbox{(Here)}\cr
10^{11}\times a(0.8\;\gev^2\leq t\leq3\,\gev^2 +
\,\omega\to3\pi)=&1\,631\pm46\;\hbox{(Narison},\;\tau+e^+e^-)\cr
10^{11}\times a(0.8\;\gev^2\leq t\leq3\,\gev^2 +\,
\omega\to3\pi)=&1\,675\pm65\;\hbox{(Narison},\;e^+e^-)\cr
10^{11}\times a(0.8\;\gev^2\leq t\leq3\,\gev^2 +\,
\omega\to3\pi)=&1\,618\pm97\;\hbox{(CLY},\;e^+e^-)\cr}
\equn{(4.13)}$$
The compatibility between the  results, using 
different methods of 
evaluation for many pieces, is  reasonable.     
\bigskip
\noindent4.1.2. {\sl The region $3\;\gev^2\leq t\leq 4.6^2\;\gev^2$}
\medskip
This is another region where the availability of recent precise data\ref{12} 
in the neighbourhood of the 
$\bar{c}c$ threshold, previously poorly known,   
permits a reliable evaluation. 
As a byproduct, we get an experimental validation 
of QCD calculations.
\medskip
\goodbreak

\setbox1\vbox{\hsize 4.0cm \noindent $3\;\gev^2\leq t\leq2^2\;\gev^2$\hb
\hrule\medskip}
\box1
We use perturbative QCD here and get
$$10^{11}\times a(3 - 2^2\,\gev^2)=120\pm0.8\,(\lambdav)\pm0.8\,(\hbox{Cond.)}.$$

\bigskip
\goodbreak
\setbox1\vbox{\hsize 4.cm \noindent $2^2\;\gev^2\leq t\leq3^2\;\gev^2$\hb
\hrule\medskip}
\box1
We have now very good recent experimental data. 
So we present two determinations:
$$\eqalign{10^{11}\times a(2^2 - 3^2\,\gev^2)=200\pm1\,(\lambdav)&\quad \hbox{(QCD)}\cr
10^{11}\times a(2^2 - 3^2\,\gev^2)=210\pm3 {\rm (St.)}\pm14{\rm (Sys.)}&\quad \hbox{(Exp. BES)}\cr}$$
We only give the error due to $\lambdav$ here 
because that due to the condensates 
is negligible. When integrating the BES data we 
have used the trapezoidal rule. 
If instead we fitted a horizontal line, we would have obtained
$$10^{11}\times a(2^2 - 3^2\,\gev^2)=207\pm2 {\rm (St.)}\pm13{\rm (Sys.)}\quad \hbox{(Exp. BES)}$$ 
The BES\ref{12} purely experimental result 
and the QCD calculation 
are compatible, but one has to take into account  the 
{\sl systematic} errors of the 
first. This shows clearly the importance of systematic variations in 
$e^+e^-$ annihilations data. We take as our preferred value for the sum 
of the two intervals that obtained from the QCD calculations:
$$10^{11}\times a(3\;\gev^2\leq t\leq3^2\,\gev^2)=320\pm2\pm1=320\pm3.\equn{(4.14)}$$
 
\goodbreak
\medskip
\setbox1\vbox{\hsize 4.4cm \noindent $3^2\;\gev^2\leq t\leq4.6^2\;\gev^2$\hb
\hrule\medskip}
\box1
We give here the results in units of $10^{-11}$. We separate 
the contribution of the $J/\psi$, $\psi'$, 
that we calculate in the n.w.a., and the rest. For the first we have,
$$\eqalign{62.0\pm 4.0&\quad J/\psi\cr
14.8\pm1.3&\quad \psi'\cr}$$
\medskip

For the remainder we have the following possibilities:
$$\eqalign{91\pm0.4\,(\lambdav)&\quad uds;\;\hbox{(QCD; $3^2\;\gev^2\leq t\leq4.6^2\;\gev^2$)}\cr
4.0\pm0.4&\quad \psi'',\psi''',\psi^{IV}\;\hbox{(N, r.d.a.)}\cr}$$
\smallskip
\noindent{\sc Total:} $172\pm4\qquad$ (N; QCD+r.d.a.)
\medskip
$$\eqalign{91\pm0.4\,(\lambdav)&\quad uds;\;\hbox{(QCD;  $3^2\;\gev^2\leq t\leq4.6^2\;\gev^2$)}\cr
46.8 - 28.6 \to 38\pm10&\quad\hbox{$\bar{c}c$. (AY, NRQCD)}\cr}$$
\smallskip
\noindent{\sc Total:} $206\pm11\qquad$ (AY; QCD+NRQCD)
\medskip
$$\eqalign{54.7\pm0.3\,(\lambdav)& \quad\hbox{(QCD)}: 3.0^2\leq t\leq 3.7^2\cr
56\pm0.3\pm3& \quad\hbox{(Exp., BES)}: 3.7^2\leq t\leq 4.6^2\cr}$$
\smallskip
\noindent{\sc Total:} $188\pm4\pm3\qquad$ (Exp., BES+QCD)
\medskip
Here N refers to the paper of Narison\ref{7}, and AY is in ref.~4. 
BES are the experimental data from ref.~12. The first error 
for them is the statistical, the second the systematic one. 

This region merits a somewhat detailed discussion, as there is 
a certain controversy 
about it. We have 
made the calculation in three different manners. 
First, we separate the $u,d,s$ quarks contribution, 
that can be evaluated using perturbative QCD. 
The contribution of the $\bar{c}c$ states is then evaluated saturating 
it by the $\psi$ resonances, in the r.d.a.; this is the result labeled (N, r.d.a.). 
This saturation procedure does not produce a good description.

\topinsert{
\setbox0=\vbox{\hsize16.4truecm{\epsfxsize 15.truecm\epsfbox{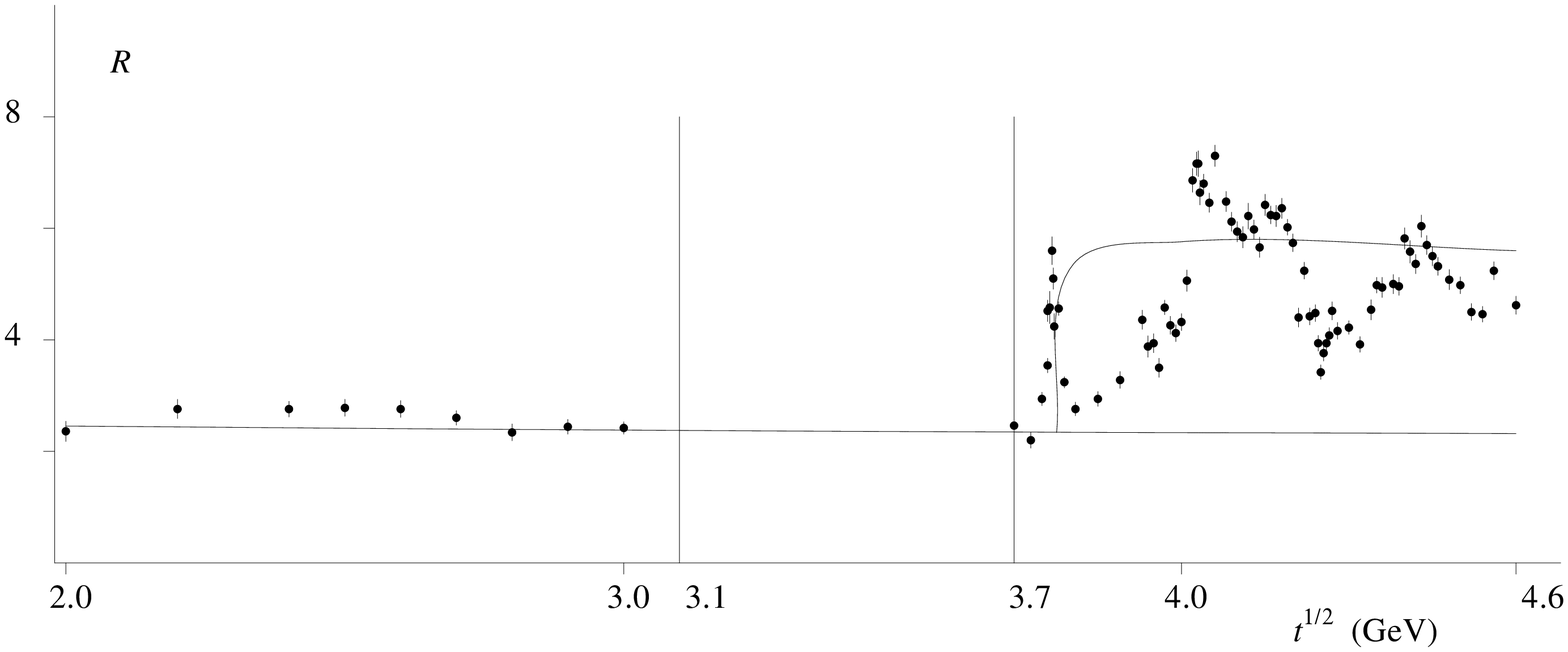}}\hfil} 
\setbox6=\vbox{\hsize 13.8truecm\captiontype\figurasc{Figure 10. }{Plot of 
BES experimental 
data and QCD for the $u,\,d,\,s$ quarks (lower $t$) 
and the same plus NRQCD for the $c$ quark contribution, from $t=3.74^2\;\gev^2$ 
to $4.6^2\;\gev^2$  
at the right. Only {\sl systematic} errors shown for experimental data. 
Statistical ones are even smaller.}\hb
\vskip.1cm} 
\medskip
\centerline{\tightboxit{\box0}}
\medskip
\centerline{\box6}
\medskip
}
\endinsert
 
In a second method one separates also the $u,d,s$ contribution; but the 
 $\bar{c}c$ one is treated differently. 
If a resonance is below the channel for open charm production, which is set 
at $t=4m^2_c$ (with $c$ the pole mass of the $c$ quark), then it is treated 
as a bound state, in the n.w.a. Above $\bar{c}c$ threshold, one uses nonrelativistic QCD 
(see AY, ref.~4 and ref.~22). 
The two values reported above for such a calculation (AY, NRQCD) 
are for two values of the $c$ quark mass: $m_c=1.750\,\gev$, in 
which case only the $J/\psi$ should be taken to be 
below threshold, and $m_c=1.866\,\gev$ and then both $J/\psi$ and $\psi'$ are 
to be added below threshold. This last gives the smallest number (28.6).
 The result of the calculation is taken as
the average  of both numbers, with half the difference as the estimated error.
In \fig~10 one can see the BES\ref{12} data and the predictions of  
QCD and NRQCD, the last for $m_c=1.87\;\gev$.

The third method, which is the one that yields our preferred number,
$$10^{11}\times a(3^2\;\gev^2\leq t\leq4.6^2\,\gev^2)=
188\pm4\pm3,\equn{(4.15)}$$
is obtained by using QCD for $u, d, s$ quarks {\sl plus} 
$J/\psi,\,\psi'$ below $t=3.7^2$, and experimental data (BES\ref{12}) above that energy.

All three methods give overlapping results, within errors, 
with the r.d.a. below experiment, and with an 
underestimated error, and with the NRQCD calculation reproducing  better  
the data. 
This NRQCD calculation depends strongly on the mass of the $c$ quark and, in fact, one can 
turn the argument backwards and {\sl predict} 
$m_c$ by requiring equality with the experimental figure. If we do so, we find 
$$m_c\simeq1.89\;\gev,$$
a very reasonable estimate consistent with the  two loop result,\ref{23} 
correct  to  $O(\alpha_s^4)$, which gives   
$m_c=1.866\pm0.20\;\gev$.

\bigskip\goodbreak
\noindent4.1.3. {\sl The region $t\geq4.6^2\;\gev^2$}
\medskip
The results from this region have not changed noticeably, but we give them for completeness.
\medskip\goodbreak
\setbox1\vbox{\hsize 4.5cm \noindent $4.6^2\;\gev^2\leq t\leq11.2^2\;\gev^2$\hb
\hrule\medskip}
\box1
For the first $\Upsilonv$ resonances, and in units of $10^{-11}$,
$$\eqalign{0.55\pm0.03&\quad \Upsilonv\cr
0.18\pm0.01&\quad \Upsilonv'.\cr}$$
Then,
$$88.8\pm1.0\,(\lambdav)\quad udsc:\;\hbox{(QCD)},\;4.6^2\;\gev^2\leq t\leq11.2^2\;\gev^2$$

$$\eqalign{0.22\pm0.04& \quad \bar{b}b: \hbox{(N, n.w.a.)}, \Upsilonv'',\Upsilonv''',\dots\cr
0.53\pm0.08& \quad \bar{b}b:\hbox{(AY, NRQCD)}\cr}$$
Adding this, we get

\noindent{\sc Total:} $90\pm1\qquad$ (N; QCD+n.w.a.)

\noindent{\sc Total:} $90\pm1\qquad$ (AY; QCD+NRQCD)

The notation is like for the $c$ threshold region. 
The error in the (AY, NRQCD) evaluation is due to the 
error in the QCD parameter, $\lambdav$, and the $b$ quark pole mass, 
for which we have taken\ref{23} $m_b=5.00\pm0.10\;\gev$.  
Both figures are essentially identical and we thus take
$$10^{11}\times a(4.6^2\;\gev^2\leq t\leq11.2^2\,\gev^2)=90\pm1.
\eqno{(4.16)} $$

\bigskip
\setbox1\vbox{\hsize 5truecm \noindent $11.2^2\;\gev^2\leq t\to\infty\;\gev^2$\hb
\hrule\medskip}
\box1
The use of QCD is  mandatory here. The contribution above $\bar{t}t$ 
threshold is negligible, so we calculate with $n_f=5$ and  get,
$$10^{11}\times a(11.2^2\;\gev^2\leq t\to \infty)=21\pm0.1\,(\lambdav).
\equn{(4.17)}$$

\booksubsection{4.2. The whole  $a^{(2)}({\rm h.v.p.})$}
 Our final result for the $O(\alpha^2)$ hadronic 
contribution to $a_\mu$ is then
$$10^{11}\times a^{(2)}(\hbox{h.v.p.})=\cases{
6\,909\pm64\quad (e^+e^-\,+\,\tau\,+\,\hbox{spacel.})\cr
6\,889\pm96\quad (e^+e^-\,+\,\hbox{spacel.})\cr}
\equn{(4.18)}$$
 
To compare with other evaluations we have to add the contribution 
 $(43\pm4)\times10^{-11}$  of some of   
the radiative decays of the $\rho$, $\omega$, $\phi$  
(see \sect~5.2)  that  other authors include. This 
comparison is shown, for a few representative calculations,\fnote{A more 
complete list of evaluations, including some of the very earliest ones, 
may be found in the paper of Narison, ref.~7.}  in Table~3.

\bigskip
\setbox0=\vbox{\petit
\medskip
\setbox1=\vbox{\offinterlineskip\hrule
\halign{
&\vrule#&\strut\hfil#\hfil&\quad\vrule\quad#&
\strut\quad#\quad&\quad\vrule#&\strut\quad#\cr
 height2mm&\omit&&\omit&&\omit&\cr 
&\hfil\phantom{l} Authors\hfil&&\hfil$10^{11}\times a(\hbox{h.v.p.})$\hfil&
&\hfil Comments\hfil& \cr
 height1mm&\omit&&\omit&&\omit&\cr
\noalign{\hrule} 
height1mm&\omit&&\omit&&\omit&\cr
&\phantom{\Big|}KNO&&\hfil$7\,068\pm174$\hfil&&\hfil$e^+e^-$ only\hfil& \cr
\noalign{\hrule}
&\phantom{\Big|}CLY&&\hfil$7\,100\pm116$\hfil&&\hfil$e^+e^-\,+\,$ spacel.\phantom{l}\hfil& \cr
\noalign{\hrule}
&\phantom{\Big|}CLY-II&&\hfil$7\,070\pm116$\hfil&&\hfil$e^+e^-\,+\,$ sp. + $\pi\pi$ ph. shifts\phantom{l}\hfil& \cr
\noalign{\hrule}
&\phantom{\Big|}ADH&&\hfil$7\,011\pm94$\hfil&&\hfil$e^+e^-\,+\,\tau$\hfil&\cr
\noalign{\hrule}
&\phantom{\Big|}BW&&\hfil$7\,026\pm160$\hfil&&\hfil$e^+e^-$\hfil&\cr
\noalign{\hrule}
&\phantom{\Big|}AY&&\hfil$7\,113\pm103$\hfil&&\hfil$e^+e^-\,+\,$ spacel.\hfil&\cr
\noalign{\hrule}
&\phantom{\Big|}DH&&\hfil$6\,924\pm62$\hfil&&\hfil$e^+e^-\,+\,\tau$\hfil&\cr
\noalign{\hrule}
&\phantom{\Big|}J&&\hfil$6\,974\pm105$\hfil&& &\cr
\noalign{\hrule}
&\phantom{\Big|}N1&&\hfil$7\,031\pm77$\hfil&&\hfil$e^+e^-\,+\,\tau$\hfil&\cr
\noalign{\hrule}
&\phantom{\Big|}N2&&\hfil$7\,011\pm117$\hfil&&\hfil$e^+e^-$ only\hfil& \cr
\noalign{\hrule}
&\phantom{\Big|}TY1&&\hfil$6\,952\pm64$\hfil&&\hfil$e^+e^-\,+\,\tau\,+\,$spacel.\hfil&\cr
\noalign{\hrule}
&\phantom{\Big|}TY2&&\hfil$6\,932\pm96$\hfil&&\hfil$e^+e^-\,+\,$ spacel. only\phantom{l}\hfil&\cr
 height1mm&\omit&&\omit&&\omit&\cr
\noalign{\hrule}}
\vskip.05cm}
\centerline{\box1}
\smallskip
\centerline{\petit Table 3}
\centerrule{6cm}
\medskip
\centerline{KNO: ref.~3; BW: ref.~3; J: ref.~6; CLY, CLY-II, AY: ref.~4; N: ref.~7; DH, ADH: ref.~5.}
\medskip}
\box0
\medskip

If we had added also the other radiative contributions ($\pi\pi\gamma$, and the 
continuum ${\rm hadron}+\gamma$, cf. \subsect~5.2) we would have found 
the hadronic vacuum polarization piece, correct to 
order $\alpha^2$ and $\alpha^3$,
$$10^{11}\times a^{(2+3)}(\hbox{h.v.p.})=\cases{
7\,002\pm66\quad (e^+e^-\,+\,\tau\,+\,\hbox{spacel.})\cr
6\,982\pm97\quad (e^+e^-\,+\,\hbox{spacel.})\cr}
\equn{(4.19)}$$

\booksection{5. Higher order hadronic contributions}
\vskip-0.7truecm
\booksubsection{5.1. Hadronic light-by-light contributions}
A contribution in a class by itself is the hadronic light by light one. 
So we split
$$a(\hbox{Other hadronic, $O(\alpha^3)$})=a(\hbox{`One blob' hadronic, $O(\alpha^3)$})+
a(\hbox{Hadronic light by light}).\equn{(5.1)}$$

\topinsert
{
\setbox0=\vbox{\hsize9.truecm{\epsfxsize 7.4truecm\epsfbox{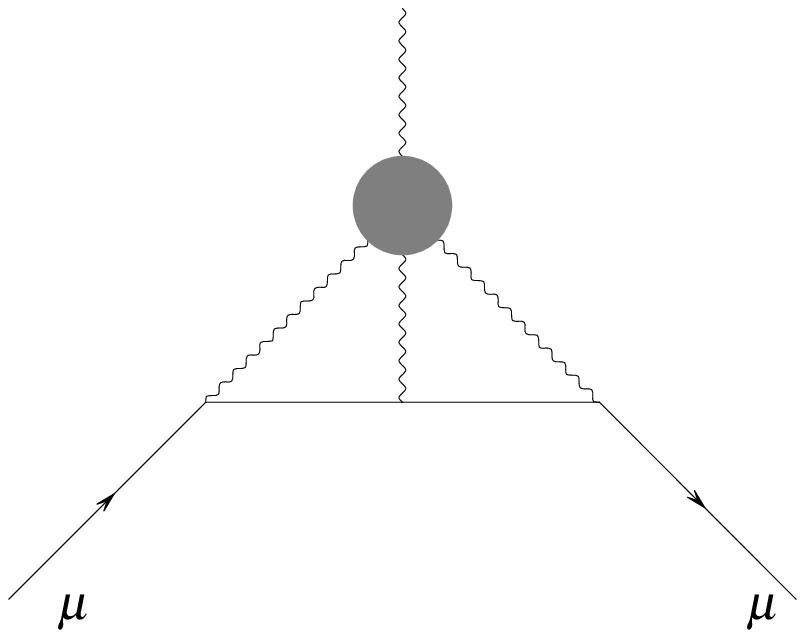}}} 
\setbox6=\vbox{\hsize 6.truecm\captiontype\figurasc{Figure 11. }{The 
hadronic light by light contributions to the muon magnetic moment.}\hb
\vskip.1cm} 
\medskip
\line{\box0\hfil\box6}
\medskip
}
\endinsert

We will start by considering the last, given diagrammatically by the graph of \fig~11.
This can be evaluated  only using {\sl models}. 
One can make a chiral model calculation, in the Nambu--Jona-Lasinio
 version or the chiral perturbation theory 
variety, with a cut-off, or one can use a constituent quark model 
in which we replace the blob in \fig~11 by a quark loop (\fig~12). 
The result depends on the cut-off (for the chiral calculation) or on the constituent mass chosen 
for the quarks.
After the correction of a sign error in the 
evaluations of ref.~24 by M.~Knecht and A.~Nyffeler (hep-ph/0111058), 
confirmed in M.~Hayakawa and T.~Kinoshita (hep-ph/0112102) we have  
$$10^{11}\times a(\hbox{Hadronic light by light})=
86\pm25\quad \hbox{Chiral calculation; BPP, HKS}.
\equn{(5.2a)}$$ 
Earlier calculations with the chiral model, using VMD to cure its divergence, gave 
(HKS, ref.~24)
$$10^{11}\times a(\hbox{Hadronic light by light})=
52\pm20\quad \hbox{Chiral calculation (HKS)}.
\equn{(5.2b)}$$

For the constituent quark model we use the results
 of Laporta and Remiddi.\ref{25} The contribution to $a_\mu$ of light by light scattering, 
with a loop with a fermion of  charge $Q_i$, and mass $m_i$ larger than the muon mass, is

$$a_{l\times l;\,i}=Q_i^4\left({\alpha}\over{\pi}\right)^3
 c_{l\times l,\,i},$$ 
where, to order $(m_\mu/m_i)^4$,

$$\eqalign{c_{l\times
l,\,i}=&\left(\dfrac{m_\mu}{m_i}\right)^2\left[\tfrac{3}{2}\zeta(3)-\tfrac{19}{16}\right]\cr
+\left(\dfrac{m_\mu}{m_i}\right)^4&\left[-\tfrac{161}{810}\log^2\dfrac{m_i}{m_\mu}
-\tfrac{16189}{48600}\log\dfrac{m_i}{m_\mu}
+\tfrac{13}{18}\zeta(3)-\tfrac{161}{9720}\pi^2-\tfrac{831931}{972000}\right]+\cdots.
\cr}$$

Taking constituent masses,
$$m_{u,d}=0.33,\quad m_s=0.50,\quad m_c=1.87\quad \gev$$
we find
$$10^{11}\times a(\hbox{Hadronic light by light})=
46\pm10\quad \hbox{(Quark const. model)}
\equn{(5.2c)}$$ 
and the error is estimated by varying $m_{u,d}$ by 10\%.

\midinsert{
\setbox0=\vbox{\hsize9.truecm{\epsfxsize 7.4truecm\epsfbox{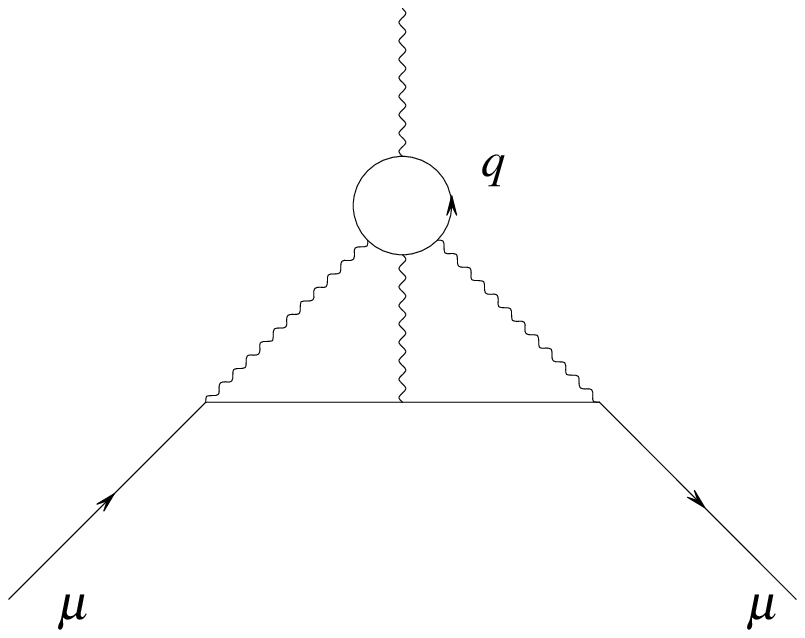}}} 
\setbox6=\vbox{\hsize 6.truecm\captiontype\figurasc{Figure 12. }{The 
light by light hadronic correction in the constituent quark model.}\hb
\vskip.1cm} 
\medskip
\line{\box0\hfil\box6}
\medskip
}\endinsert 

One could also take the estimate of the $\pi^0$ pole  from 
Hayakawa, Kinoshita and Sanda\ref{24} {\sl and} add  
the constituent quark loop, in which case we get
$$10^{11}\times a(\hbox{Hadronic light by light})=
98\pm22\quad \hbox{(Quark const. model+ pion pole)}: 
\equn{(5.2d)}$$ 
one expects the chiral calculation to be valid for small 
values of the virtual photon momenta, 
and  the constituent model to hold for large values of the same.\fnote{Strictly 
speaking, one would also need large momentum of the  external photon to get 
a really trustworthy 
evaluation with the 
constituent model.} 
Indeed,  almost a half of the contribution to $a(\hbox{Hadronic light by light})$ 
in the chiral calculation comes from a region of momenta above $0.5$ GeV, 
where the chiral perturbation theory starts to fail,
while for this range of energies, and 
at least for the imaginary part of 
(diagonal) light by light scattering,  
 the quark model reproduces reasonably well the experimental data 
(see for example ref.~26 for a recent review of this).

We will take here the figure, which comprises the relevant determinations,
$$10^{11}\times a(\hbox{Hadronic light by light})= 92\pm20 
\equn{(5.2e)}$$

\booksubsection{5.2. Photon radiation corrections to the hadronic 
vacuum polarization}
The  
$a(\hbox{`One blob' hadronic, $O(\alpha^3)$})$ corrections 
 are obtained by attaching a photon or fermion loop to the various lines in \fig~1. 
They can be further split into two pieces: the piece where both ends 
of the photon line are  attached  to the 
hadron blob, $a({\rm h.v.p.},\;\gamma)$, shown in \fig~13, and the rest. So we write,
$$a(\hbox{`One blob' hadronic, $O(\alpha^3)$})=a({\rm h.v.p.},\;\gamma)+
a(\hbox{`One blob' hadronic, rest}).\equn{(5.3)}$$
The last can be evaluated \ref{27} in terms of 
the hadronic contributions to the photon vacuum polarization, finding
$$10^{11}\times a(\hbox{`One blob' hadronic, rest})=-101\pm6.\equn{(5.4)}$$
(Note, however, that this result has not, as far as we know, been checked by 
an independent calculation).

\topinsert
{
\setbox0=\vbox{\hsize9.truecm{\epsfxsize 7.truecm\epsfbox{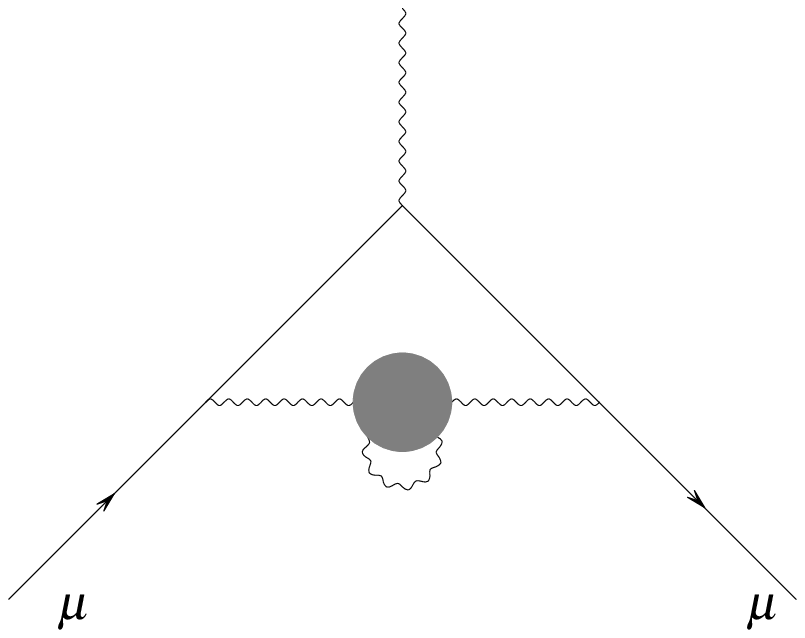}}} 
\setbox6=\vbox{\hsize 6.truecm\captiontype\figurasc{Figure 13. }{The
$O(\alpha^3)$ hadronic\hb
 correction $a({\rm h.v.p.},\;\gamma)$.}\hb
\vskip.1cm} 
\medskip
\line{\box0\hfil\box 6}
\medskip
}\endinsert
 
The only contribution that requires further discussion is 
that depicted in \fig~13, $a({\rm h.v.p.},\;\gamma)$. 
In principle, this contribution can be evaluated straightforwardly 
by a generalization of the Brodsky--de~Rafael method. 
We can write
$$a^{(2)}({\rm h.v.p.})+a({\rm h.v.p.},\;\gamma)=\int_{4m^2_\pi}^\infty \dd t\,K(t)R^{(2)}(t),
\equn{(5.5)}$$
where 
$$R^{(2)}(t)=
\dfrac{\sigma^{(0)}(e^+e^-\to {\rm hadrons})+\sigma^{(2)}(e^+e^-\to {\rm hadrons})+
\sigma^{(0)}(e^+e^-\to {\rm hadrons};\,\gamma)}{\sigma^{(0)}(e^+e^-\to\mu^+\mu^-)}.$$
The notation means that we evaluate the hadron annihilation cross section to second order in 
$\alpha$, and we add to it the first order annihilation into hadrons plus a photon.

 For energy ($t$) large enough, 
this can be calculated with the parton model, and 
leads to a correction 
 $\tfrac{3}{4}(\sum_fQ^4_f/\sum_fQ^2_f)\alpha/\pi$
 times the parton model evaluation. 
Taking then  $t\geq1.2\,\gev^2$, this is  
$(0.76\pm0.04)\times10^{-11}.$
The error is that due to $\lambdav$ and the masses of $c,\,b$ quarks. 
We have excluded the contribution of the radiative decays 
 of the $J/\psi,\,\psi'\,\upsilonv,\,\upsilonv'$ resonances 
since we have taken these into account already (we took the full $e^+e^-$ width for them).

\topinsert
{
\setbox0=\vbox{\hsize9.8truecm{\epsfxsize 8.truecm\epsfbox{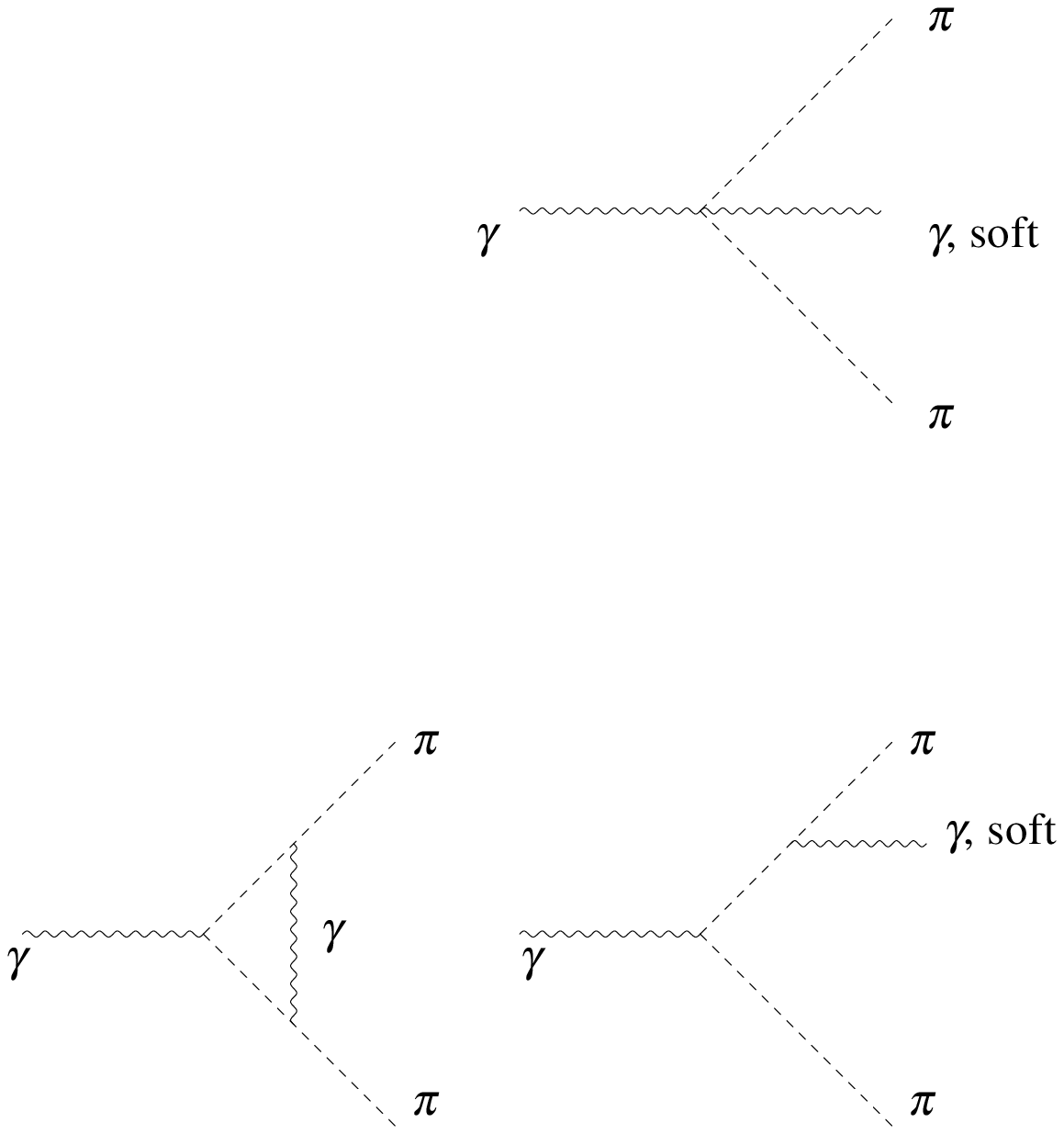}}\hfil}
\setbox1=\vbox{\hsize6.6truecm{\epsfxsize 5.truecm\epsfbox{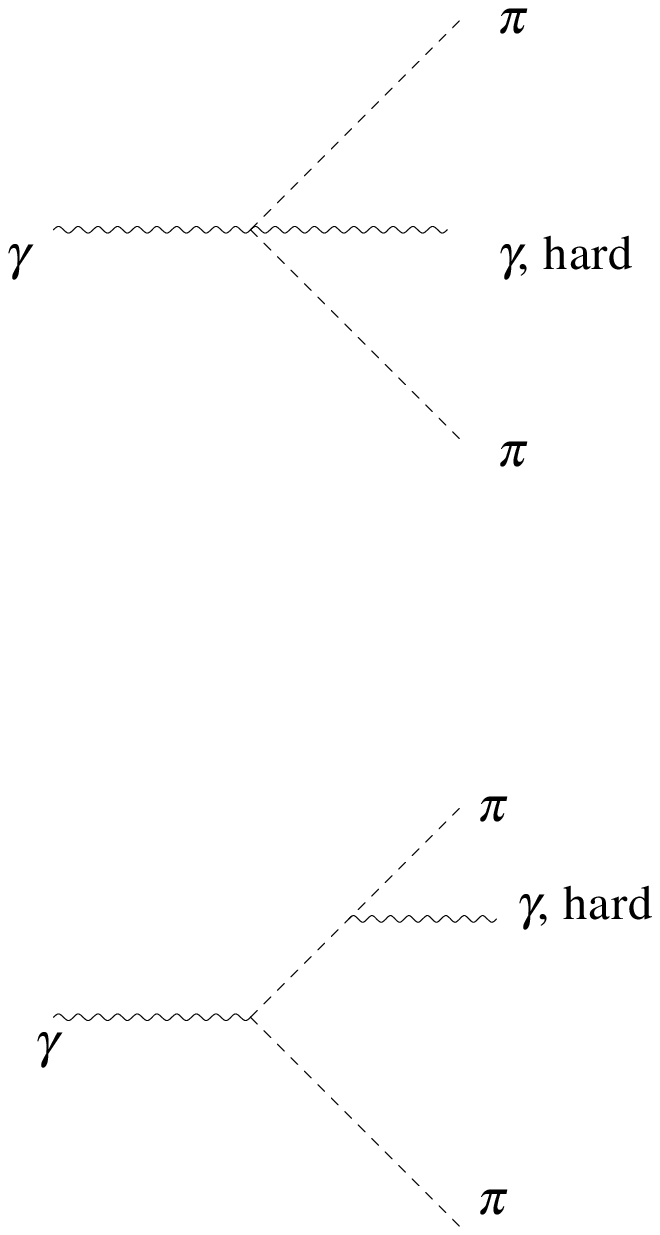}}\hfil} 
\setbox6=\vbox{\hsize 5.truecm\captiontype\figurasc{Figure 14B. }{Diagrams 
{\sl not}\hb included in the pion form factor.}\hb
\vskip.1cm} 
\setbox5=\vbox{\hsize 7.truecm\captiontype\figurasc{Figure 14A. }{Diagrams
 included in the pion form factor.}\hb
\vskip.1cm} 
\line{\kern-1truecm\box0\hfil\box1}
\line{\phantom{x}\kern1truecm\box5\hfil\box 6}
\medskip
}
\endinsert

Then comes the  contribution of small 
momenta, $t\leq 1.2\,\gev^2$. 
We start by discussing the process involving two pions.  
In our determination in \sects~3,~4 of $a^{(2)}({\rm h.v.p.})$,
we made calculations  by fitting the experimental cross section $e^+e^-\to\pi^+\pi^-$, 
which specifically excludes 
radiation of {\sl hard} photons (hard photons defined as those that
 are identified experimentally).
  Diagrammatically, this means that our evaluations of 
\sects~3, 4 included the diagrams of 
\fig~14A (where a soft photon is one that 
is not detected), but not those of \fig~14B (radiation of a hard photon). 
So, we have to include this radiation 
into $a({\rm h.v.p.},\;\gamma)$. 
This can be easily done if we consider this region to be dominated by the rho, 
hence we approximate 
$$\sigma^{(0)}(e^+e^-\to {\rm hadrons};\,\gamma)
\simeq \sigma^{(0)}(e^+e^-\to(\rho)\to\pi^+\pi^-\gamma).$$
The last can be evaluated in terms of the branching ratio for the decay 
$\rho\to\pi^+\pi^-\gamma$, which is indeed measured experimentally 
(see the review of Dolinsky et al., ref.~20)
from the reaction $e^+e^-\to\rho\to\pi^+\pi^-\gamma$. 
In the narrow width approximation for the rho, 
the contribution to $a_\mu$ is 
$$\dfrac{\gammav(\rho\to\pi^+\pi^-\gamma)}{\gammav_\rho}\,
\dfrac{3\gammav_{ee}(\rho)\hat{K}(m^2_\rho)}{\pi m_\rho}.\equn{(5.6a)}$$
In this way, we find   
$$10^{11}\times a({\rm h.v.p.},\;\pi^+\pi^-\gamma)=45\pm 7\quad \hbox{(n.w.a.)},\equn{(5.6b)}$$
and the error is that induced 
by the experimental error in the  width  $\gammav(\rho\to\pi^+\pi^-\gamma)$.

We will elaborate a bit more on  this contribution.
The final state interaction of the 
$\pi^+\pi^-$ in the state $\pi^+\pi^-\gamma$ is very strong. 
The pions are produced in an S-wave, which presents a wide enhancement\ref{15}
 in the energy region 
$E_{\pi^+\pi^-}\simeq0.6\pm0.2\;\gev$. 
However, this is only a small part of the 
contribution to the rate for $\pi\pi\gamma$. 
According to Dolinsky et al.,\ref{20} pp.~126~ff., most of the effect would be 
due to Bremsstrahlung by the pions. 
Above procedure to estimate this, in terms of the $\pi^+\pi^-\gamma$ decay 
of the $\rho$ would be exact only if the experimental cuts made for 
identifying this decay, and to measure the pion form factor were the same. 
A more accurate procedure is as follows.
We write
$$a(\pi^+\pi^-\gamma,\;t\leq1.2)=\int_{4m^2_\pi}^{1.2}\dd t\,
K(t)R_{\pi^+\pi^-\gamma}(t)
\equn{(5.7a)}$$
where
$$R_{\pi^+\pi^-\gamma}(t)=B(t,E_\gamma)R_{\pi^+\pi^-}(t)$$
and the Bremsstrahlung factor $B$ is given by\ref{28}
$$\eqalign{B(t,E_\gamma)=&\;\dfrac{8t^{1/2}\alpha}{\pi(t-4m^2_\pi)^{3/2}}
\int^{k_m}_{E_\gamma} \dfrac{\dd k}{k}I(k),\cr
I(k)=&\;k_m\left(\dfrac{t-2m^2_\pi}{2t^{1/2}}-k\right)\log\dfrac{1+\xi}{1-\xi}
-\left[k_m(\tfrac{1}{2}t^{1/2}-k)-k^2\right]\xi.
\cr}
\equn{(5.7b)}$$
Here $\xi=[(k_m-k)/(\tfrac{1}{2}t^{1/2}-k)]^{1/2}$,  $k_m=(t-4m^2_\pi)/2t^{1/2}$ is the maximum 
energy of the photon and, finally, $E_\gamma$ is the minimum energy for photon detection.

To evaluate $E_\gamma$, we have to look at the setup of experiments measuring 
the pion form factor. Typically, one takes that no (hard) photon has been emitted 
when the angle between the pion momenta differs from $\pi$ by less than a small 
given amount, $\eta_0$. 
The energy cut is, in this case, 
$$E_\gamma=\dfrac{\sqrt{t-4m^2_\pi}}{2}\,\eta_0.$$
The effective $\eta_0$ depends on the specific cuts made in each experiment; 
those in ref.~8 are covered if we take $\eta_0=0.15\pm0.05$. 
Using this we find the result, for this range of $\eta_0$,  of
$$10^{11}\times a(\pi^+\pi^-\gamma,\;t\leq1.2)=46\pm0.5\pm9.
\equn{(5.8)}$$ 

The first error corresponds 
to the error in the integral 
of $|F_\pi|^2$ and the second 
is induced by the dispersion in the values of $\eta_0$ of the various 
experiments. 
(5.8) is practically identical to the n.w.a. result, \equn{(5.6b)} (which of course is a satisfactory 
result). 
The reason for this agreement is that, 
when {\sl detecting} $\pi^+\pi^-\gamma$, the energy cut 
made, $E_\gamma=50\;\mev$ (Dolinsky et al., ref.~20), turns out to be very similar to the average 
energy cut made when measuring the pion form factor.

A similar analysis ought  to be made, in principle, for other 
radiative intermediate states like $3\pi+\gamma$ and $KK+\gamma$, 
which can be estimated in terms of the corresponding decays of 
the $\omega$ and $\phi$, but they give a contribution 
below the $10^{-11}$ level and we neglect them.

The  lowest energy contributions to $\sigma^{(0)}(e^+e^-\to\,{\rm hadrons};\gamma)$ 
are those of the intermediate states $\pi^0\gamma$ and $\eta\gamma$, 
\fig~15. At energies below the rho mass, one can evaluate the first (the only one 
that gives a sizable contribution) by relating 
the process to the 
decay $\pi^0\to2\gamma$. 
We write an effective interaction, corresponding to the vertex factor in the Feynman rules of
$$\dfrac{\ii G_\pi}{2m_\pi}\epsilon_{\mu\nu\alpha\beta}k_1^\alpha k_2^\beta;$$
with it 
$$\gammav(\pi^0\to2\gamma)=\dfrac{G^2_\pi m_\pi}{256\pi}$$
so that $G^2_\pi=4.6\times10^{-5}$. 
Then, with $e$ the electron charge,
$$\imag \piv^{\pi\gamma}(t)=
\dfrac{G^2_\pi t}{384\pi e^2 m^2_\pi}\left(1-\dfrac{m^2_\pi}{t}\right)^3.$$ 
This gives a very small contribution to $a_\mu$, about $0.76\times10^{-11}$ 
if we integrate up to $t^{1/2}\simeq 0.7\,\gev$, and  $0.96\times10^{-11}$ 
if we go  to $t^{1/2}\simeq 0.84\,\gev$ 
(the integral only grows logarithmically). 
We only integrate up to the rho, i.e., to $t^{1/2}=0.7\;\gev$ with this pointlike model.

\topinsert 
{
\setbox0=\vbox{\hsize8.2truecm{\epsfxsize 6.7truecm\epsfbox{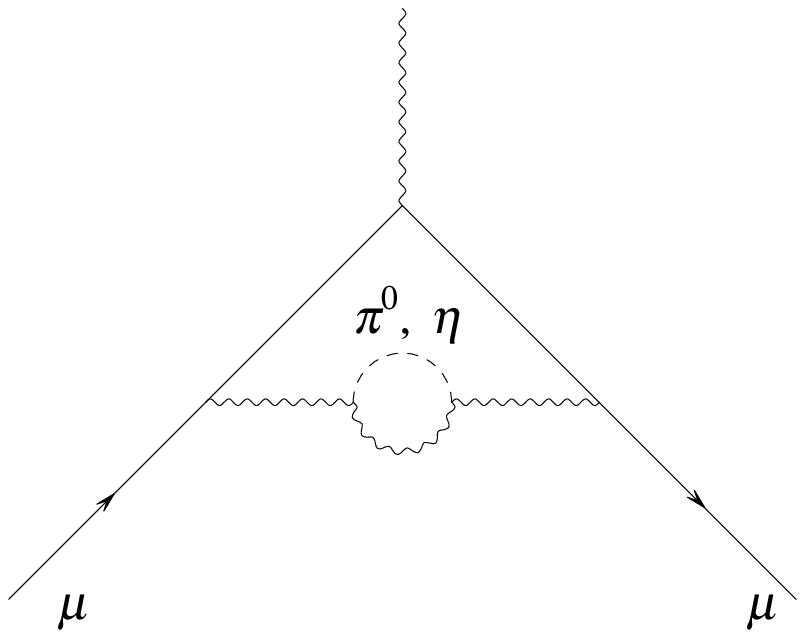}}} 
\setbox6=\vbox{\hsize 5.5truecm\captiontype\figurasc{Figure 15. }{The 
$\pi^0\gamma$, $\eta\gamma$  contribution to $a({\rm h.v.p.},\;\gamma)$.}\hb
\vskip.1cm} 
\medskip
\line{\box0\hfil\box6}
\medskip
}
\endinsert
 
Around the $\rho$ region we have to take into account 
the excitation of this resonance, which produces 
 the corresponding enhancement. 
This piece can be obtained in terms of the radiative width $\rho\to\pi^0\gamma$. 
More important is the $\omega\to\pi^0\gamma$ 
process which gives $(33\pm2)\times10^{-11}$. 
Likewise, the contribution of the $\eta\gamma$ state 
is evaluated in terms of the decay $\phi\to\eta\gamma$. 
Finally, the contribution from $\pi^0\pi^0\gamma$ is taken from ref.~29. 
Collecting all of this, we  get
$$\eqalign{
10^{11}\times a({\rm h.v.p.},\;\rho\to\pi^0\gamma)=&\;4\pm1\cr
10^{11}\times a({\rm h.v.p.},\;\rho\to\eta\gamma)=&\;1.1\pm0.4\cr
10^{11}\times a({\rm h.v.p.},\;\omega\to\pi^0\gamma)=&\;33\pm2\cr
10^{11}\times a({\rm h.v.p.},\;\phi\to\eta\gamma)=&\;5\pm1;\cr
\hbox{\sc Total}:\quad&\;43\pm4\cr}
\equn{(5.9a)}$$
we have included the lower energy contribution of $\pi^0\gamma$ into 
$a({\rm h.v.p.},\;\rho\to\pi^0\gamma)$ and, 
because we are relying on models, we  added the errors {\sl linearly}.
For the $\pi\pi\gamma$ states, 
\smallskip
$$\eqalign{
10^{11}\times a({\rm h.v.p.},\;\pi^+\pi^-\gamma)=&\;46\pm9\cr
10^{11}\times a({\rm h.v.p.},\;\pi^0\pi^0\gamma)=&\;2\pm0.3;\cr
}
\equn{(5.9b)}$$
and, for the high energy piece,
$$\eqalign{
10^{11}\times a({\rm hadrons}+\gamma,\;t\geq1.2\,\gev^2)=&
1\pm0.5.\cr 
\cr}
\equn{(5.9c)}$$
Adding other contributions  that are below the $10^{-11}$ level
($\epsilon(700)\gamma\sim0.7\times 10^{-11}$, etc.) we get the total 
  effect of the states ${\rm hadrons}+\gamma$,
$$10^{11}\times a({\rm hadrons}+\gamma)=93\pm11.
\equn{(5.10)}$$

\booksection{6. Conclusions}
We present first, for ease of reference, Table~4 with a summary of the 
results obtained for $a{\rm (h.v.p.)}$ in 
the previous sections. 

\topinsert{
\setbox0=\vbox{
\medskip
\setbox1=\vbox{\petit \offinterlineskip\hrule
\halign{
&\vrule#&\strut\hfil\quad#\quad\hfil&\vrule#&\strut\hfil\quad#\quad\hfil&
\vrule#&\strut\hfil\quad#\quad\hfil&\vrule#&\strut\hfil\quad#\quad\hfil\cr
 height2mm&\omit&&\omit&&\omit&&\omit&\cr 
&\hfil Channel\hfil&&\hfil Energy range\hfil&
&\hfil Method of calculation\hfil&
&\hfil ${\hbox{Contribution to}\atop{\textstyle 10^{11}\times a({\rm h.v.p.})}}$\hfil& \cr
 height1mm&\omit&&\omit&&\omit&&\omit&\cr
\noalign{\hrule} 
height1mm&\omit&&\omit&&\omit&&\omit&\cr
&$\pi^+\pi^-$&&\vphantom{\Big|}$t\leq 0.8\,\gev^2$&
&\hfil Fit to $e^+e^-\,+\,\tau\,+\,{\rm spacel.}$ data\hfil&&
$4\,774\pm51$& \cr
\noalign{\hrule} 
height1mm&\omit&&\omit&&\omit&&\omit&\cr
&\vphantom{\Big|}$\pi^+\pi^-$ 
\phantom{\big|}&&\phantom{\Big|}$0.8\leq t\leq 1.2\,\gev^2$&
&\hfil Fit to exp. $e^+e^-$ data \hfil&&$230\pm5$& \cr
\noalign{\hrule} 
height1mm&\omit&&\omit&&\omit&&\omit&\cr
&$3\pi$&&\vphantom{\Big|}$t\leq 1.2\,\gev^2$&
&\hfil B.--W. + const. fit to $e^+e^-$  data\hfil&&$438\pm12$& \cr
\noalign{\hrule} 
height1mm&\omit&&\omit&&\omit&&\omit&\cr
&\vphantom{\Big|}$2K$&&$t\leq 1.2\,\gev^2$&
&\hfil B.--W. + const. fit to $e^+e^-$  data\hfil&&$314\pm13$& \cr
\noalign{\hrule}
height1mm&\omit&&\omit&&\omit&&\omit&\cr
&\vphantom{\Big|}$4\pi,\;5\pi,\;\eta\pi,\dots$&
&$t\leq 1.2\;\gev^2$&&\hfil Fit to $e^+e^-$  data \hfil&&$24\pm5$& \cr
\noalign{\hrule}
height1mm&\omit&&\omit&&\omit&&\omit&\cr
&\vphantom{\Big|}Inclusive&&$1.2\,\gev^2\leq t\leq2\,\gev^2$&
&\hfil  Fit to $e^+e^-$  data\hfil&&$270\pm27$& \cr
 height1mm&\omit&&\omit&&\omit&&\omit&\cr
\noalign{\hrule}
height1mm&\omit&&\omit&&\omit&&\omit&\cr
&\vphantom{\Big|}$J/\psi,\,\psi';\,\upsilonv,\,\upsilonv'$ && &&\hfil N.w.a. \hfil&&$77.5\pm4.4$&
\cr
\noalign{\hrule}height1mm&\omit&&\omit&&\omit&&\omit&\cr
&\vphantom{\Big|}Inclusive&&$3.7^2\,\gev^2\leq t\leq 4.6^2\,\gev^2$&
&\hfil Fit to experimental $e^+e^-$  data \hfil&&$56\pm3$& \cr
\noalign{\hrule}height1mm&\omit&&\omit&&\omit&&\omit&\cr
&\vphantom{\Big|}Inclusive; $uds$ &&$2\,\gev^2\leq t\leq 3.7^2\gev^2$&
&\hfil Perturbative QCD \hfil&&$615\pm9$& \cr
\noalign{\hrule}height1mm&\omit&&\omit&&\omit&&\omit&\cr
&\vphantom{\Big|}Inclusive; $udsc$ &&$4.6^2\,\gev^2\leq t\leq 11.2\,\gev^2$&
&\hfil  Perturbative QCD \hfil&&$89\pm1$& \cr
\noalign{\hrule}height1mm&\omit&&\omit&&\omit&&\omit&\cr
&\vphantom{\Big|}$b$ quark&&$10.1^2\,\gev^2\leq t\leq 11.2^2\,\gev^2$&
&\hfil Nonrelativistic QCD \hfil&&$0.5\pm0.1$& \cr
\noalign{\hrule}height1mm&\omit&&\omit&&\omit&&\omit&\cr
&\vphantom{\Big|}Inclusive&&$11.2^2\,\gev^2\leq t\leq \infty$&
&\hfil Perturbative QCD \hfil&&$21\pm0.1$& \cr
\noalign{\hrule}height1mm&\omit&&\omit&&\omit&&\omit&\cr
&\vphantom{\Big|}$\gamma+$ hadrons&&\hfil Full range\hfil&&\hfil Various methods \hfil&&$93\pm11$& \cr
\noalign{\hrule}}
\vskip.05cm}
\centerline{\box1}
\smallskip
\centerline{\petit Table~4}
\centerrule{6cm}
\medskip
\noindent{Summary of contributions to $a^{(2+3)}$, with what we consider the 
more reliable methods,  
as used in the present work. ``B.--W. + const." means a Breit--Wigner fit, including the correct 
phase space factors, plus a constant; note that only for the four narrow resonances
$J/\psi,\,\psi';\,\upsilonv,\,\upsilonv'$ 
we use the n.w.a. The errors are 
uncorrelated except those for QCD calculations 
(that have to be added linearly) and those for the $2\pi$ states, 
for whose treatment we refer to the text. 
The errors given include statistical, systematic
and (estimated) theoretical errors.  For the details of the final states $\gamma+$ hadrons
 we refer to \eqs~(5.9).}
\medskip}
\box0
\bigskip
}\endinsert

Taking into account  all  
contributions, and errors,  we 
complete the best values for the h.v.p. piece, and the whole hadronic 
part of the anomaly:
$$10^{11}\times a^{(2+3)}({\rm h.v.p.})=\cases{
7\,002\pm66\quad(e^+e^-\,+\,\tau\,+\,{\rm spacel.})
\cr
6\,982\pm97\quad (e^+e^-\,+\,{\rm spacel.}),\cr
}
\equn{(6.2)}$$
and adding the other radiative and light-by-light corrections,
$$10^{11}\times a(\hbox{Hadronic})=
\cases{
6\,993\pm69\quad(e^+e^-\,+\,\tau\,+\,{\rm spacel.})\cr
6\,973\pm99\quad(e^+e^-\,+\,{\rm spacel.})\cr
}
\equn{(6.3)}$$ 
\equn{(6.3)} is of course the main outcome of the present paper. 
Because, even after adding systematic and theoretical errors 
the evaluation including $\tau$ decay data is more precise, 
we may consider it to provide the best result for $a(\hbox{Hadronic})$  available at present.

We can add to (6.3) the pure electroweak contributions and present the result 
as the standard model prediction for $a_\mu$: 
$$10^{11}\times a_\mu=
116\,591\,849\pm69\quad(e^+e^-\,+\,\tau\,+\,{\rm spacel.})
\equn{(6.4)}$$

\topinsert{ 
\setbox0=\vbox{\hsize16.4truecm
\epsfxsize 12truecm\epsfbox{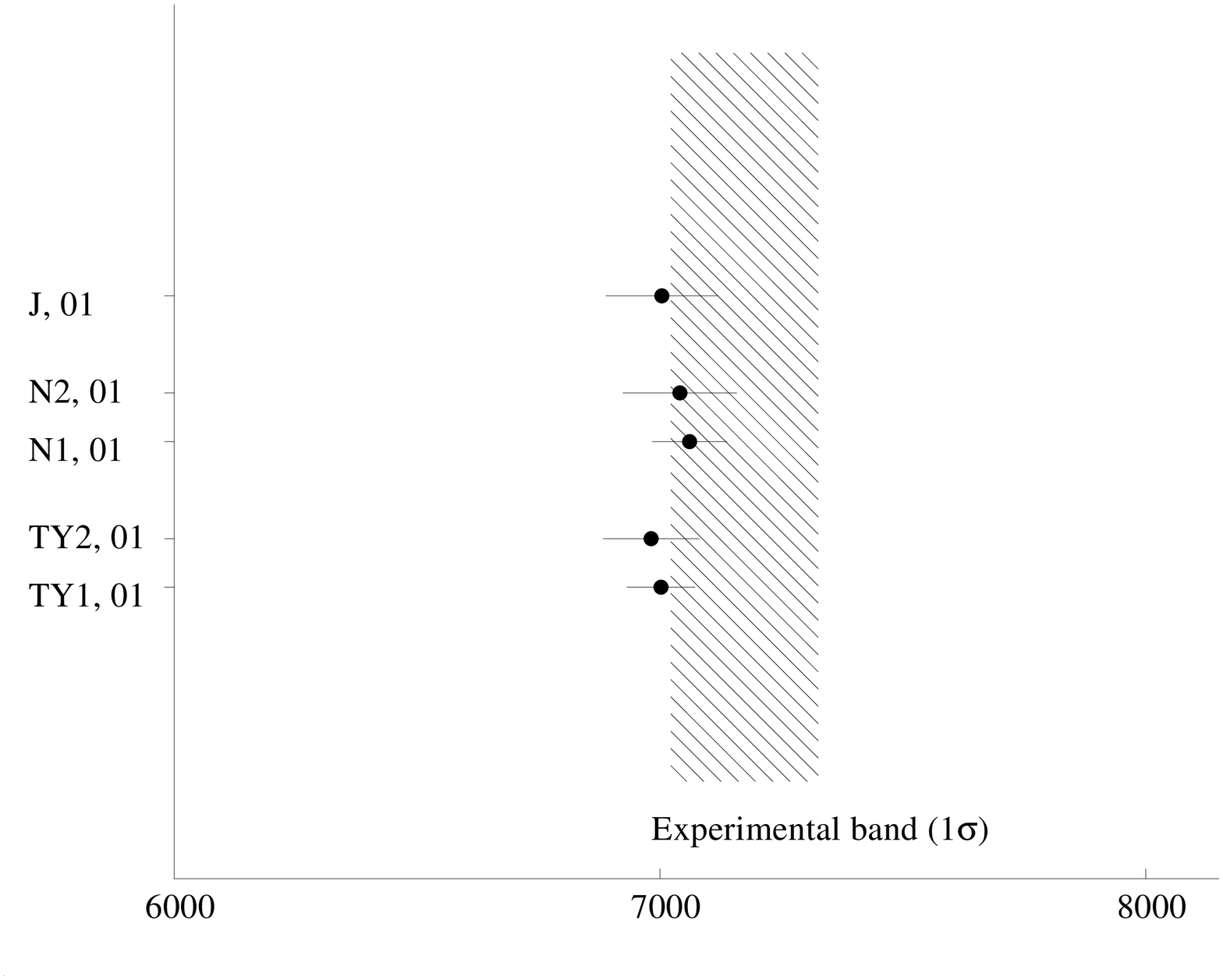}}
\centerline{\box0}
\setbox1=\vbox{\hsize13  truecm\figurasc{Figure 16. }{Theoretical results on 
 $a({\rm Hadronic})\times10^{-11}$, and experiment.}}
\bigskip
\centerline{\box1}
\centerrule{5truecm}
\medskip
{\petit\noindent J: ref.~6: N1: ref.~7, data from $e^+e^-\,+\,\tau$. 
N2: id., $e^+e^-$ only. T1, T2: this paper with data from $e^+e^-\,+\,\tau$ 
or data from $e^+e^-$ only, respectively (including syst. and th. 
errors).}  
}\endinsert   
\bigskip  

We will next compare our results with other recent evaluations of the 
same quantities 
in \fig~16, 
together with the experimental band.  
(They  are shown incorporating the 
contribution of the $\pi^+\pi^-\gamma$ and $\pi^0\pi^0\gamma$ channels).

A further point to emphasize is the importance of using, in the low energy region, 
parametrizations of $F_\pi(t)$ compatible with unitarity and analyticity. 
Only in this way we can incorporate data on $F_\pi(t)$ for spacelike $t$ 
into the fits. 
As discussed in \subsect~3.2, these data not only provide 
a substantial shift for $a_\mu$ (of $39\times10^{-11}$ in the 
evaluation with $e^+e^-$ data only) 
but, by so doing, allow compatibility of these with the results from $\tau$ decay, 
hence allowing a combination of the two in a meaningful way: 
this permits an important reduction of the errors of the calculation.\fnote{Or, 
put conversely, not using data at spacelike $t$ for $F_\pi$ implies a hidden error 
of about $40\times10^{-11}$.}

To finish this section, we can add a few words on prospects for improvements. 
In our view they are rather dim 
in the sense that it is not easy to see how one could 
get an error estimate clearly below the $70\times10^{-11}$ mark, 
when taking into account systematic and theoretical errors. 
In fact, the central values have 
moved little, and the errors have 
not improved much, since 1985. 
It is true that experiments planned or in progress can clear further the 
region between $1.2$ and $3\;\gev^2$. 
However, a serious improvement of the very important low energy region 
 for  
 $\pi\pi$ is unlikely:  as our evaluations show, 
one can get a fit to {\sl all} data relevant for the hadronic component of 
$a_\mu$, with a \chidof\ of unity
and verifying {\sl all} theoretical constraints,  
with an error of at least $51\times10^{-11}$, already for $t\leq0.8\,\gev^2$
(see \equn{(3.19)}). 
In this respect the improvement obtained by adding $\tau$ decay data, although not negligible,  
is minor: 
statistical errors are smaller, but  
theoretical ones are increased.

\vfill\eject

\booksection{Acknowledgments}
The authors are indebted to G.~Cvetic, F.~Jegerlehner and S.~Narison for
 several illuminating discussions and, for the last two, for information concerning 
 their work on the subject. 
Discussions with J.~Vermaseren (who helped with 
a critical reading of the manuscript) are 
acknowledged for  several matters; in particular, 
those concerning the $O(\alpha^3)$ hadronic 
contributions have  been very helpful.

It is also a pleasure to record our indebtedness to 
Achim Stahl (Opal) and Andreas H\"ocker (Aleph) for providing us with 
the corresponding data listings, and for very useful information about them. 
Thanks are also due to S.~I.~Eidelman for the same reason ($4\pi$ data).

 The financial support of CICYT is  gratefully acknowledged.
\booksection{References}
{\petit
\item{1.-}{\prajnyp{H.N. Brown et al.}{Phys. Rev. Letters}{86}{2001}{2227}.}
\item{2.-}{\prajnyp{J. Bailey et al.}{Nucl. Phys.}{B150}{1979}{1}.}
\item{3.-}{(KNO): \prajnyp{T. Kinoshita, B. Nizi\'c and Y. Okamoto}{Phys. Rev.}{D32}{1985}{736}; 
(DM): \prajnyp{S. Dubnicka and L.~Martinovic}{Phys. Rev.}{D42}{1990}{7884}; 
(BW): \prajnyp{D. H. Brown and W. A. Worstell}{Phys. Rev.}{D54}{1996}{3237};.}
\item{4.-}{(CLY): \prajnyp{A. Casas, C. L\'opez and F. J. Yndur\'ain}{Phys. Rev.}{D32}{1985}{736}; 
(AY): \prajnyp{K. Adel and F.~J.~Yndur\'ain}{Rev. Acad. Ciencias (Esp.)}{92}{1998}{113} 
(hep-ph/9509378).}
\item{5.-}{(ADH): \prajnyp{R. Alemany, M. Davier and A. H\"ocker}{Eur. Phys. J.}{C2}{1998}{123};
\prajnyp{(DH): M. Davier and A.~H\"ocker}{Phys. Letters}{B435}{1998}{419}.}
\item{6.-}{(EJ): \prajnyp{S. Eidelman and F.~Jegerlehner}{Z. Phys.}{C67}{1995}{585}; 
(J): {\sc F.~Jegerlehner,} DESY 01-028 (hep-ph/0104304), 
 Symposium in honor of A.~Sirlin, New York University, 2000.}
\item{7.-}{\/(N): {\sc S. Narison}, PM/01-13 (hep-ph/0103199).}
\item{8.-}{\/Novosibirsk, $\rho$ region:\prajnyp{L.~M.~Barkov et al.}{Nucl. Phys.}{B256}{1985}{365}; 
 {\sc R. R. Akhmetsin et al.} Budker INP 99-10 (hep-ex/9904027)} 
\item{9.-}{\/Novosibirsk, $\omega$ and $\phi$ region, $KK$ and $3\pi$: (CMD2)
  \prajnyp{R. R. Akhmetshin et al.}{Phys. Letters}{B466}{1999}{385 and 392}; 
ibid., {\bf B434} (1998) 426 and ibid., {\bf B476} (2000) 33. (SND) 
 \prajnyp{M.~N.~Achasov et al.}{Phys. Rev.}{D63}{2001}{072002};    
{\sc M.~N.~Achasov et al.,} {\sl Phys. Letters} {\bf B462} (1999) 365 and   
 Preprint Budker INP 98-65, 1998 (hep-ex/9809013). 
$\phi\to2\pi$:  \prajnyp{M.~N.~Achasov et al.}{Phys. Letters}{B474}{2000}{188}.}
\item{10.-}{Aleph: \prajnyp{R. Barate et al.}{Z.Phys.}{C76}{1997}{15}; 
Opal: \prajnyp{K.~Ackerstaff et al.}{Eur.Phys.J.}{C7}{1999}{571}.}
\item{11.-}{NA7:  \prajnyp{S. R. Amendolia et al.}{Nucl. Phys.}{B277}{1986}{168}.}
\item{12.-}{BES: {\sc J. Z. Bai et al}, hep-ex/0102003.}
\item{13.-}{\prajnyp{V. W. Hughes and T. Kinoshita}{Rev. Mod. Phys.}{71}{1999}{S133}.}
\item{14.-}{\prajnyp{S. J. Brodsky and E. de Rafael}{Phys. Rev.}{168}{1968}{1620}.}
\item{15.-}{\/{\sc B. R. Martin, D. Morgan and G.~Shaw}, {\sl 
Pion-Pion Interactions in particle Physics}, Academic Press, New York 1976.}
\item{16.-}{Particle Data Group: \prajnyp{D. E. Groom et al.}{Eur. Phys. J.}{C15}{2000}{1}.}
\item{17.-}{$F_\pi$, spacelike data: \prajnyp{C.~J.~Bebek et al.}{Phys. Rev.}{D17}{1978}{1693}; 
$F_\pi$, timelike data: 
\prajnyp{D.~Bollini et al.}{Nuovo Cimento Lett.}{14}{1977}{418}; 
\prajnyp{A.~Quenzer et al.}{Phys. Letters}{76B}{1978}{512}. 
$\pi\pi$ phase shifts: \prajnyp{B.~Hyams et al}{Nucl. Phys.}{B64}{1973}{134} and  
\prajnyp{S.~D.~Protopopescu et al.}{Phys. Rev.}{D7}{1973}{1279}.}
\item{18.-}{$a^1_1$. (CGL):  \prajnyp{G.~Colangelo,
J.~Gasser and H.~Leutwyler}{Nucl. Phys.}{B603}{2001}{125}; 
(ACGL): {\sc B.~Ananthanarayan, G.~Colangelo, J.~Gasser and H.~Leutwyler},
hep-ph/0005297;
(ABT): 
\prajnyp{G.~Amor\'os, J.~Bijnens and P.~Talavera}{Nucl. Phys.}{B585}{2000}{329} and 
Erratum, LU TP 00-11, 2001.  
$\langle r^2_\pi\rangle,\,c_\pi$: 
\prajnyp{G. Colangelo, M.~Finkelmeir and R.~Urech}{Phys. Rev.}{D54}{1996}{4403}.}
\item{19.-}{\prajnyp{G. J. Gounnaris and J. J. Sakurai}{Phys. Rev. Letters}{21}{1968}{244}.}
\item{20.-}{$KK$\/: \prajnyp{P. M. Ivanov et al.}{Phys. Letters}{107B}{1981}{297}; $3\pi$: 
\prajnyp{A.~Cordier et al.}{Nucl. Phys.}{B172}{1980}{13}; $4\pi$ and more: 
\prajnyp{G.~Cosme et al.}{Nucl. Phys.}{B152}{1979}{215}. For a review, see  
\prajnyp{S. Dolinsky et al.}{Phys. Reports}{C202}{1991}{99}, and work quoted there. 
All these references 
give results for energies below $t=2\,\gev^2$. 
Between $2$ and $9\;\gev^2$, see 
\prajnyp{C. Bacci et al.}{Phys. Letters}{86B}{1979}{234}.}
\item{21.-}{\prajnyp{S. Bethke}{J. Phys.}{G26}{2000}{R27}; 
{\sc J.~Santiago and F. J. Yndur\'ain},  FTUAM 01-01, 2001 (hep-ph/0102247);
{\sc D. Strom}, ``Electroweak measurements on the $Z$ resonance", 
Talk presented at the 5th Int. Symposium on Radiative Corrections, RadCor2000, 
Carmel, Ca., September 2000.}
\item{22.-}{\prajnyp{K. Adel and F.~J.~Yndur\'ain}{Phys. Rev.}{92}{1998}{113}. 
For a clear exposition of NRQCD calculations see \prajnyp{N.~Brambilla, A.~Pineda, 
J.~Soto and A.~Vairo}{Nucl. Phys.}{B566}{2000}{275}.}
\item{23.-}{\prajnyp{S.~Titard and 
F.~J.~Yndur\'ain}{Phys. Rev.}{D49}{1994}{6007}, to one loop;
 \prajnyp{A. Pineda and F.~J.~Yndur\'ain}{Phys. Rev.}{D58}{1998}{094022} 
and ibid., {\bf D61}, 077505 (2000), to two loops.}
\item{24.-}{\/(BPP): 
\prajnyp{J. Bijnens, E. Pallante and J. Prades}{Nucl. Phys.}{B474}{1996}{379}; 
(HKS, HK): \prajnyp{M.~Hayakawa, T.~Kinoshita and A.~I.~Sanda}{Phys. Rev.}{D54}{1996}{3137}; 
\prajnyp{M.~Hayakawa and T.~Kinoshita}{Phys. Rev.}{D57}{1998}{465}.}
\item{25.-}{\prajnyp{S. Laporta and E. Remiddi}{Phys. Letters}{B301}{1993}{440}.}
\item{26.-}{\/{\sc F.~Barreiro}, {\sl Photon Structure}, 
Plenary talk, Proc. Int. Europhysics Conf. on 
Highe Energy Physics, Tampere, 1999   (Huitu et al. Eds.). Published by IOP, Bristol, U.~K.}
\item{27.-}{\prajnyp{B. Krause}{Phys. Letters}{B390}{1997}{392}.}
\item{28.-}{\prajnyp{P. Singer}{Phys. Rev.}{130}{1963}{2441} and Erratum, 
{\bf 161}, 1694 (1967).}
\item{29.-}{\prajnyp{M. N. Achasov et al.}{Phys. Letters}{B486}{2000}{29}.}
\item{}{}
}
\bookendchapter

\bye